\documentclass[useAMS,usenatbib]{mn2e}

\usepackage[a4paper,margin=2cm]{geometry} 

\usepackage{graphicx}
\usepackage{lscape}
\usepackage{amssymb,amsmath,mathrsfs}
\usepackage{subfig}
\usepackage{color}
\usepackage{threeparttable}
\usepackage{url}

\renewcommand{\tabcolsep}{4.5pt}

\title[H\,{\normalsize \it I} in nearby galaxies II]
  {H\,{\Large\bf I} emission and absorption in nearby, gas-rich galaxies II. -- sample completion and detection of intervening absorption in NGC\,5156}
\author[S. N. Reeves et al.]
{S. N.~Reeves,$^1$$^,$$^2$$^,$$^3$\thanks{E-mail: sarah@physics.usyd.edu.au}
E. M.~Sadler,$^1$$^,$$^2$ 
J. R.~Allison,$^2$$^,$$^3$ 
B. S.~Koribalski,$^3$
\newauthor 
S. J.~Curran,$^4$ 
M. B.~Pracy,$^1$ 
C. J. Phillips,$^3$ 
H. E. Bignall,$^5$
and C. Reynolds$^3$
\\
$^1$Sydney Institute for Astronomy, School of Physics A28, The University of Sydney, NSW 2006, Australia\\
$^2$ARC Centre of Excellence for All-Sky Astrophysics (CAASTRO)\\
$^3$Australia Telescope National Facility, CSIRO Astronomy and Space Science, PO Box 76, Epping, NSW 1710, Australia\\
$^4$School of Chemical and Physical Sciences, Victoria University of Wellington, PO Box 600, Wellington 6140, New Zealand\\
$^5$International Centre for Radio Astronomy Research, Curtin University, Building 610, 1 Turner Avenue, Bentley WA 6102, Australia}
\date{Released 2016 Xxxxx XX}

\pagerange{\pageref{firstpage}--\pageref{lastpage}} \pubyear{2016}

\def\LaTeX{L\kern-.36em\raise.3ex\hbox{a}\kern-.15em
    T\kern-.1667em\lower.7ex\hbox{E}\kern-.125emX}

\begin{document}

\label{firstpage}

\maketitle

\begin{abstract}
We present the results of a survey for intervening 21cm \mbox{H\,{\sc i}} absorption in a sample of 10 nearby, gas-rich galaxies selected from the \mbox{H\,{\sc i}} Parkes All-Sky Survey (HIPASS). 
This follows the six HIPASS galaxies searched in previous work and completes our full sample. 
In this paper we searched for absorption along 17 sightlines with impact parameters between 6 and 46 kpc, making one new detection. 
We also obtained simultaneous \mbox{H\,{\sc i}} emission-line data, allowing us to directly relate the absorption-line detection rate to the \mbox{H\,{\sc i}} distribution. 
From this we find the majority of the non-detections in the current sample are because sightline does not intersect the \mbox{H\,{\sc i}} disc of the galaxy at sufficiently high column density, but that source structure is also an important factor.

The detected absorption-line arises in the galaxy NGC\,5156 ($z = 0.01$) at an impact parameter of 19 kpc. 
The line is deep and narrow with an integrated optical depth of 0.82 km s$^{-1}$. 
High resolution Australia Telescope Compact Array (ATCA) images at 5 and 8 GHz reveal that the background source is resolved into two components with a separation of 2.6 arcsec (500 pc at the redshift of the galaxy), with the absorption likely occurring against a single component. 
We estimate that the ratio of the spin temperature and covering factor, $T_{\mathrm{S}}/f$, is approximately 950 K in the outer disc of NGC\,5156, but further observations using VLBI would allow us to accurately measure the covering factor and spin temperature of the gas.
\end{abstract}


\section{Introduction}
\label{introduction}

Studies of the 21 cm transition of neutral atomic hydrogen (\mbox{H\,{\sc i}}) provide a unique view of the gas in and around galaxies (see e.g. \citealt{2001ASPC..240..657H,2008glv..book.....K,2008AJ....136.2563W}). 
While \mbox{H\,{\sc i}} emission-line studies are inhibited by a rapid drop-off in detectability with redshift, the detectability of the \mbox{H\,{\sc i}} absorption-line is essentially independent of distance (set instead by the flux of the background source used to search for absorption). 
This allows us to study neutral hydrogen to much higher redshifts than is possible in emission. 
With the arrival of the next generation of radio telescopes, including the Australian Square Kilometre Array Pathfinder \citep[ASKAP,][]{2008ExA....22..151J}, MeerKAT \citep{2012AfrSk..16..101B}, and Apertif \citep{2009wska.confE..70O}, it will become possible to conduct the first large, blind absorption-line surveys, allowing us to study the evolution of neutral hydrogen over a wide range of cosmic times. 
FLASH (`The First Large Absorption Survey in \mbox{H\,{\sc i}}') is the planned \mbox{H\,{\sc i}} absorption-line survey with ASKAP, and will search for \mbox{H\,{\sc i}} absorption along 150,000 sightlines in order to investigate the evolution of neutral hydrogen over redshifts $z = 0.5-1.0$. 
However, if we wish to derive physical galaxy properties from absorption-line data, we need to know the expected detection rate of intervening \mbox{H\,{\sc i}} absorption, and how this varies with distance from the centre of the galaxy (i.e. impact parameter).

\citet{2010MNRAS.408..849G} conducted a systematic study to address this question. 
Using a combination of new observations and available literature results, they estimate that the detection rate of \mbox{H\,{\sc i}} absorption is around 50 per cent, for impact parameters less than 20 kpc, and integrated optical depths greater than 0.1 km s$^{-1}$. 
The addition of more recent results \citep{2011ApJ...727...52B,2014ApJ...795...98B,2013MNRAS.428.2198S,2015MNRAS.453.1268Z} gives a similar overall detection rate.
While detection rates may differ somewhat between individual surveys, a common result is that detections of \mbox{H\,{\sc i}} absorption at impact parameters greater than 20 kpc appear to be extremely rare.

In contrast, detections of neutral gas through optical Lyman-$\alpha$ studies have frequently been made at impact parameters of several tens of kpc, and in some cases exceeding 100 kpc (see e.g. \citet{2011MNRAS.416.1215R} and references therein), showing that the gas disc extends well beyond what we see even in 21cm observations. 
This is only possible, however, because the detectability of Lyman-$\alpha$ absorption does not depend on the spin temperature of the gas, meaning that these studies are sensitive to very high spin temperature gas which would not be detectable in \mbox{H\,{\sc i}} absorption.

In \citealt{2015MNRAS.450..926R} (hereafter Paper I), we investigated the detection rate of intervening absorption in a sample of nearby ($z < 0.04$) gas-rich galaxies, selected from the \mbox{H\,{\sc i}} Parkes All-Sky Survey \citep[HIPASS,][]{2001MNRAS.322..486B,2004MNRAS.350.1195M,2006MNRAS.371.1855W}. 
By targeting nearby galaxies we were also able to map the \mbox{H\,{\sc i}} emission, allowing us to directly relate the absorption-line detection rate to the extended \mbox{H\,{\sc i}} distribution. 
While four of the six sightlines intersected the \mbox{H\,{\sc i}} disc, no absorption-lines were detected. 
We attributed this to the background sources becoming resolved or extended, thus reducing the continuum flux and dramatically affecting the absorption-line sensitivity.

Our detection rate in Paper I was low compared to previous surveys (although this is difficult to interpret accurately given the small sample size). 
However, while many previous surveys have targeted quasar sightlines, ours represented an unbiased sample of radio sources (a mixture of radio galaxies and quasars, in a ratio consistent with the overall radio source population). 
We therefore suggested that the differences in detection rate may be (at least partly) due to the differences in source type (since quasars are more likely to remain unresolved, and might therefore be expected to produce a higher detection rate), and that our results may be more representative of the expected detection rate for future large, blind absorption-line surveys. 
In this paper we build on our previous work, expanding our sample with an additional 10 galaxies. 
With the expanded sample we aim to better establish: (i) the influence of background source structure and source type on detection rate, (ii) the expected detection rate for future blind surveys, and (iii) how detection rate varies as a function of impact parameter.

Throughout this work we assume a flat $\Lambda$CDM cosmology, with $\Omega_{\mathrm{M}}$ = 0.27, $\Omega_{\Lambda}$ = 0.73, and $H_{\mathrm{0}}$ = 71 km s$^{-1}$ Mpc$^{-1}$. 
All uncertainties refer to the 68.3 per cent confidence interval, unless otherwise stated.

\section{Observations and data reduction}
\label{data}

To ensure a uniform sample, and consistency in the derived quantities, all aspects of the sample selection, observations, data reduction, and analysis are performed as described in Paper I, unless otherwise specified. 
Therefore, in order to avoid unnecessary repetition, we describe only the most important points in relation to these aspects of the work here, and refer the reader to Paper I for a more detailed description.

\begin{table*}
\begin{minipage}{\linewidth}
\centering
\caption{\mbox{H\,{\sc i}} and optical properties of the target galaxies in our sample. 
Columns (1)-(6) are the \mbox{H\,{\sc i}} properties as given in the HIPASS BGC \citep{2004AJ....128...16K}. 
Column (1) is the HIPASS name. 
Columns (2) and (3) are the HIPASS J2000 right ascension and declination. HIPASS positions have typical uncertainties of $\leq$ 1.5 arcmin for signal-to-noise ratios of $\geq$ 10. 
Columns (4) and (5) are the systemic and local group velocities. 
Column (6) is the optical/IR identification \citep{2005MNRAS.361...34D}. 
Columns (7)-(9) are the optical properties as given in the NASA Extragalatic Database (NED). 
Column (7) is the optical morphological classification. 
Columns (8) and (9) are the optical J2000 right ascension and declination.}
\label{table:hipass}
\begin{threeparttable}
\begin{tabular}{@{} lllrrlllll @{}} 
\hline
HIPASS Name & RA$_{\mathrm{HIPASS}}$ & Dec$_{\mathrm{HIPASS}}$ & $v_{\mathrm{sys}}$ & $v_{\mathrm{LG}}$ & Optical ID & Morph. Type & RA$_{\mathrm{opt}}$ & Dec$_{\mathrm{opt}}$ \\
& (J2000) & (J2000) & (km s$^{-1}$) & (km s$^{-1}$)& & & (J2000) & (J2000) \\
\hline
J0309-41 & 03 09 42 & -41 01 08 & 955 & 809 & ESO\,300-G\,014 & SAB(s)m & 03 09 37.87 & -41 01 49.7\tnote{$1$} \\ 
J0316-35 & 03 16 52 & -35 32 19 & 1570 & 1436 & ESO\,357-G\,012 & SB(s)d & 03 16 52.81 & -35 32 25.9\tnote{$2$} \\ 
J0541-35 & 05 41 01 & -35 42 43 & 1264 & 1035 & ESO\,363-G\,015 & SA(s)d & 05 41 00.87 & -35 42 27.1\tnote{$3$} \\ 
J0319-49 & 03 19 25 & -49 36 34 & 1028 & 853 & IC\,1914 & SAB(s)d	& 03 19 25.24 & -49 35 59.0\tnote{$4$} \\ 
J1415-43 & 14 15 05 & -43 58 06 & 1875 & 1684 & IC\,4386/7 & SAB(s)dm pec? & 14 15 03.07 & -43 57 45.2\tnote{$3$} \\ 
J0310-53 & 03 10 02 & -53 20 05 & 1072 & 893 & NGC\,1249 & SB(s)cd & 03 10 01.23 & -53 20 08.7\tnote{$4$} \\ 
J0419-54 & 04 19 56 & -54 56 58 & 1504 & 1287 & NGC\,1566 & SAB(s)bc & 04 20 00.42 & -54 56 16.1\tnote{$4$} \\ 
J0610-34 & 06 10 11 & -34 06 56 & 747 & 505 & NGC\,2188 & SB(s)m edge-on & 06 10 09.53 & -34 06 22.3\tnote{$3$} \\ 
J1328-48 & 13 28 41 & -48 54 09 & 2988 & 2762 & NGC\,5156 & SB(r)b & 13 28 44.09 & -48 55 00.5\tnote{$4$} \\ 
J2200-43 & 22 00 27 & -43 07 41 & 2269 & 2258 & NGC\,7162A & SAB(s)m & 22 00 35.74 & -43 08 22.4\tnote{$5$} \\ 
\hline
\end{tabular}
\begin{tablenotes}
\footnotesize{
\item[] {Optical position references:}
\item[$1$] {\citet{1996MNRAS.278.1025L}}
\item[$2$] {\citet{1998AJ....116....1D}}
\item[$3$] {\citet{1982euse.book.....L}}
\item[$4$] {\citet{2006AJ....131.1163S}}
\item[$5$] {\citet{1990MNRAS.243..692M}}
}
\end{tablenotes}
\end{threeparttable}
\end{minipage}
\end{table*}

\begin{table*}
\begin{minipage}{\linewidth}
\centering
\caption{Properties of the background continuum sources used to search for absorption. 
Column (1) is the identification (ID) used throughout this work. 
Columns (2)-(6) are the radio properties as given in the SUMSS catalogue \citep{2003MNRAS.342.1117M}. 
Column (2) is the SUMSS name. 
Columns (3) and (4) are the SUMSS J2000 right ascension and declination.
Columns (5) and (6) are the 843 MHz peak and integrated fluxes. 
Columns (7) and (8) are the angular and linear separations of the radio source and the foreground galaxy.}
\label{table:sumss}
\begin{threeparttable}
\begin{tabular}{@{} llllrrrr@{}} 
\hline
Source Name\tnote{$*$} & SUMSS Name & RA$_{\mathrm{SUMSS}}$ & Dec$_{\mathrm{SUMSS}}$ & $S_{\mathrm{peak,843}}$ & $S_{\mathrm{int,843}}$ & Ang. Sep. & Imp. Param. \\
& & (J2000) & (J2000) & (mJy beam$^{-1}$) & (mJy) & (arcmin) & (kpc) \\
\hline
C-ESO\,300-G\,014-1 & J030946-410456 & 03 09 46.24 & -41 04 56.6 & 113.5 & 114.5 & 3.5 & 11.5 \\
C-ESO\,300-G\,014-2 & J030941-410006 & 03 09 41.24 & -41 00 06.8 & 110.2 & 112.1 & 1.8 & 6.1 \\
C-ESO\,357-G\,012 & J031705-353440 & 03 17 05.10 & -35 34 40.3 & 67.7 & 71.8 & 3.4 & 19.6 \\

C-ESO\,363-G\,015-1 & J054104-354640 & 05 41 04.72 & -35 46 40.3 & 67.5 & 67.5 & 4.3 & 18.1 \\
C-ESO\,363-G\,015-2 & J054100-354634 & 05 41 00.16 & -35 46 34.2 & 59.9 & 66.0 & 4.1 & 17.4 \\
C-IC\,1914 & J031955-493551 & 03 19 55.29 & -49 35 51.1 & 210.7 & 222.9 & 4.9 & 17.0 \\
C-IC\,4386-1 & J141517-435922 & 14 15 17.99 & -43 59 22.5 & 81.7 & 90.2 & 3.1 & 21.5 \\
C-IC\,4386-2 & J141455-440101 & 14 14 55.11 & -44 01 01.8 & 24.9 & 25.7 & 3.6 & 24.5 \\
C-NGC\,1249-1 & J030954-532339 & 03 09 54.31 & -53 23 39.2 & 48.5 & 55.6 & 3.7 & 13.3 \\
C-NGC\,1249-2 & J030938-532417 & 03 09 38.05 & -53 24 17.3 & 27.2 & 32.9 & 5.4 & 19.6 \\
C-NGC\,1249-3 & J030917-531756 & 03 09 17.15 & -53 17 56.9 & 42.5 & 48.3 & 6.9 & 25.3 \\
C-NGC\,1566-1 & J042015-545345 & 04 20 15.32 & -54 53 45.2 & 58.1 & 58.9 & 3.3 & 17.3 \\
C-NGC\,1566-2 & J042007-545143 & 04 20 07.48 & -54 51 43.4 & 35.3 & 52.6 & 4.7 & 24.4 \\
C-NGC\,2188 & J061019-340258 & 06 10 19.16 & -34 02 58.7 & 95.9 & 130.9 & 3.9 & 8.1 \\
C-NGC\,5156 & J132846-485638 & 13 28 46.53 & -48 56 38.7 & 790.3 & 823.2 & 1.7 & 18.8 \\
C-NGC\,7162A-1 & J220040-430950 & 22 00 40.68 & -43 09 50.1 & 46.4 & 50.8 & 1.7 & 15.7 \\
C-NGC\,7162A-2 & J220057-431138 & 22 00 57.05 & -43 11 38.8 & 15.8 & 17.4 & 5.1 & 46.5 \\
\hline
\end{tabular}
\begin{tablenotes}
\footnotesize{
\item[$*$] {Throughout this work we refer to the continuum sources by the ID listed in Column (1). This is the galaxy name, with a `C' (for `continuum') prepended, allowing us to easily identify the galaxy-continuum source pairs. 
Where there are multiple sources in a single field we have added a suffix, to number the different sources. 
In addition, if a source resolves into multiple components at higher resolution then we add a second suffix (a lowercase letter) to indicate this (e.g. C-ESO\,363-G\,015-2a, C-ESO\,363-G\,015-2b). 
We note that this differs slightly from our convention in paper I, to allow for the fact that there may be multiple sources per field in the present sample.}
}
\end{tablenotes}
\end{threeparttable}
\end{minipage}
\end{table*}

\subsection{Sample selection}

Our main sample in this paper consists of 10 radio source-galaxy pairs with impact parameters less than 25 kpc. 
Radio source-galaxy pairs were selected by cross-matching galaxies from the HIPASS Bright Galaxy Catalogue \citep[HIPASS BGC,][]{2004AJ....128...16K} with radio continuum sources from the 843 MHz Sydney University Molonglo Sky Survey \citep[SUMSS,][]{2003MNRAS.342.1117M}. 
In addition to the impact parameter cutoff, we imposed a minimum (integrated) flux criterion of $S_{843}$ $>$ 50 mJy, since the flux of the background source (and not the redshift of the galaxy) is what determines the absorption-line sensitivity. 
Redshift data is not available for the radio sources in our sample, but we have assumed all of the continuum sources are at higher redshift than the target galaxies (i.e. genuine background sources). 
This is supported by redshift analysis of other similar radio source samples \citep{1998AJ....115.1693C,2010A&ARv..18....1D} and of previous surveys for intervening absorption in the local Universe \citep{2011ApJ...742...60D}.

\begin{figure*}
\includegraphics[width=0.3\linewidth]{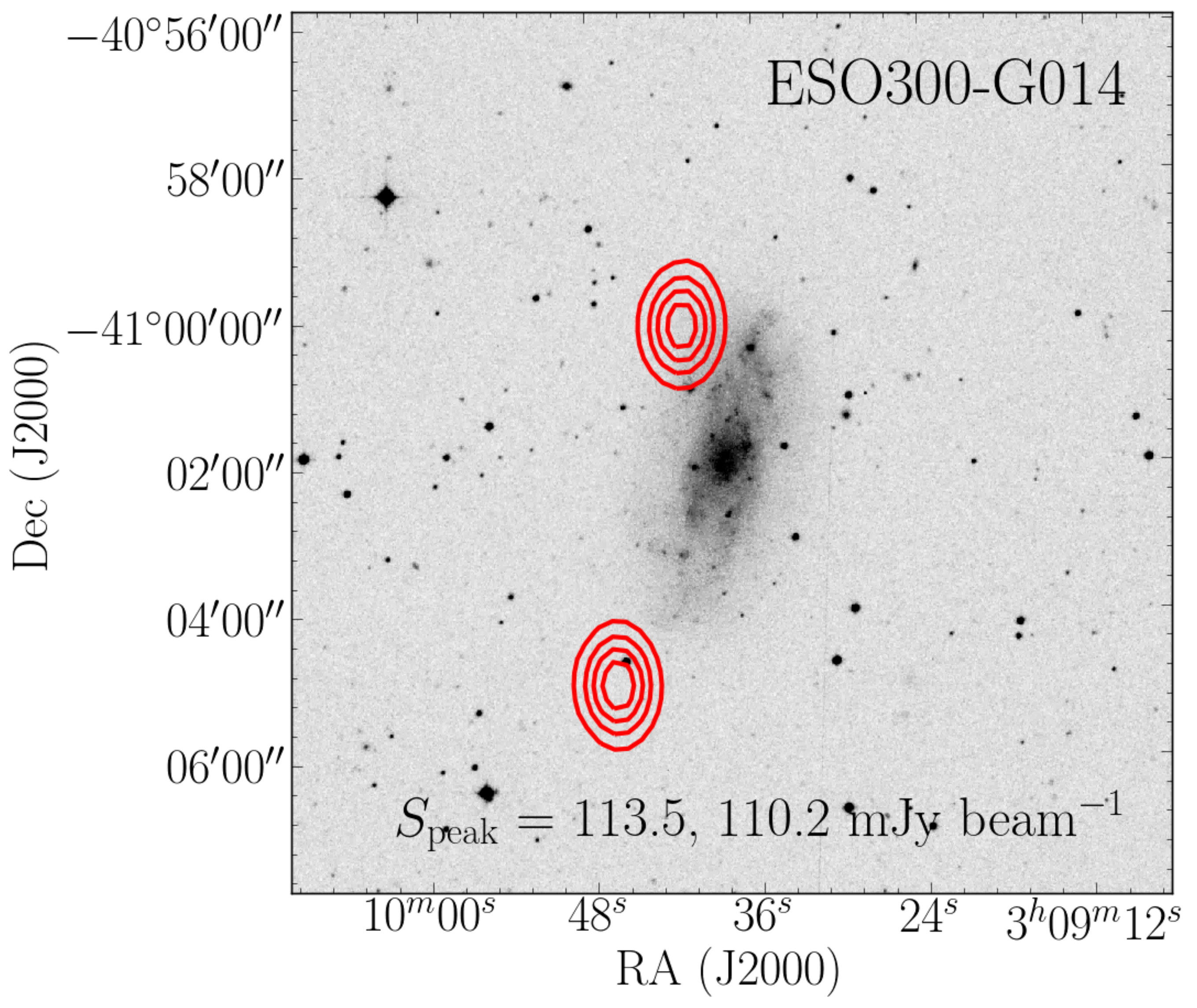}
\includegraphics[width=0.3\linewidth]{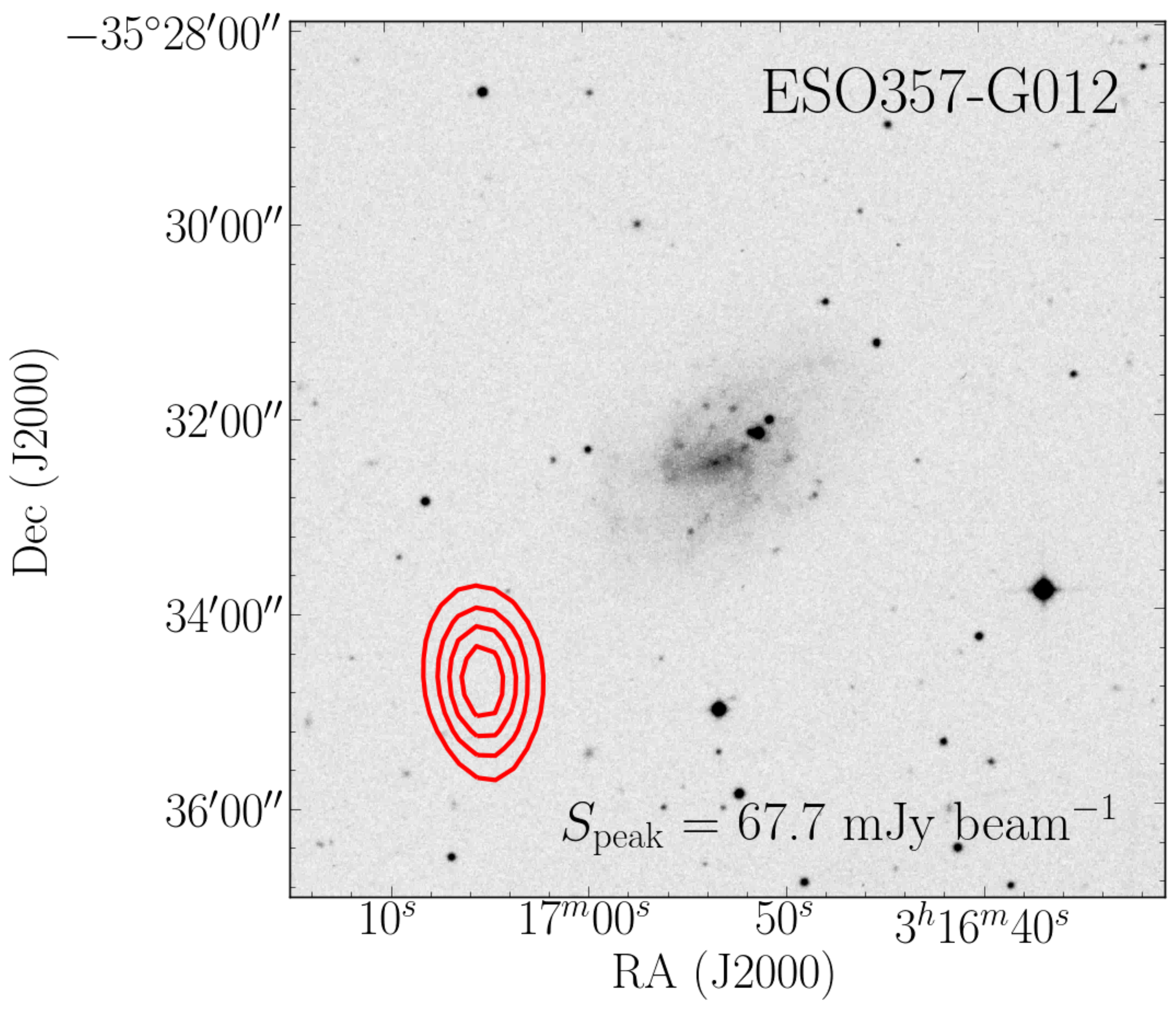}
\includegraphics[width=0.3\linewidth]{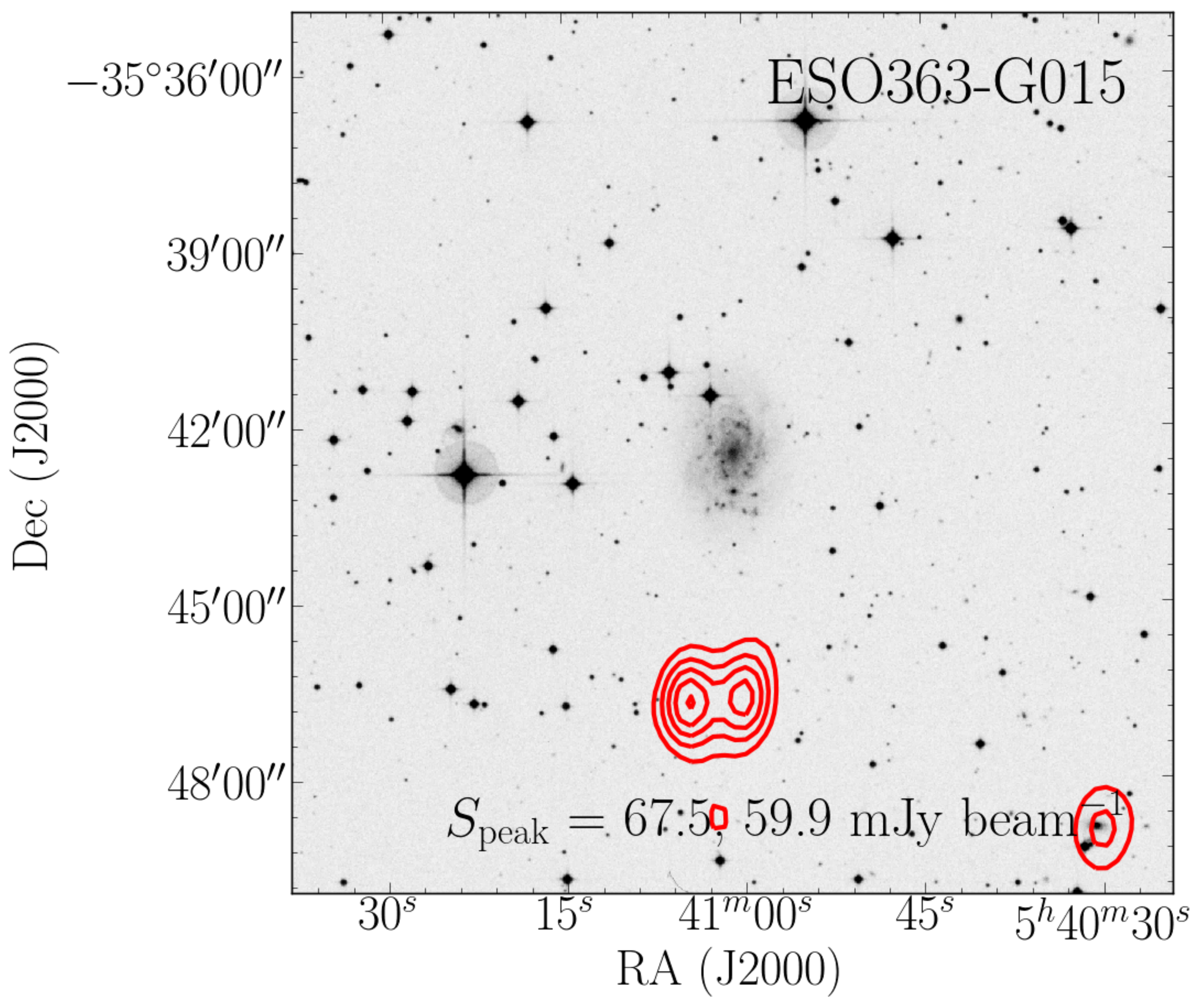}
\includegraphics[width=0.3\linewidth]{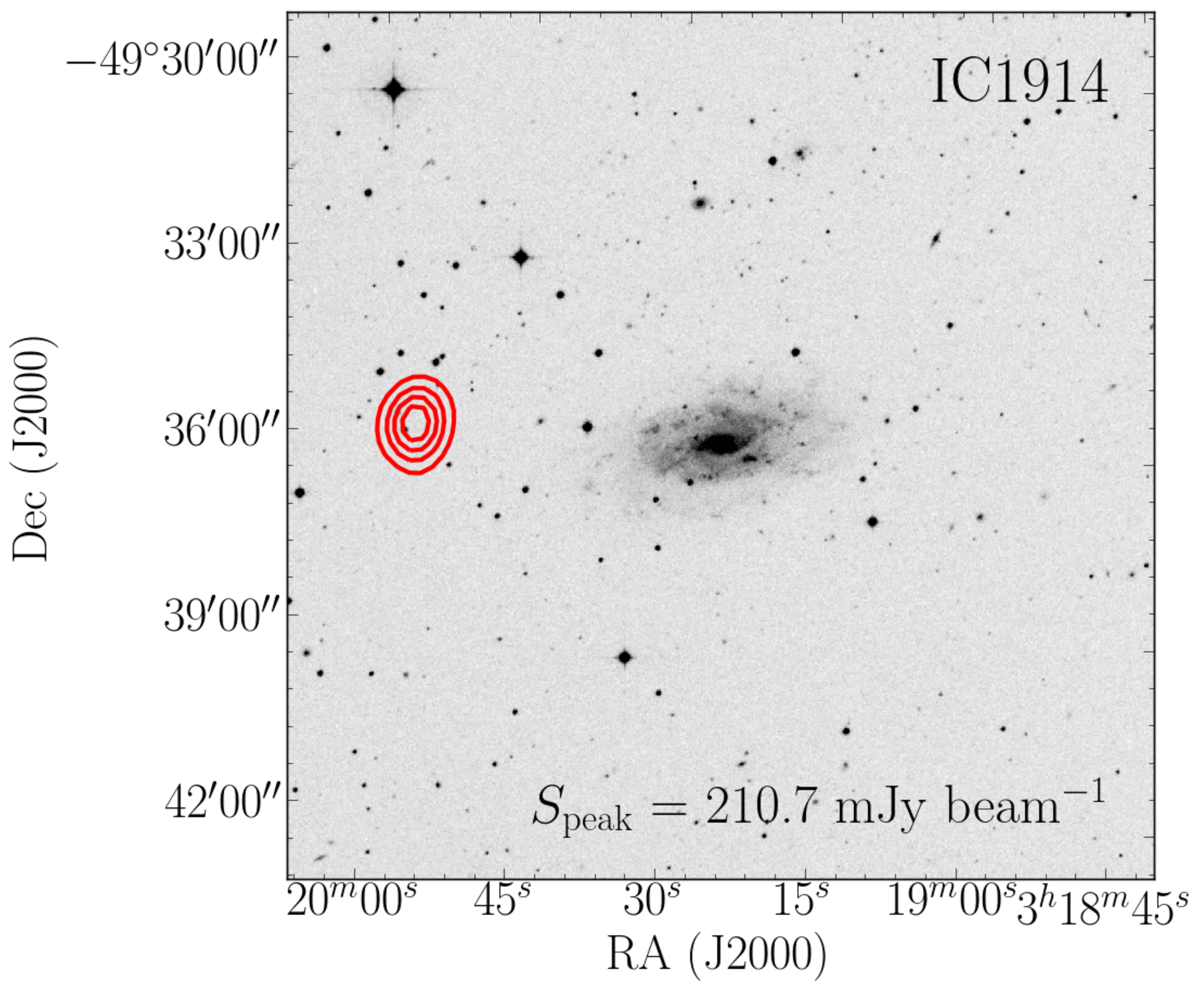}
\includegraphics[width=0.3\linewidth]{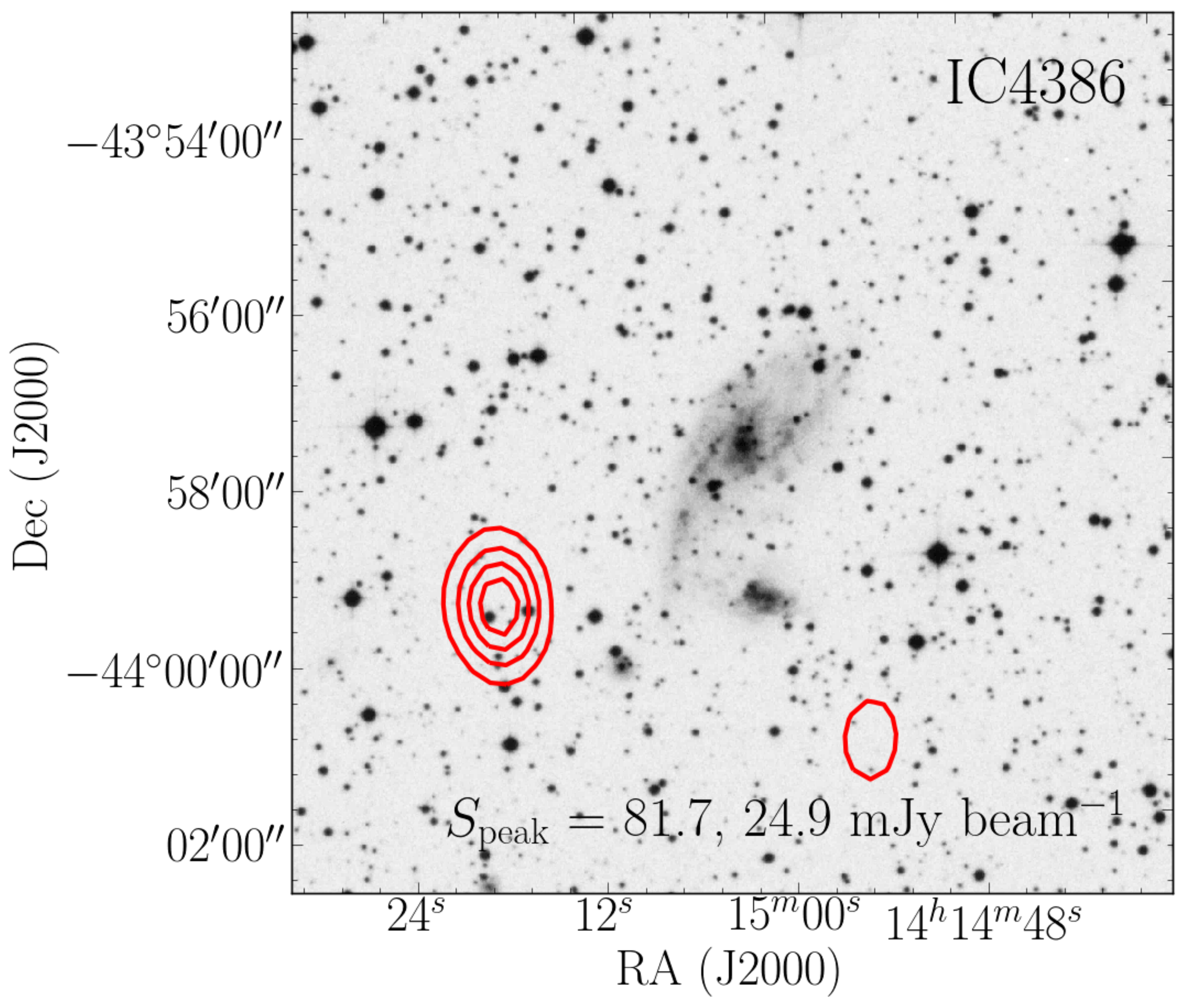}
\includegraphics[width=0.3\linewidth]{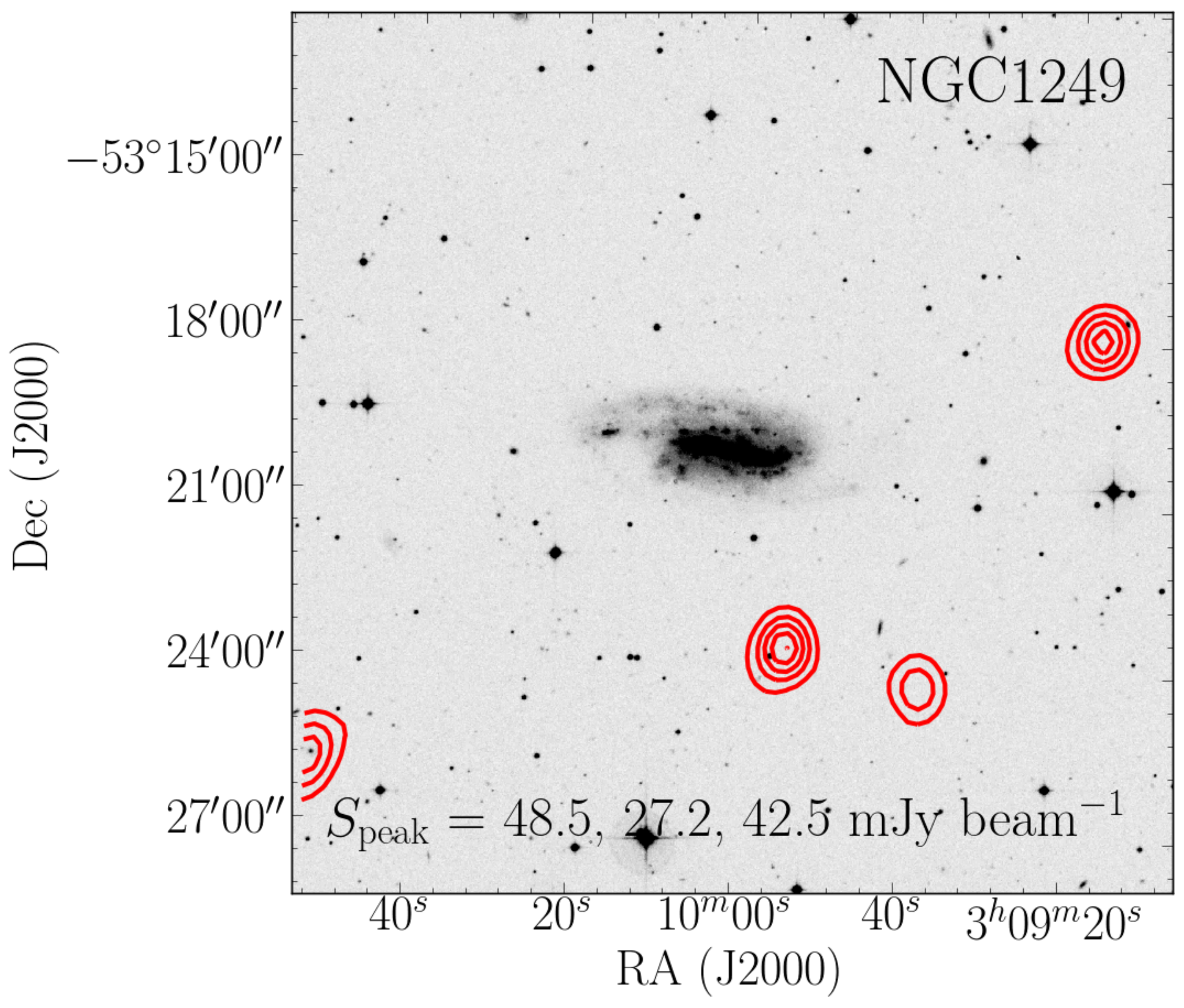}
\includegraphics[width=0.3\linewidth]{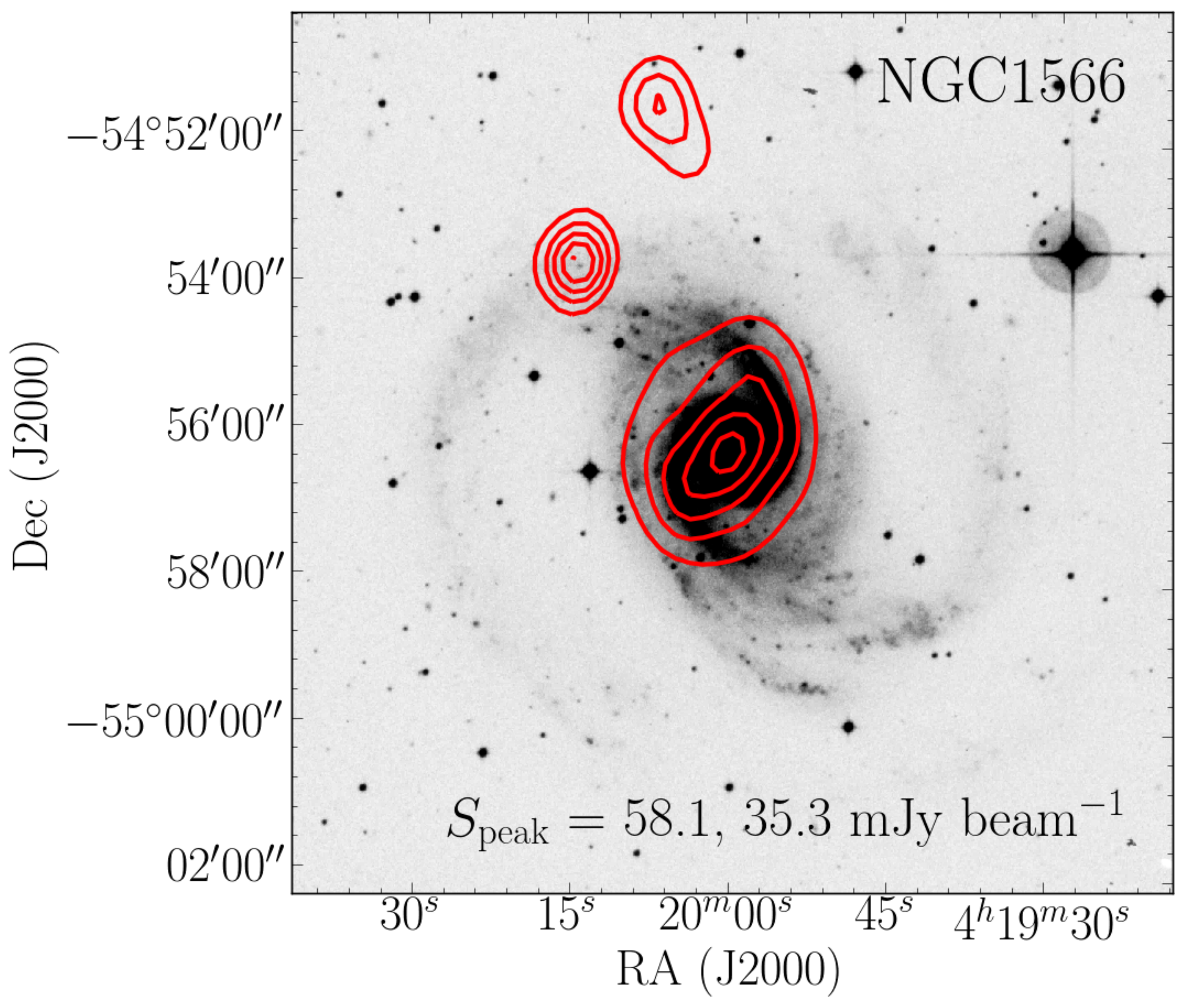}
\includegraphics[width=0.3\linewidth]{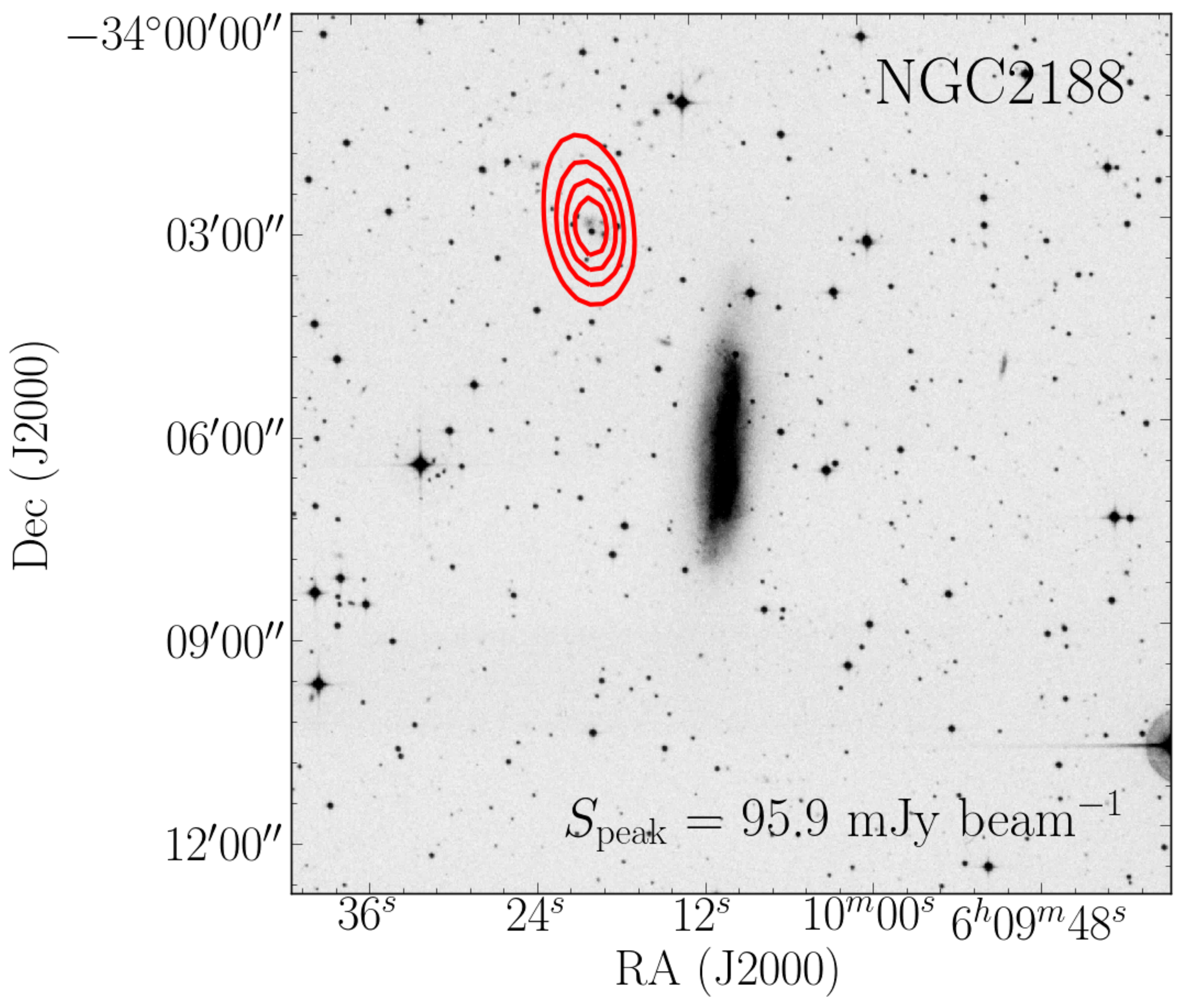}
\includegraphics[width=0.3\linewidth]{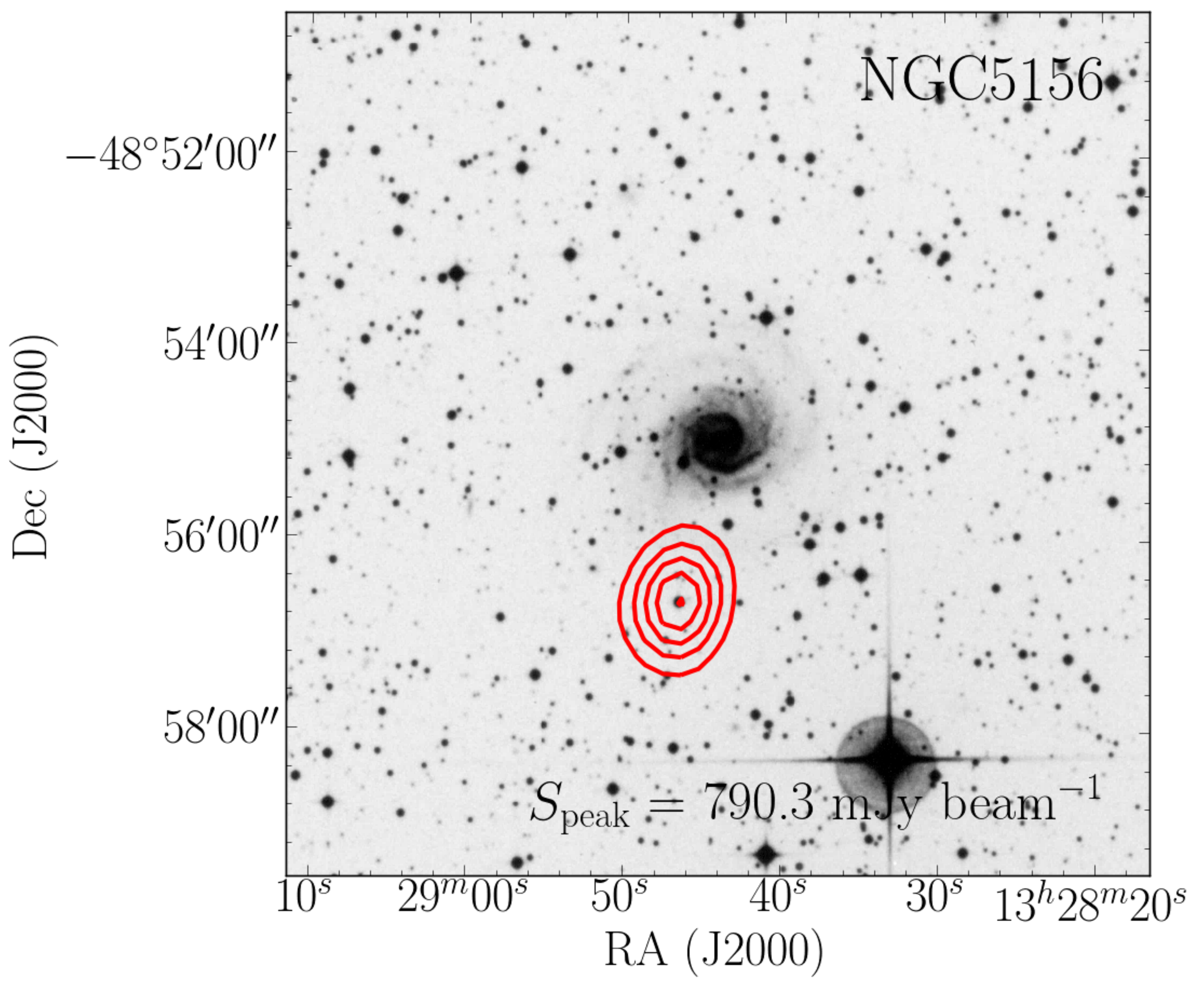}
\includegraphics[width=0.3\linewidth]{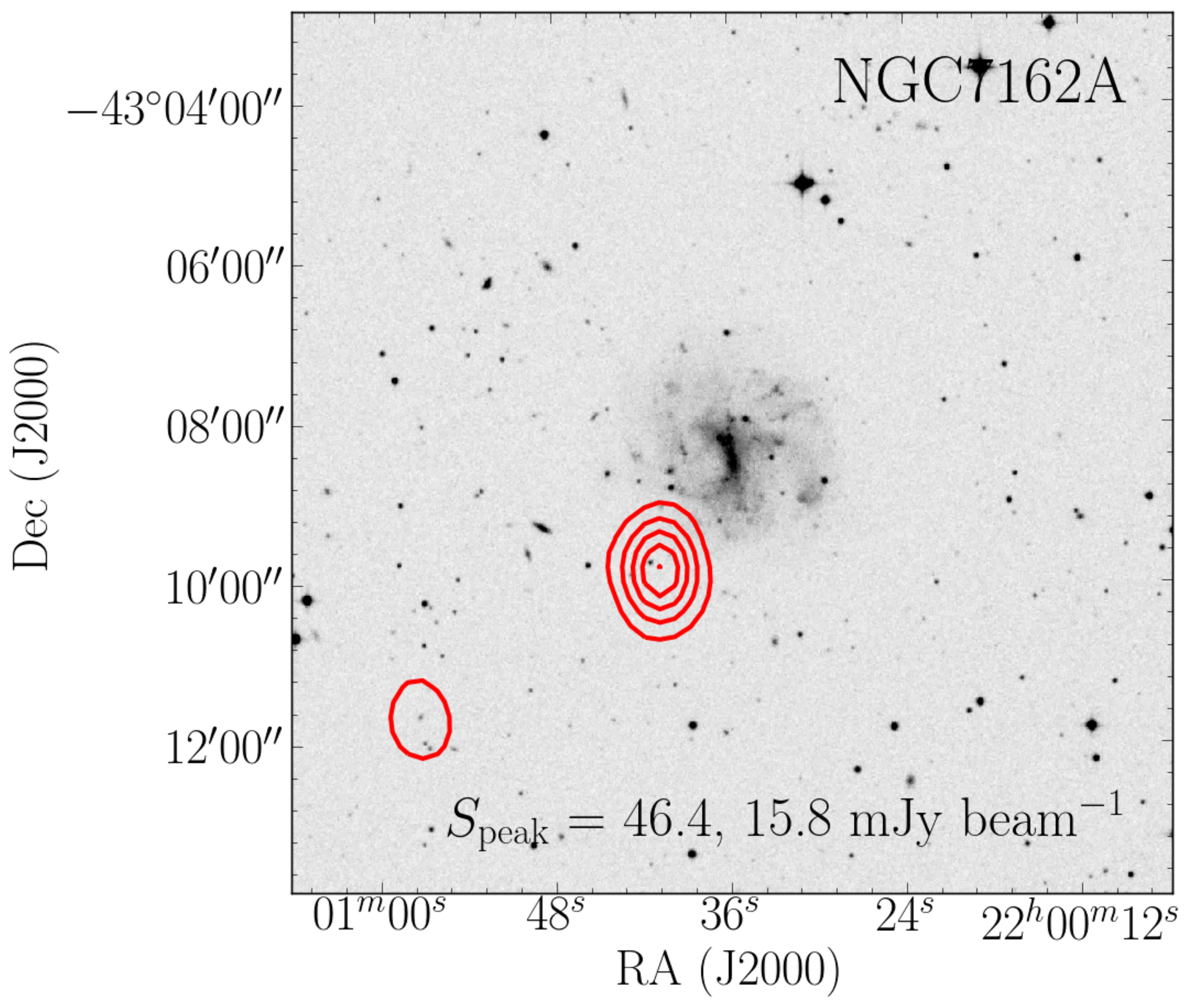}
\caption{SuperCOSMOS B-band images \citep{2001MNRAS.326.1279H} of the six target galaxies, with SUMSS 843 MHz radio continuum contours overlaid \citep{2003MNRAS.342.1117M}. 
The SUMSS peak flux is given in the bottom right corner. 
Radio contours start at 90 per cent of the peak flux and decrease in 10 per cent increments.} 
\label{figure:targets}
\end{figure*}

The sample selection described above is identical to that in Paper I, but with a slightly higher impact parameter cutoff. 
In addition, a number of the selected fields contained additional continuum sources which, although they do not meet the above flux or impact parameter criteria, have been included in order to increase the sample size. 
Therefore, in total our sample in this paper consists of 17 sightlines with impact parameters between 6 and 46 kpc, and 843 MHz continuum fluxes of 17-823 mJy. 
Combined with the sightlines searched in Paper I, our full sample consists of 23 sightlines. 
Images of the targets are shown in Figure \ref{figure:targets}. 
\mbox{H\,{\sc i}} and optical properties of the target galaxies in our sample are given in Table \ref{table:hipass}, and properties of the background radio sources, along with the angular and linear separations, are given in Table \ref{table:sumss}.

Throughout this paper we refer to the objects observed in the pilot sample \citep{2015MNRAS.450..926R} as sample A, and the new objects observed for this paper as sample B (see Table \ref{table:samples}). 
In most cases, the results presented in this paper refer only to the galaxies of sample B, however, where relevant, we consider the combined results of the two samples.

\subsection{ATCA \mbox{H\,{\sc i}} observations}

Australia Telescope Compact Array (ATCA) \mbox{H\,{\sc i}} observations were carried out in 10 $\times$ 12 hour periods between October 2012 and August 2013, using the 750 m arrays, which give an angular scale sensitivity of a few arcseconds to $\sim$10-20 arcmin. 
The target sources were observed in 40 minute scans, interleaved with 10 minute scans of a nearby phase calibrator. 
The ATCA primary calibrator, PKS 1934-638, was observed for about 10 minutes at both the beginning and end of each observation for calibration of the absolute flux scale. 
Observations were conducted at night time in order to minimise the effects of solar interference, especially on the shorter baselines.

For the frequency setup we have used the Compact Array Broadband Backend \citep[CABB,][]{2011MNRAS.416..832W} 64M-32k configuration, which gives a bandwidth of 64 MHz with 32 kHz channels (spectral resolution of $\sim$6.7 km s$^{-1}$ at the frequency of the redshifted \mbox{H\,{\sc i}} line). 
The noise level achieved in a 12 hour observation ($\sim$2-3 mJy per channel) is sufficient to detect absorption in DLA-strength systems ($N_{\mathrm{HI}} \gtrsim 2 \times 10^{20}$ cm$^{-2}$) against the background sources we have selected, assuming the source is unresolved.
A summary of the observations is given in Table \ref{table:atca_obs}.

Follow-up observations using the 6 km arrays were conducted for two sources, NGC\,1566 and NGC\,5156. 
Our initial observations of NGC\,1566 showed a high \mbox{H\,{\sc i}} column density along the sightline searched, but no \mbox{H\,{\sc i}} absorption. 
These observations at higher spatial resolution were therefore designed to resolve out any remaining \mbox{H\,{\sc i}} emission and determine whether the strong emission-line could be hiding any absorption. 
For NGC\,5156 we wished to re-observe the new absorption-line detected in the 750 array observations with higher spectral spectral resolution, in order to spectrally resolve the line profile. 
The follow-up observations for NGC\,5156 were therefore conducted using the CABB 1M-0.5k configuration, which provides 1 MHz bandwidth with 0.5 kHz spectral resolution (or about 0.1 km s$^{-1}$ at the frequency of the \mbox{H\,{\sc i}} line). 

All data reduction was carried out in {\sc miriad} \citep{1995ASPC...77..433S} using a purpose-built data reduction pipeline, and following standard data reduction procedures for continuum and spectral-line data. 
For each of the targets in our sample we have produced an \mbox{H\,{\sc i}} data cube and 1.4 GHz radio continuum image. 
For the 1M-0.5k observations of NGC\,5156, the full spectral resolution of 0.1 km s$^{-1}$ was higher than we required, so when Fourier transforming the data we produced a data cube binned to a resolution of 1 km s$^{-1}$, which provided an optimal balance between spectral resolution and signal-to-noise per channel.

As in Paper I, we have produced three sets of data-cubes and continuum images for the 750 array data, using different weighting schemes. 
In doing so we are able to span the range of resolutions possible with the 750 arrays, and thus optimise for the detection of either emission or absorption. 
Low resolution cubes (with a synthesised beam of $\sim$60 arcsec) were made using natural weighting, and excluding the baselines to antenna 6, medium resolution cubes ($\sim$20 arcsec) were made using natural weight with the antenna 6 baselines included, and high resolution cubes ($\sim$5 arcsec) were made using uniform weighting.

\mbox{H\,{\sc i}} moment maps were produced from the low resolution cubes and spectra were extracted from all three sets of cubes at the position of the radio continuum source. 
In the case that a background source was resolved into multiple components at higher resolution (see Figure \ref{figure:continuum_and_spectra}), we have extracted a separate spectrum for each component.

\begin{table}
\centering
\caption{List of galaxies observed as part of samples A and B for this survey.}
\label{table:samples}
\begin{tabular}{@{} ll @{}} 
\hline
Sample A & Sample B \\
(Reeves+15) & (this work) \\
\hline
ESO\,150-G\,005 & ESO\,300-G\,014 \\
ESO\,345-G\,046 & ESO\,357-G\,012 \\
ESO\,402-G\,025 & ESO\,363-G\,015 \\
IC\,1954 & IC\,1914 \\
NGC\,7412 & IC\,4386 \\
NGC\,7424 & NGC\,1249 \\
& NGC\,1566 \\
& NGC\,2188 \\
& NGC\,5156 \\
& NGC\,7162A \\
\hline
\end{tabular}
\end{table}

\begin{table}
\centering
\caption{Summary of the ATCA observations for the galaxies observed in sample B (this paper).}
\label{table:atca_obs}
\begin{threeparttable}
\begin{tabular}{@{} lllrl @{}} 
\hline
Target galaxy & Obs. Date & Array\tnote{$*$} & Int. time\\
 & & & (h) & \\
\hline
ESO\,300-G\,014 & 2012 Oct 26 & 750B &  7.73 \\
ESO\,357-G\,012 & 2012 Oct 30 & 750B & 8.10 \\
ESO\,363-G\,015 & 2012 Oct 31 & 750B & 8.51 \\
IC\,1914 & 2012 Oct 27 & 750B & 8.15 \\
IC\,4386 & 2013 Feb 02 & 750C & 8.01 \\
NGC\,1249 & 2012 Oct 29 & 750B & 7.70 \\
NGC\,1566 & 2012 Oct 28 & 750B & 7.97 \\
NGC\,1566 (follow-up) & 2013 Jun 10 & 6C & 9.24 \\
NGC\,2188 & 2012 Nov 01 & 750B & 8.10 \\
NGC\,5156 & 2013 Feb 03 & 750C & 7.61 \\
NGC\,5156 (follow-up) & 2013 Nov 12 & 6A & 8.15 \\
NGC\,7162A & 2013 Aug 02 & 750D & 7.68 \\
\hline
\end{tabular}
\begin{tablenotes}
\footnotesize{
\item[$*$] {Range of baseline lengths:}
\item[] {750B: 61 -- 4500 m; 750C: 46-5020 m; 750D: 31 -- 4469 m; 6A: 337-5939 m; 6C: 153-6000 m}
}
\end{tablenotes}
\end{threeparttable}
\end{table}

\section{H\,{\sevensize\bf I} distribution in the target galaxies}
\label{results_part1}

\subsection{H\,{\sevensize\bf I} emission maps}
\label{results_part1:hi_maps}

\mbox{H\,{\sc i}} moment maps (total intensity, velocity, and velocity dispersion) for each of the galaxies are presented in Figure \ref{figure:moment_maps}.
In general we find fairly symmetric, regular rotating discs --- although some galaxies, such as IC\,4386 and NGC\,2188, show notable asymmetries in their total intensity and velocity maps, which could indicate recent interactions. 

The measured \mbox{H\,{\sc i}} masses are given in Table~\ref{table:HI_properties_derived}, along with the HIPASS \mbox{H\,{\sc i}} masses for comparison. 
We find a range of \mbox{H\,{\sc i}} masses, between $\sim$4.0 $\times$ 10$^{8}$ and 5.6 $\times$ 10$^{9}$ M$_{\odot}$ (with an estimated uncertainty of 5 per cent). 
Our values are consistent with those from HIPASS -- or slightly lower since with an interferometer we resolve out some of the more diffuse \mbox{H\,{\sc i}} flux which is detectable with Parkes.

\begin{figure*}
\includegraphics[width=0.3\linewidth]{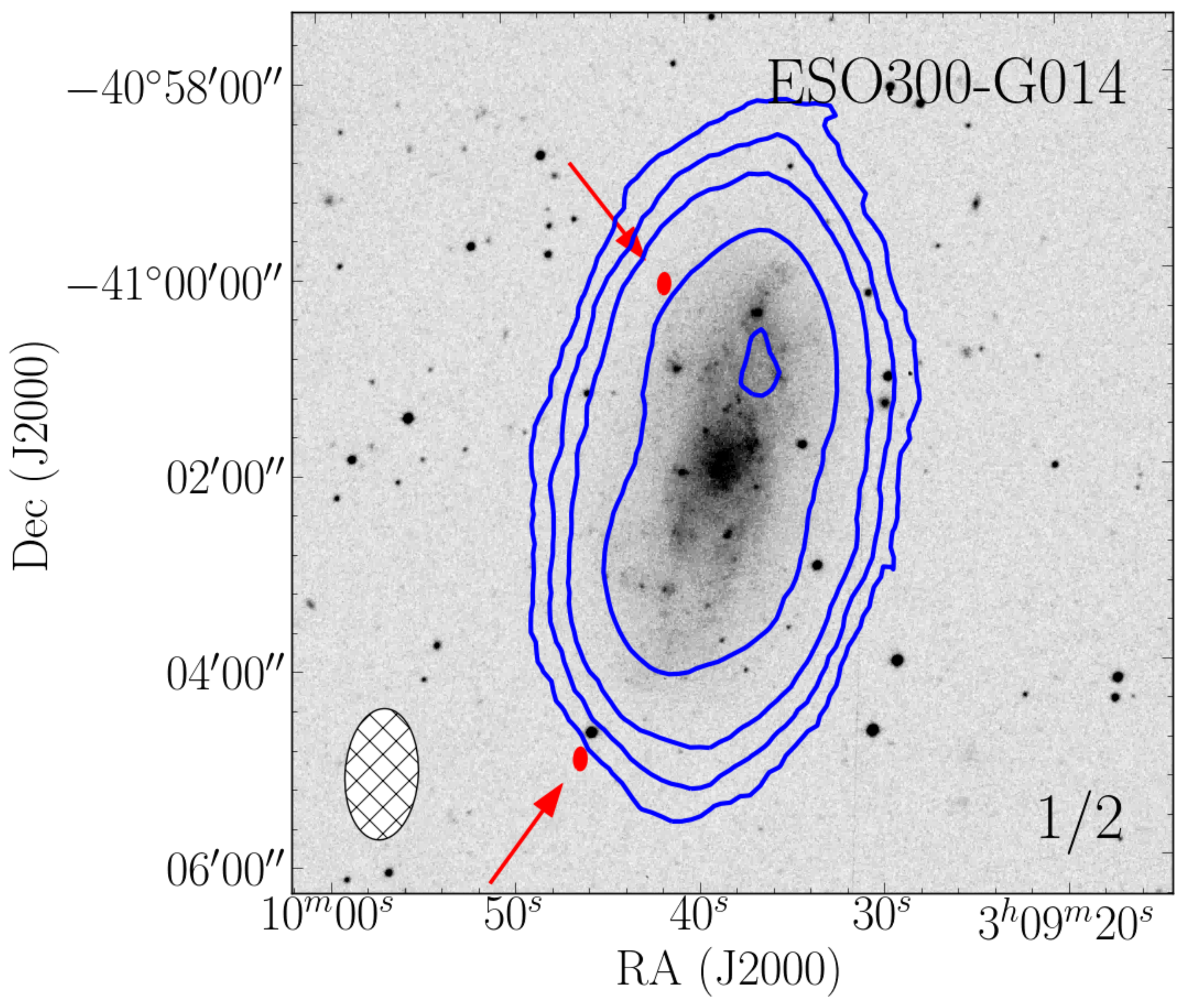}
\includegraphics[width=0.3\linewidth]{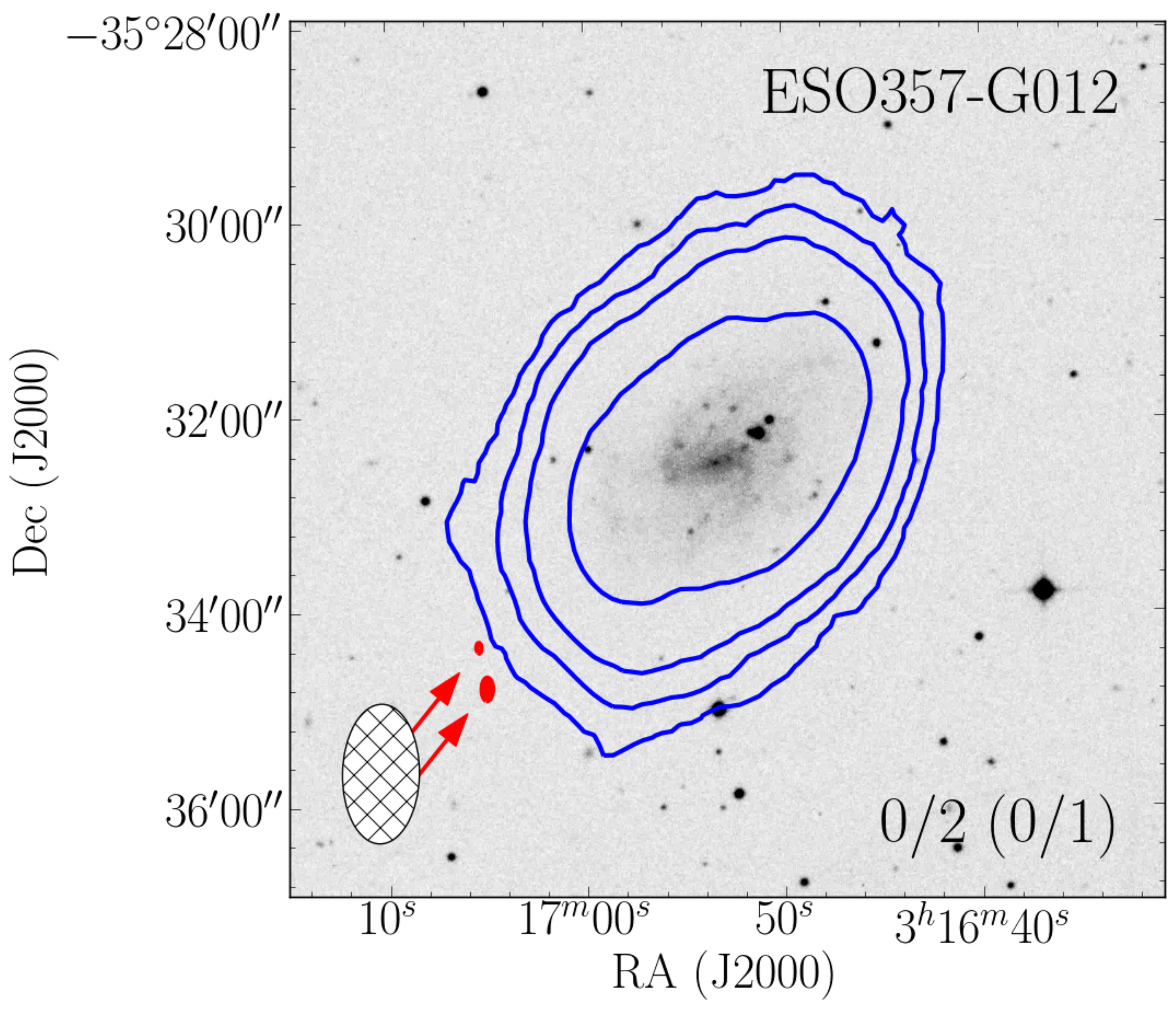} 
\includegraphics[width=0.3\linewidth]{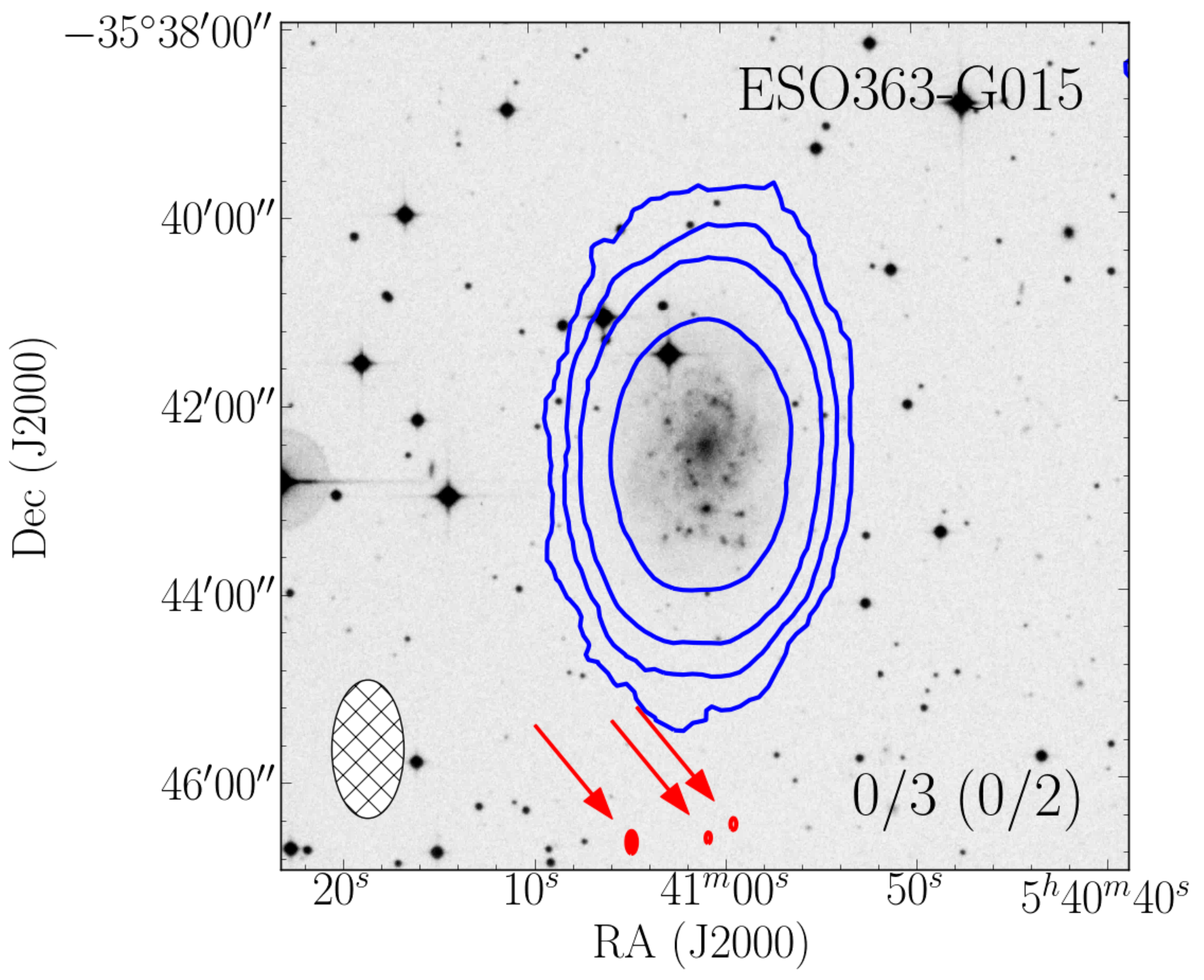}
\includegraphics[width=0.3\linewidth]{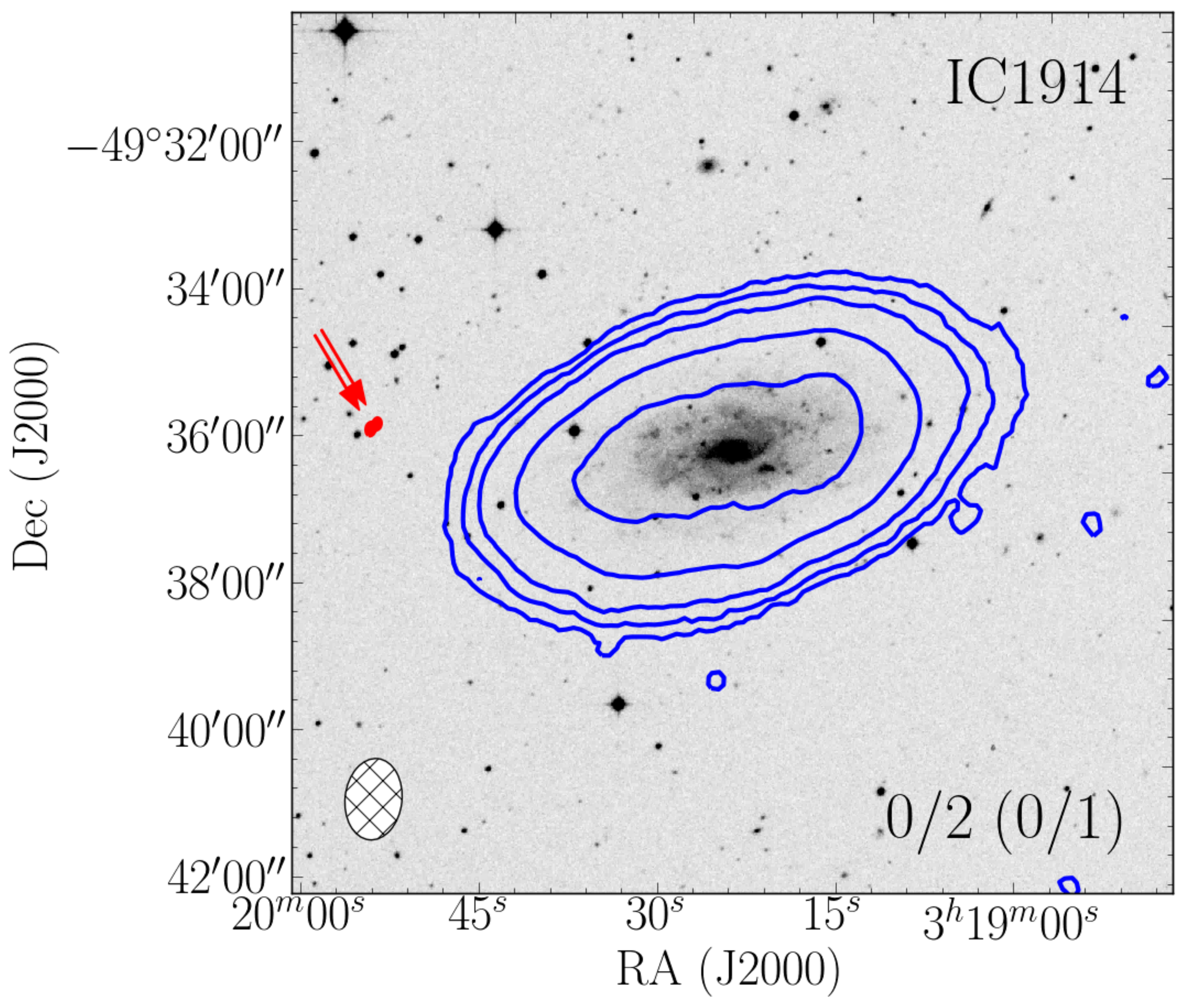}
\includegraphics[width=0.3\linewidth]{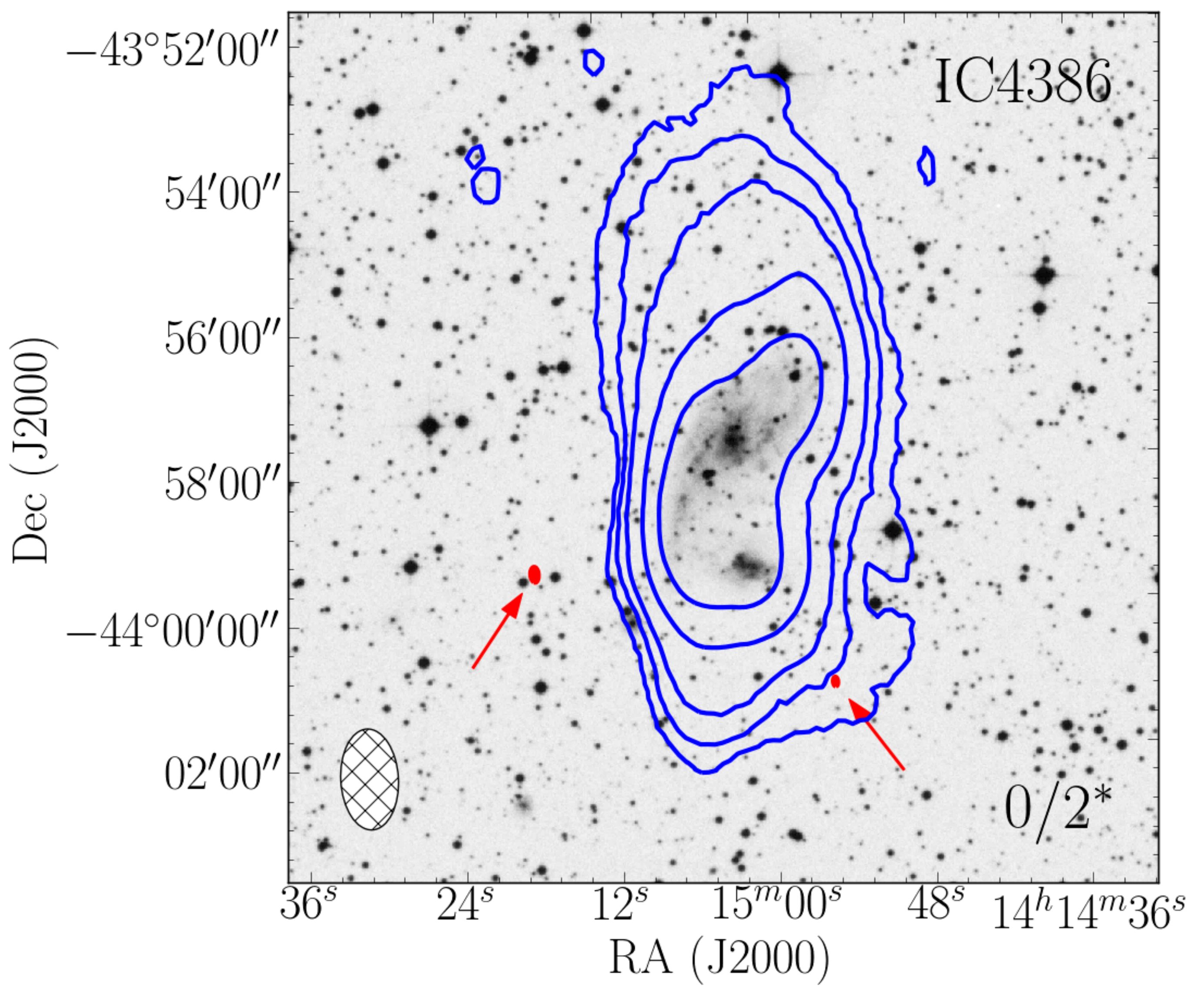}
\includegraphics[width=0.3\linewidth]{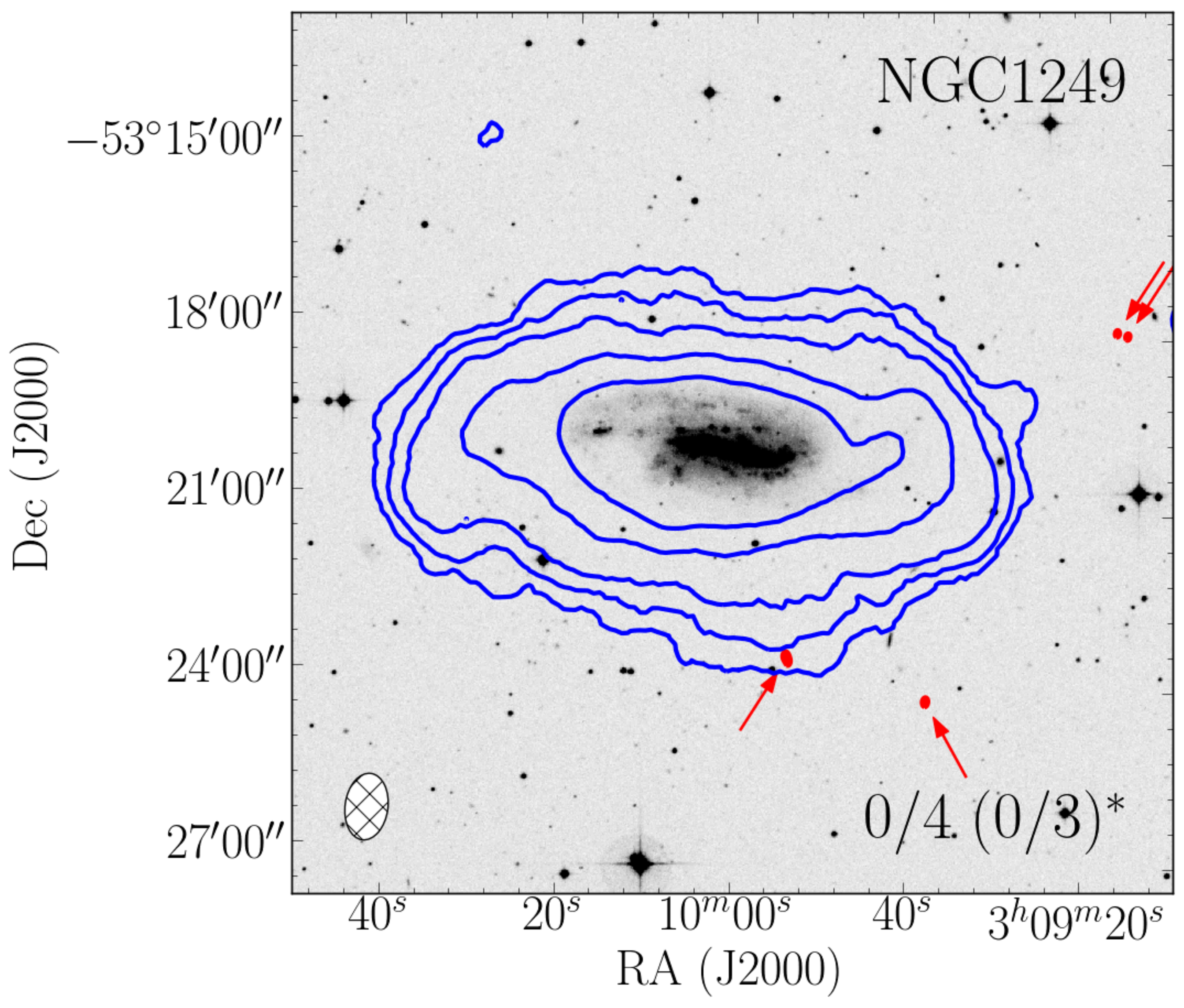}
\includegraphics[width=0.3\linewidth]{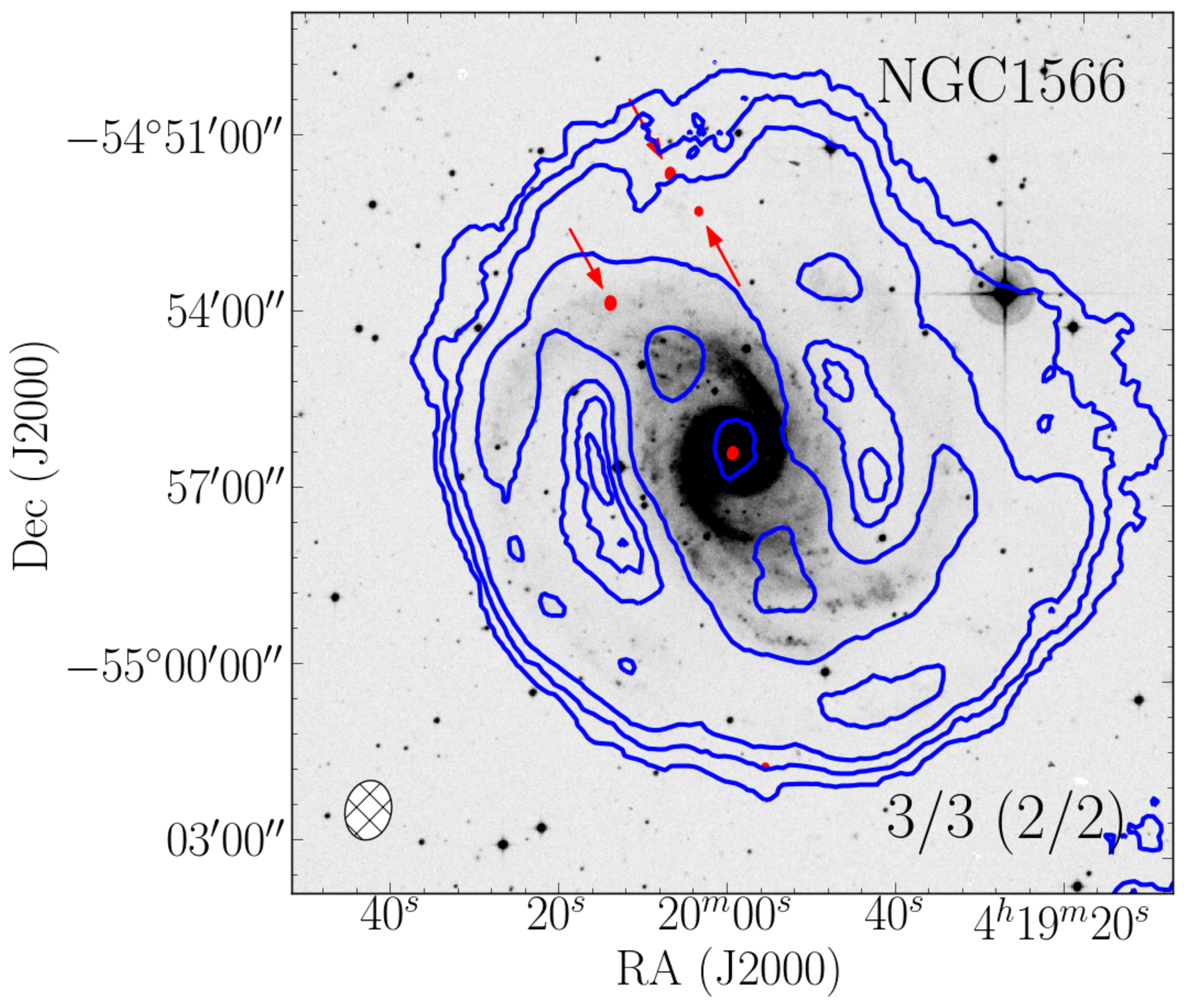}
\includegraphics[width=0.3\linewidth]{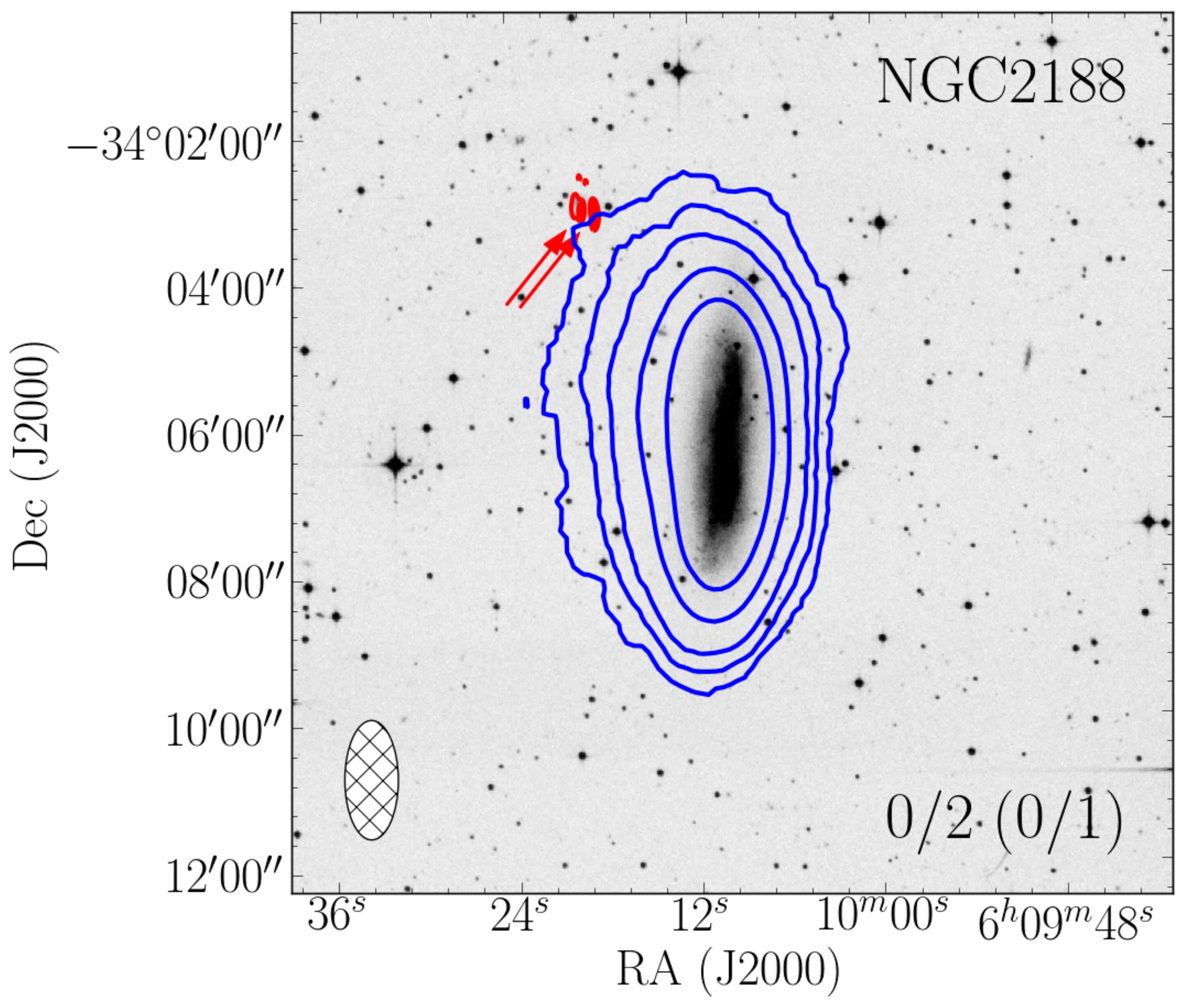}
\includegraphics[width=0.3\linewidth]{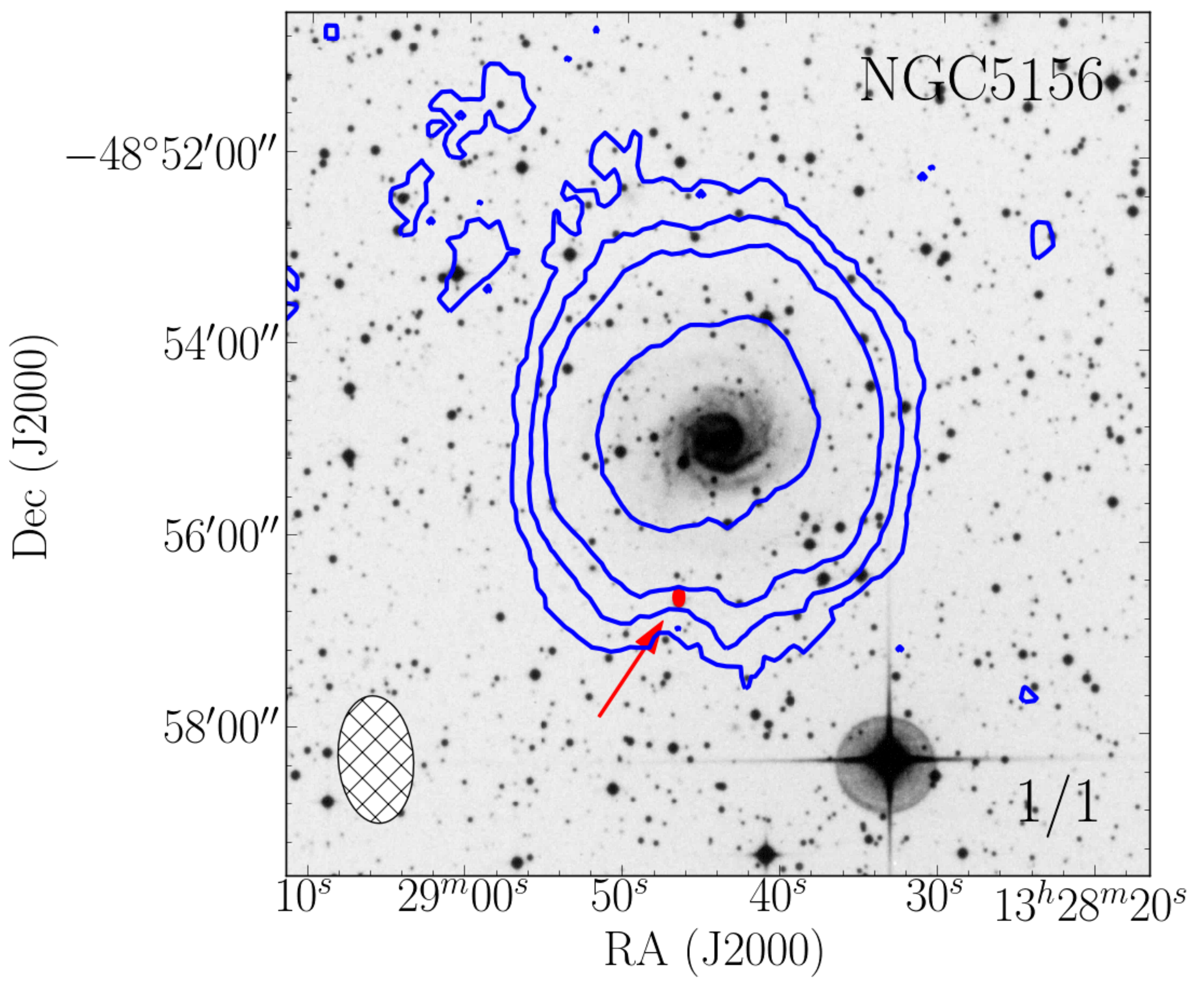}
\includegraphics[width=0.3\linewidth]{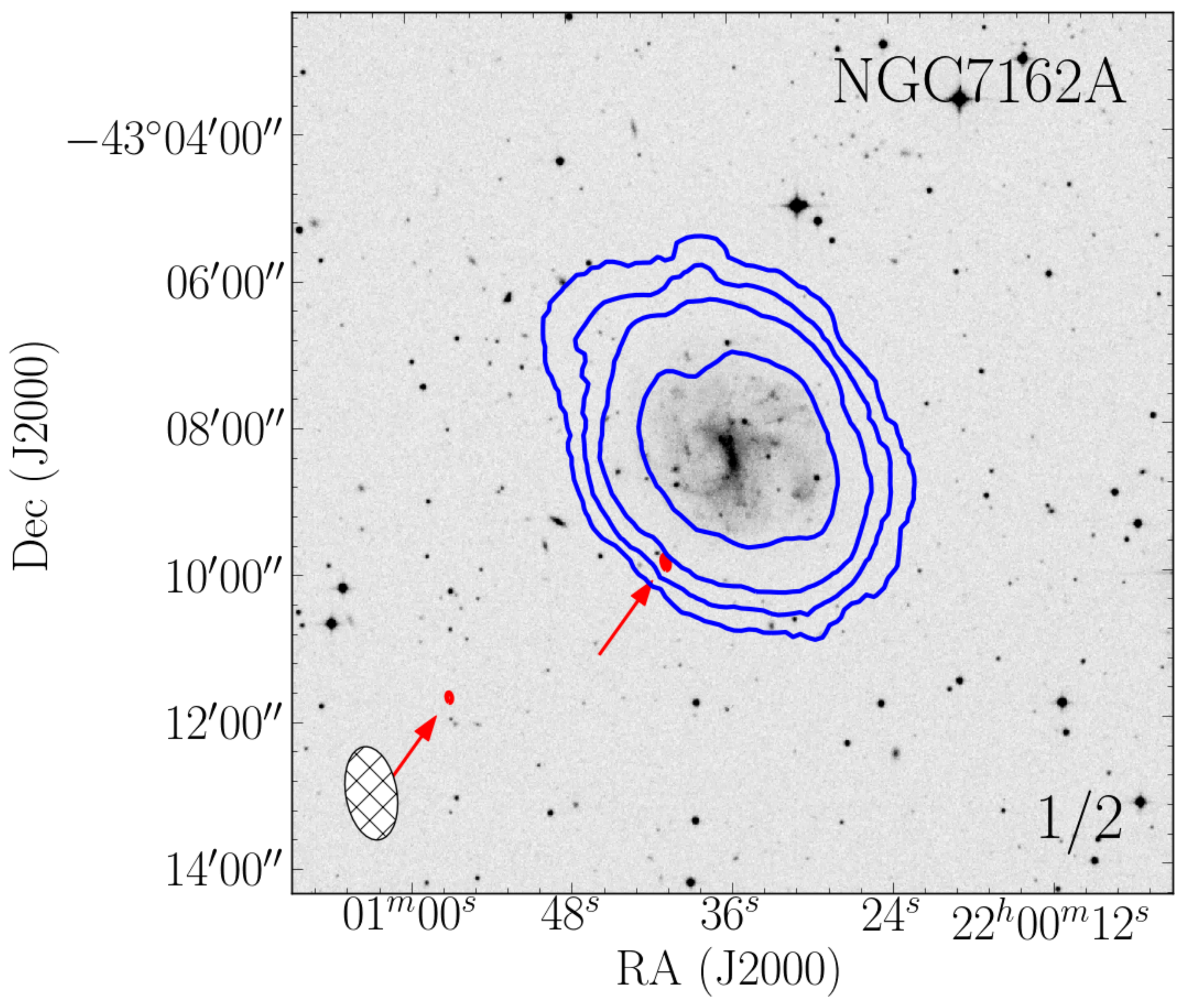}
\caption[]{SuperCOSMOS B-band images \citep{2001MNRAS.326.1279H} with \mbox{H\,{\sc i}} contours (blue) overlaid. 
The \mbox{H\,{\sc i}} contour levels are $N_{\mathrm{HI}}$ = 3 $\times$ 10$^{19}$, (1, 2, 5) $\times$ 10$^{20}$, and 1 $\times$ 10$^{21}$ cm$^{-2}$. 
The high resolution 1.4 GHz radio continuum emission is also overlaid (red contours), with an arrow indicating the location of the background continuum source(s) (which are offset from the centre of the galaxy). 
The number in the bottom right-hand corner shows how many of the individual sightlines intersect the disc at a column density greater than 1 $\times$ 10$^{20}$ cm$^{-2}$ (with the number of sightlines at the lower SUMSS resolution, from which these sources were selected, given in brackets, if different). 
A star indicates that the column density is borderline along one or more of the remaining sightlines. 
The synthesised beam for the \mbox{H\,{\sc i}} maps ($\sim$60 arcsec) is shown in the bottom left corner, and the synthesised beam for the continuum images (not shown) is $\sim$5 arcsec.}
\label{figure:overlay_maps}
\end{figure*}

In Figure \ref{figure:overlay_maps} we show the \mbox{H\,{\sc i}} distribution overlaid on the SuperCOSMOS optical image \citep{2001MNRAS.326.1279H} for each galaxy. 
The radio continuum emission contours are also plotted, allowing us to examine the location of the background source(s) relative to the \mbox{H\,{\sc i}} disc. 
We find that five of the 17 sightlines in our sample (or 6 of the 23 individual components seen at higher resolution) intersect the \mbox{H\,{\sc i}} disc of the galaxy at column densities above $\sim$1-2 $\times$ 10$^{20}$ cm$^{-2}$ (the limit of our absorption-line sensitivity). 
We discuss this further in Section \ref{discussion}.

\begin{table*}
\begin{minipage}{\linewidth}
\centering
\caption{Derived \mbox{H\,{\sc i}} parameters of the target galaxies (listed in Table \ref{table:hipass}). 
Column (1) is the optical galaxy ID. 
Column (2) is the \mbox{H\,{\sc i}} mass measured from the ATCA data, assuming $D = v_{\mathrm{LG}}/H_{0}$ (with an estimated uncertainty of $\pm$ 5 per cent). 
Column (3) is the \mbox{H\,{\sc i}} mass from HIPASS. 
The HIPASS masses have been re-scaled using a Hubble constant of H$_{0}$ = 71 km s$^{-1}$ Mpc$^{-1}$, as elsewhere in this work, and the mean uncertainty on the HIPASS values is 15 per cent (assuming that the uncertainty on the integrated \mbox{H\,{\sc i}} flux is the dominant source of error). 
Columns (4) and (5) are the ellipse parameters (ellipticity and position angle) derived from the \mbox{H\,{\sc i}} maps, used to produce the azimuthally averaged radial profiles shown in Figure \ref{figure:radial_profiles}. 
Column (6) is the optical disc size ($R_{25}$, taken from the RC3 catalogue \citealt{1991rc3..book.....D}). 
Column (7) is the \mbox{H\,{\sc i}} disc size, derived from the radial profiles. 
Column (8) is the ratio of the \mbox{H\,{\sc i}} and optical disc sizes.}
\label{table:HI_properties_derived}
\begin{tabular}{@{} lrrrrrrr @{}} 
\hline
& \multicolumn{2}{c}{log$_{10}$M$_{\mathrm{HI}}$} & & & & & \\
& ATCA & HIPASS &  Ellipticity & PA & R$_{\mathrm{opt}}$ & R$_{\mathrm{HI}}$ & R$_{\mathrm{HI}}$/R$_{\mathrm{opt}}$ \\
& (M$_{\odot}$) & (M$_{\odot}$) & & (deg) & (kpc) & (kpc) & \\
\hline
ESO\,300-G\,014 & 8.87 & 8.91 & 0.49 & $-$9.9 & 8.3 & 9.9 & 1.2 \\
ESO\,357-G\,012 & 9.31 & 9.25 & 0.30 & $-$35.7 & 8.1 & 14.6 & 1.8 \\
ESO\,363-G\,015 & 8.87 & 8.90 & 0.40 & $-$3.3 & 5.2 & 9.3 & 1.8 \\
IC\,1914 & 9.16 & 9.21 & 0.45 & $-$68.4 & 6.6 & 12.2 & 1.8 \\
IC\,4386 & 9.75 & 9.85 & 0.57 & $-$8.3 & 10.1 & 26.0 & 2.6 \\
NGC\,1249 & 9.52 & 9.58 	& 0.57 & 87.9 & 8.9 & 18.9 & 2.1 \\
NGC\,1566 & 9.92 & 10.04 & 0.02 & 135.0 & 21.8 & 26.7 & 1.2 \\
NGC\,2188 & 8.64 & 8.59 & 0.55 & $-$0.4 & 4.5 & 6.4 & 1.4 \\
NGC\,5156 & 9.75 & 9.79 	& 0.11 & $-$3.4 & 13.1 & 22.4 & 1.7 \\
NGC\,7162A & 9.63 & 9.78 & 0.31 & 35.2 & 12.1 & 22.0 & 1.8 \\
\hline
\end{tabular}
\end{minipage}
\end{table*}

\subsection{Radial H\,{\sevensize\bf I} profiles}
\label{results_part1:radial_profiles}

\begin{figure*}
\includegraphics[width=\textwidth]{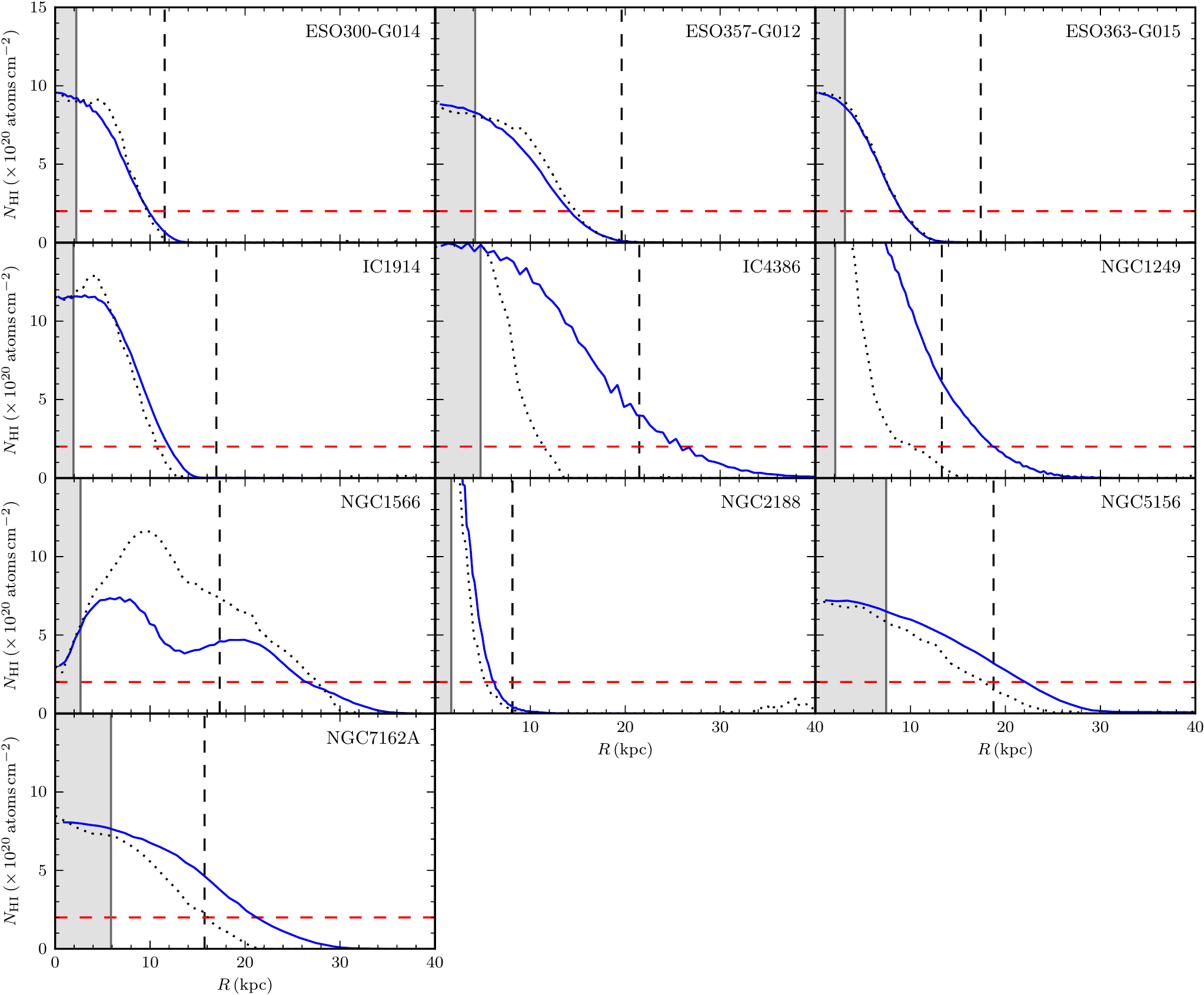}
\caption{Radial \mbox{H\,{\sc i}} profiles of the galaxies in our sample, derived from the ATCA \mbox{H\,{\sc i}} total intensity maps. 
The blue (solid) curves show the azimuthally averaged profile, and the black (dotted) curves the profile along the axis towards the continuum source. 
The impact parameter of the continuum source is shown by the dashed vertical line. 
For fields with multiple background sources, the continuum axis profile is just shown for the first source (as given in Table \ref{table:sumss}), as is the dotted line indicating the impact parameter of the sightline. 
The grey shaded region indicates semi-major axis of the synthesised beam and dashed horizontal line indicates the DLA limit ($N_{\mathrm{HI}}$ = 2 $\times$ 10$^{20}$ cm$^{-2}$).}
\label{figure:radial_profiles}
\end{figure*}

In Figure \ref{figure:radial_profiles} we show the radial \mbox{H\,{\sc i}} profiles of the galaxies in our sample. 
For each galaxy we show two versions of the profile -- the azimuthally averaged profile over the whole gas disc, and the profile along the axis to the continuum source. 
The ellipse parameters used to derive the azimuthally averaged profiles are given in Table \ref{table:HI_properties_derived}. 
For fields with multiple background sources, the continuum axis profile is just shown for the first source (as listed in Table \ref{table:sumss}), which was the main target of our observations (while other sources were simply found by chance in the same field, and did not necessarily meet all of the specified selection criteria).

As in Paper I we see a variety of different profile shapes -- from relatively flat profiles as in NGC\,5156 and NGC\,7162A, to much steeper profiles such as NGC\,2188 and IC\,1914. 
Comparing the azimuthally averaged profiles to the profiles in the direction of the continuum source, we find that the two profiles are almost identical for half of the galaxies in our sample. 
Only three galaxies show dramatically different profiles -- IC\,4386 and NGC\,1249 (where the profile towards the continuum source drops off much more steeply, as the source is not located along the major axis), and NGC\,1566 (due to local variations in the \mbox{H\,{\sc i}} distribution from the spiral arms of the galaxy).

We have also measured the size of the \mbox{H\,{\sc i}} disc (from the azimuthally-averaged profiles) at a column density of $N_{\mathrm{HI}}$ = 2 $\times$ 10$^{20}$ cm$^{-2}$. 
The measured \mbox{H\,{\sc i}} disc sizes are presented in Table \ref{table:HI_properties_derived}, alongside the optical disc sizes ($R_{25}$, taken from the literature). 
There is a large range in disc sizes -- the \mbox{H\,{\sc i}} discs range from 6.4 to 26.7 kpc, with \mbox{H\,{\sc i}}-to-optical disc ratios of 1.2 to 2.6. 
Typically though, we find the \mbox{H\,{\sc i}} disc to be around 10-20 kpc in radius, and around twice the size of the optical disc.

\section{H\,{\sevensize\bf I} emission and absorption in the spectra towards the background sources}
\label{results_part2}

\subsection{Spectra and continuum images}
\label{results_part2:continuum_images_and_spectra}

In Figure \ref{figure:continuum_and_spectra} we show the 1.4 GHz radio continuum images of each of the background sources in our sample, and below each the spectrum towards that source (at all three resolutions). 
As in Paper I, we apply a Bayesian analysis to all spectra, using the {\sc multi-nest} algorithm \citep{2008MNRAS.384..449F,2009MNRAS.398.1601F}, to determine objectively whether there is \mbox{H\,{\sc i}} emission and/or absorption present.

The development and testing of this algorithm are described by \citet{2012PASA...29..221A}, and the application to real spectral-line data by \citet{2014MNRAS.440..696A} and references therein. 
Since then, the software has been upgraded and is now capable of fitting simultaneously for a combination of both emission and absorption. 
It also now includes a number of additional line profiles such as the `Busy Function' \citep{2014MNRAS.438.1176W}, which we used in fitting the double-horned emission-line profile seen along the sightline towards C-IC\,4386. 
All other spectra were fit (or upper limits calculated) assuming a gaussian line-profile.

Throughout this work we use the statistic referred to as the $R$-value to give a measure of the significance of any detected spectral lines. 
For a full explanation of this statistic, and further details on the application of this algorithm to our data, we refer the reader to our previous work \citep{2015MNRAS.450..926R} as well as the above references.

\subsection{H\,{\sevensize\bf I} emission and absorption along the target sightlines}
\label{results_part2:spectra_results}

\subsubsection{\mbox{H\,{\sc i}} emission}

\mbox{H\,{\sc i}} emission-lines were detected along 10 of the 17 sightlines in the low resolution spectra, 6 sightlines in the medium resolution spectra, and 2 sightlines in the high resolution spectra. 
This indicates that more than half of the sightlines in our sample intersect the \mbox{H\,{\sc i}} disc at some column density. 
Where an emission-line was detected in the higher resolution spectra, a detection was also always made at the lower resolutions (but the reverse is obviously not always true). 
Where emission was detected only in the low resolution spectra the column density of the gas is typically a few $\times$ 10$^{20}$ cm$^{-2}$ -- but for sightlines with detections in the medium or high resolution spectra, this implies the presence of gas with column densities of up to $N_{\mathrm{HI}}$ = 10$^{21}$-10$^{22}$ cm$^{-2}$.

Table \ref{table:emission_line_results} gives the best-fitting parameters for each of the detected  emission-lines, as well as the derived \mbox{H\,{\sc i}} column density along these sightlines. 
We note that, since the beam in the low and medium resolution cubes is much larger than that in the high resolution cubes, the number of sightlines with emission-line detections $\gtrsim$1-2 $\times$ 10$^{20}$ cm$^{-2}$ is slightly higher than we might predict based on the \mbox{H\,{\sc i}} shown in Figure \ref{figure:overlay_maps}. 
This is simply a consequence of the fact that the beam is much larger at lower resolution, and that the emission-lines seen in the extracted spectra represent a weighted average over this beam (or put another way, the much smaller beam of the high resolution continuum images means we cannot predict the column density along a given sightline quite as accurately as it might appear we can from Figure~\ref{figure:overlay_maps}).

\subsubsection{\mbox{H\,{\sc i}} absorption}

Despite the fact that almost one-third of the sightlines searched intersect the \mbox{H\,{\sc i}} disc of the foreground galaxy at column densities above $\sim$1-2 $\times$ 10$^{20}$ cm$^{-2}$, we have detected only one absorption-line in our sample. 
The absorption arises in the galaxy NGC\,5156, a barred spiral galaxy at a redshift of $z = 0.01$. 
The sightline intersects the disc at an impact parameter of 19 kpc, and the absorption-line has an integrated optical depth of 0.82 km s$^{-1}$. 
We note that this is the highest impact parameter of any intervening \mbox{H\,{\sc i}} absorption-line detected to date (though, as mentioned in Section \ref{introduction}, this is still much lower than what has been seen in optical Lyman-$\alpha$ studies). 
The background source, C-NGC\,5156, is by far the brightest in our sample, with a 1.4 GHz flux of around 400 mJy beam$^{-1}$, so given that we have made only one detection in our sample it is not surprising that it is along this sightline.

To test for absorption below the noise, we have also stacked the spectra from all of the non-detections to search for a statistical detection of absorption along these sightlines. 
Despite the large number of sightlines stacked we did not find any evidence for absorption. 
We calculate a 5-$\sigma$ upper limit of $\tau_{\mathrm{peak}}$ $\lesssim$ 0.015, which corresponds to a column density of $N_{\mathrm{HI}}$ $\lesssim$ 1.5 $\times$ 10$^{20}$.

Table \ref{table:absorption_line_results} presents the best-fitting parameters for the detected absorption-line, as well as upper limits for each of the non-detections. 
We discuss the detected absorption-line in greater detail in Section \ref{results_part3} and investigate the reasons for the low detection rate in our sample in Section \ref{discussion}.
 
\subsubsection{Spin temperature of the gas}

We can also use the combined emission- and absorption-line data to investigate the spin temperature of the gas along the sightlines studied. 
VLBI data is required to resolve the background sources in order to measure the covering factor and thus determine an accurate spin temperature. 
However, the ATCA data can still be used to estimate the \emph{ratio} of the spin temperature to the covering factor ($T_{\mathrm{S}}/f$). 
For the detected absorption-line we estimate the ratio of the spin temperature and covering factor is $T_{\mathrm{S}}/f$ $\approx$ 950 K (something we would be able to refine with improved VLBI observations -- see Section \ref{results_part3}).

For the non-detections, we can still use the combined data (\mbox{H\,{\sc i}} column density from the emission-line data and the upper limit on the optical depth from the absorption-line data) to put some constraints on the value of $T_{\mathrm{S}}/f$.  
We find that the lower limit is typically $T_{\mathrm{S}}/f$ $\sim$10-100 K (see Table \ref{table:absorption_line_results}) --- however, for sightlines where emission was detected in the medium or high resolution cubes (implying column densities of $\sim$10$^{21}$-10$^{22}$ cm$^{-2}$), we obtain lower limits of up to several thousand K.

This is consistent with the lower limits of a few hundred K found by \citet{1992ApJ...399..373C} from similar emission- and absorption-line data, as well as estimates from recent Ly-$\alpha$ studies which typically find spin temperatures of a few tens to a few hundred K \citep{2010ApJ...713..131B,2014ApJ...795...98B,2013MNRAS.428.2198S,2014MNRAS.438.2131K}. 
We stress, however, that, since our limits are based on non-detections, they are not very restrictive, and absorption-line detections would be required to obtain more accurate estimates of the value of $T_{\mathrm{S}}/f$ along these sightlines.

For the sightlines where the lower limit on $T_{\mathrm{S}}/f$ is very high, we suggest that this can be explained by a clumpy gas medium, like that suggested by \citet{2012ApJ...749...87B}. 
In this scenario we have high column density clumps ($\sim$100 pc in size) embedded within a more diffuse gas. 
This would result in a high average column density in the large beam of the lower resolution cubes but, if the narrow sightline towards the background source misses these small, dense clumps, would still give an absorption-line non-detection, as observed. 
Evidence for such structures is also seen in other recent \mbox{H\,{\sc i}} absorption-line studies \citep{2013MNRAS.431.3408C,2013MNRAS.428.2198S,2010ApJ...713..131B,2011ApJ...727...52B,2014ApJ...795...98B}.

\begin{table*}
\begin{minipage}{\linewidth}
\centering
\caption{The best-fitting parameters estimated for the detected \mbox{H\,{\sc i}} emission-lines. 
Columns (1) and (2) are the sightline and cube resolution at which the line was detected. 
Columns (3), (4), and (5) are the velocity, width, and peak flux of the line.
Column (6) is the integrated \mbox{H\,{\sc i}} line flux. 
Column (7) is the derived \mbox{H\,{\sc i}} column density. 
Column (8) is the $R$-value of the detected line (see paper I).}
\label{table:emission_line_results}
\begin{tabular}{@{} llrrrrrr @{}} 
\hline
Sightline & Cube & Velocity ($cz$) & S$_{\mathrm{peak}}$ & Width ($\Delta cz$) & $\int S\,dv$ & N$_{\mathrm{HI}}$ & $R$-value \\
 & resolution & (km s$^{-1}$) & (mJy beam$^{-1}$) & (km s$^{-1}$) & (mJy beam$^{-1}$ km s$^{-1}$) & ($\times$10$^{20}$ cm$^{-2}$) &  \\
\hline
\vspace{+1mm}
C-ESO\,300-G\,014-1 & Low & 1014.7$^{+1.9}_{-2.0}$ & 11.2$^{+1.3}_{-1.2}$ & 39.3$^{+5.3}_{-4.8}$ & 439.4$^{+45.0}_{-46.0}$ & 1.4$^{+0.1}_{-0.1}$ & 58.29$ \pm $0.06 \\
C-ESO\,300-G\,014-2 & Low & 915.8$^{+0.4}_{-0.4}$ & 58.4$^{+1.2}_{-1.2}$ & 45.4$^{+1.1}_{-1.1}$ & 2650.3$^{+51.3}_{-52.1}$ & 8.1$^{+0.2}_{-0.2}$ & 1973.42$ \pm $0.07 \\
C-ESO\,300-G\,014-2 & Medium & 911.3$^{+2.1}_{-2.0}$ & 11.8$^{+1.1}_{-1.1}$ & 47.1$^{+5.8}_{-5.3}$ & 555.3$^{+49.9}_{-50.7}$ & 12.1$^{+1.1}_{-1.1}$ & 85.14$ \pm $0.06 \\
\vspace{+1mm}
C-ESO\,300-G\,014-2 & High & 920.4$^{+7.3}_{-6.7}$ & 4.4$^{+1.8}_{-1.4}$ & 48.8$^{+25.9}_{-18.1}$ & 214.5$^{+71.6}_{-67.0}$ & 54.7$^{+18.2}_{-17.1}$ & 3.92$ \pm $0.04 \\
\vspace{+1mm}
C-ESO\,357-G\,012 & Low & 1508.4$^{+2.8}_{-2.8}$ & 8.6$^{+1.2}_{-1.2}$ & 42.0$^{+7.1}_{-5.8}$ & 359.8$^{+48.7}_{-46.2}$ & 1.0$^{+0.1}_{-0.1}$ & 32.71$ \pm $0.05 \\
\vspace{+1mm}
C-IC\,4386-2 & Low & 1807.1$^{+3.2}_{-3.2}$ & 10.6$^{+0.7}_{-0.6}$ & 108.6$^{+7.4}_{-6.9}$ & 1149.2$^{+69.6}_{-69.3}$ & 3.2$^{+0.2}_{-0.2}$ & 183.38$ \pm $0.06 \\
C-NGC\,1249-1 & Low & 1040.7$^{+1.6}_{-1.6}$ & 14.3$^{+1.3}_{-1.2}$ & 39.9$^{+4.3}_{-4.0}$ & 571.8$^{+48.6}_{-46.4}$ & 2.1$^{+0.2}_{-0.2}$ & 101.08$ \pm $0.06 \\
\vspace{+1mm}
C-NGC\,1249-1 & Medium & 1048.0$^{+3.0}_{-3.2}$ & 5.0$^{+1.4}_{-1.3}$ & 25.7$^{+8.7}_{-6.7}$ & 128.6$^{+31.4}_{-30.6}$ & 2.9$^{+0.7}_{-0.7}$ & 6.43$ \pm $0.05 \\
C-NGC\,1566-1 & Low & 1415.0$^{+0.1}_{-0.2}$ & 142.9$^{+1.9}_{-2.0}$ & 25.1$^{+0.4}_{-0.4}$ & 3583.5$^{+43.9}_{-44.9}$ & 13.9$^{+0.2}_{-0.2}$ & 5000.22$ \pm $0.07 \\
C-NGC\,1566-1 & Medium & 1413.2$^{+0.5}_{-0.5}$ & 34.6$^{+2.1}_{-2.0}$ & 20.5$^{+1.6}_{-1.5}$ & 706.9$^{+37.0}_{-35.4}$ & 15.5$^{+0.8}_{-0.8}$ & 310.68$ \pm $0.06 \\
\vspace{+1mm}
C-NGC\,1566-1 & High & 1415.9$^{+2.4}_{-2.4}$ & 10.9$^{+1.7}_{-1.6}$ & 35.3$^{+6.7}_{-6.2}$ & 383.2$^{+56.8}_{-53.7}$ & 130.3$^{+19.3}_{-18.2}$ & 29.26$ \pm $0.05 \\
C-NGC\,1566 (JUN13) & - & 1413.2$^{+2.3}_{-2.4}$ & 5.5$^{+0.9}_{-0.9}$ & 28.6$^{+5.0}_{-4.4}$ & 156.1$^{+24.4}_{-24.3}$ & 16.6$^{+2.6}_{-2.6}$ & 18.80$ \pm $0.05 \\
C-NGC\,1566-2 & Low & 1428.7$^{+0.5}_{-0.5}$ & 42.0$^{+1.6}_{-1.5}$ & 30.9$^{+1.4}_{-1.3}$ & 1297.5$^{+44.0}_{-44.8}$ & 5.0$^{+0.2}_{-0.2}$ & 630.20$ \pm $0.07 \\
C-NGC\,1566-2a & Medium & 1427.3$^{+2.5}_{-2.6}$ & 8.2$^{+1.2}_{-1.1}$ & 36.5$^{+6.1}_{-5.3}$ & 300.6$^{+40.0}_{-40.3}$ & 6.6$^{+0.9}_{-0.9}$ & 30.59$ \pm $0.05 \\
\vspace{+1mm}
C-NGC\,1566-2b & Medium & 1431.4$^{+1.7}_{-1.6}$ & 11.5$^{+1.4}_{-1.3}$ & 30.6$^{+4.5}_{-4.3}$ & 351.5$^{+39.0}_{-38.8}$ & 7.7$^{+0.9}_{-0.9}$ & 54.22$ \pm $0.06 \\
\vspace{+1mm}
C-NGC\,2188 & Low & 743.9$^{+4.3}_{-4.4}$ & 6.8$^{+0.9}_{-0.9}$ & 83.4$^{+17.1}_{-14.8}$ & 562.3$^{+73.8}_{-71.5}$ & 1.5$^{+0.2}_{-0.2}$ & 54.42$ \pm $0.05 \\
C-NGC\,5156 & Low & 2933.1$^{+1.7}_{-1.7}$ & 30.7$^{+3.1}_{-2.5}$ & 69.7$^{+6.3}_{-5.7}$ & 2147.7$^{+127.0}_{-122.9}$ & 6.5$^{+0.4}_{-0.4}$ & 207.46$ \pm $0.08 \\
\vspace{+1mm}
C-NGC\,5156 & Medium & 2942.7$^{+7.0}_{-6.9}$ & 7.9$^{+1.4}_{-1.2}$ & 95.1$^{+8.2}_{-14.0}$ & 728.9$^{+112.6}_{-109.4}$ & 14.3$^{+2.2}_{-2.2}$ & 18.90$ \pm $0.07 \\
C-NGC\,7162A-1 & Low & 2247.7$^{+1.3}_{-1.3}$ & 22.5$^{+1.1}_{-1.1}$ & 59.1$^{+3.8}_{-3.5}$ & 1331.3$^{+63.1}_{-61.4}$ & 4.6$^{+0.2}_{-0.2}$ & 372.45$ \pm $0.06 \\
C-NGC\,7162A-1 & Medium & 2242.7$^{+3.9}_{-4.4}$ & 5.9$^{+2.2}_{-2.0}$ & 30.7$^{+17.2}_{-10.7}$ & 181.7$^{+56.1}_{-52.4}$ & 4.4$^{+1.3}_{-1.3}$ & 5.07$ \pm $0.05 \\
\hline
\end{tabular}
\end{minipage}
\end{table*}

\begin{table*}
\begin{minipage}{\linewidth}
\centering
\caption{Absorption-line parameters and estimates of $T_{\mathrm{S}}/f$. 
Column (1) is the sightline searched. 
Columns (2)-(6) are the values derived from the absorption-line data (3-$\sigma$ upper limits for the non-detections). 
Column (2) is the rms-noise level. 
Column (3) is the peak 1.4 GHz continuum flux of the background source. 
Column (4) is the peak optical depth of the line.
Column (5) is the integrated optical depth. For non-detections we have calculated the upper limit assuming a gaussian profile with a line-width of 10 km s$^{-1}$. 
Column (6) is the absorption-line \mbox{H\,{\sc i}} column density (assuming $T_{\mathrm{S}}$ = 100 K and $f$ = 1.0). 
Columns (7) and (8) are the relevant emission-line values used to estimate $T_{\mathrm{S}}/f$.
Column (7) is the emission-line \mbox{H\,{\sc i}} column density. 
Column (8) is the weighting scheme from which the emission-line column density was derived. 
Column (9) is the limit for $T_{\mathrm{S}}/f$ calculated from the combined emission- and absorption-line data. 
If an emission-line was detected at multiple spatial resolutions, we have calculated a separate limit for $T_{\mathrm{S}}/f$ at each resolution. 
For the deeper absorption-line observations of C-NGC\,1566-1 and C-NGC\,5156 (made in June 2013 and November 2013, respectively) the limit on $T_{\mathrm{S}}/f$ is calculated using the emission-line column density derived from the original (750 array) observations, which we have denoted by square brackets.
}
\label{table:absorption_line_results}
\begin{tabular}{@{} lrrrrrrlr @{}} 
\hline
&  \multicolumn{5}{l}{\mbox{H\,{\sc i}} absorption (high resolution spectra)} & \multicolumn{2}{l}{\mbox{H\,{\sc i}} emission} & \multicolumn{1}{l}{Combined}\\
\hline
Sightline & $\sigma_{\mathrm{chan}}$ & S$_{\mathrm{peak,1.4}}$ & $\tau_{\mathrm{peak}}$ & $\int \tau\,dv$ & N$_{\mathrm{HI}}$(abs) & N$_{\mathrm{HI}}$(em) & Cube & T$_{\mathrm{S}}/f$ \\
 & (mJy beam$^{-1}$) & (mJy beam$^{-1}$) & (per cent) & (km s$^{-1}$) & ($\times$10$^{20}$ cm$^{-2}$) & ($\times$10$^{20}$ cm$^{-2}$) & resolution & (K) \\
\hline
\vspace{+1mm}
C-ESO\,300-G\,014-1 & 2.56 & 66.7$ \pm $1.8 & $<$0.12 & $<$1.22 & $<$2.2 & 1.4$^{+0.1}_{-0.1}$ & Low & $>$61 \\
C-ESO\,300-G\,014-2 & 2.74 & 53.5$ \pm $1.4 & $<$0.15 & $<$1.63 & $<$3.0 & 8.1$^{+0.2}_{-0.2}$ & Low & $>$274 \\
C-ESO\,300-G\,014-2 & 2.74 & 53.5$ \pm $1.4 & $<$0.15 & $<$1.63 & $<$3.0 & 12.1$^{+1.1}_{-1.1}$ & Medium & $>$406 \\
\vspace{+1mm}
C-ESO\,300-G\,014-2 & 2.74 & 53.5$ \pm $1.4 & $<$0.15 & $<$1.63 & $<$3.0 & 54.7$^{+18.2}_{-17.1}$ & High & $>$1841 \\
\vspace{+1mm}
C-ESO\,357-G\,012-1a & 2.35 & 21.4$ \pm $1.5 & $<$0.33 & $<$3.49 & $<$6.4 & 1.0$^{+0.1}_{-0.1}$ & Low & $>$15 \\
\vspace{+1mm}
C-ESO\,357-G\,012-1b & 2.05 & 5.6$ \pm $0.6 & $<$1.11 & $<$11.70 & $<$21.3 & 1.0$^{+0.1}_{-0.1}$ & Low & $>$5 \\
\vspace{+1mm}
C-ESO\,363-G\,015-1 & 2.36 & 39.4$ \pm $0.6 & $<$0.18 & $<$1.90 & $<$3.5 & - & - & - \\
\vspace{+1mm}
C-ESO\,363-G\,015-2a & 2.37 & 12.4$ \pm $1.2 & $<$0.57 & $<$6.09 & $<$11.1 & - & - & - \\
\vspace{+1mm}
C-ESO\,363-G\,015-2b & 2.44 & 11.4$ \pm $1.0 & $<$0.64 & $<$6.81 & $<$12.4 & - & - & - \\
\vspace{+1mm}
C-IC\,1914-1a & 2.46 & 58.4$ \pm $1.0 & $<$0.13 & $<$1.34 & $<$2.4 & - & - & - \\
\vspace{+1mm}
C-IC\,1914-1b & 2.33 & 38.9$ \pm $1.0 & $<$0.18 & $<$1.91 & $<$3.5 & - & - & - \\
\vspace{+1mm}
C-IC\,4386-1 & 2.48 & 81.1$ \pm $0.6 & $<$0.09 & $<$0.97 & $<$1.8 & - & - & - \\
\vspace{+1mm}
C-IC\,4386-2 & 2.46 & 20.0$ \pm $0.4 & $<$0.37 & $<$3.91 & $<$7.1 & 3.2$^{+0.2}_{-0.2}$ & Low & $>$45 \\
C-NGC\,1249-1 & 2.36 & 16.2$ \pm $0.6 & $<$0.44 & $<$4.64 & $<$8.5 & 2.1$^{+0.2}_{-0.2}$ & Low & $>$25 \\
\vspace{+1mm}
C-NGC\,1249-1 & 2.36 & 16.2$ \pm $0.6 & $<$0.44 & $<$4.64 & $<$8.5 & 2.9$^{+0.7}_{-0.7}$ & Medium & $>$34 \\
\vspace{+1mm}
C-NGC\,1249-2 & 2.34 & 14.1$ \pm $0.5 & $<$0.50 & $<$5.28 & $<$9.6 & - & - & - \\
\vspace{+1mm}
C-NGC\,1249-3a & 2.47 & 8.2$ \pm $0.5 & $<$0.90 & $<$9.54 & $<$17.4 & - & - & - \\
\vspace{+1mm}
C-NGC\,1249-3b & 2.46 & 7.5$ \pm $0.6 & $<$0.99 & $<$10.46 & $<$19.1 & - & - & - \\
C-NGC\,1566-1 & 2.48 & 23.9$ \pm $0.8 & $<$0.31 & $<$3.30 & $<$6.0 & 13.9$^{+0.2}_{-0.2}$ & Low & $>$231 \\
C-NGC\,1566-1 & 2.48 & 23.9$ \pm $0.8 & $<$0.31 & $<$3.30 & $<$6.0 & 15.5$^{+0.8}_{-0.8}$ & Medium & $>$258 \\
C-NGC\,1566-1 & 2.48 & 23.9$ \pm $0.8 & $<$0.31 & $<$3.30 & $<$6.0 & 130.3$^{+19.3}_{-18.2}$ & High & $>$2164 \\
C-NGC\,1566-1 (JUN13) & 1.31 & 32.0$ \pm $0.9 & $<$0.12 & $<$1.30 & $<$2.4 & [13.9$^{+0.2}_{-0.2}$] & Low & $>$584 \\
C-NGC\,1566-1 (JUN13) & 1.31 & 32.0$ \pm $0.9 & $<$0.12 & $<$1.30 & $<$2.4 & [15.5$^{+0.8}_{-0.8}$] & Medium & $>$655 \\
\vspace{+1mm}
C-NGC\,1566-1 (JUN13) & 1.31 & 32.0$ \pm $0.9 & $<$0.12 & $<$1.30 & $<$2.4 & [130.3$^{+19.3}_{-18.2}$] & High & $>$5484 \\
C-NGC\,1566-2a & 2.57 & 13.3$ \pm $0.5 & $<$0.58 & $<$6.16 & $<$11.2 & 5.0$^{+0.2}_{-0.2}$ & Low & $>$45 \\
\vspace{+1mm}
C-NGC\,1566-2a & 2.57 & 13.3$ \pm $0.5 & $<$0.58 & $<$6.16 & $<$11.2 & 6.6$^{+0.9}_{-0.9}$ & Medium & $>$59 \\
C-NGC\,1566-2b & 2.21 & 5.2$ \pm $0.3 & $<$1.27 & $<$13.42 & $<$24.5 & 5.0$^{+0.2}_{-0.2}$ & Low & $>$21 \\
\vspace{+1mm}
C-NGC\,1566-2b & 2.21 & 5.2$ \pm $0.3 & $<$1.27 & $<$13.42 & $<$24.5 & 7.7$^{+0.9}_{-0.9}$ & Medium & $>$32 \\
\vspace{+1mm}
C-NGC\,2188-1a & 2.55 & 13.0$ \pm $0.6 & $<$0.59 & $<$6.25 & $<$11.4 & 1.5$^{+0.2}_{-0.2}$ & Low & $>$13 \\
\vspace{+1mm}
C-NGC\,2188-1b & 2.31 & 8.8$ \pm $0.0 & $<$0.79 & $<$8.37 & $<$15.3 & 1.5$^{+0.2}_{-0.2}$ & Low & $>$10 \\
C-NGC\,5156 & 3.12 & 388.6$ \pm $4.7 & 0.15$^{+0.40}_{-0.05}$ & 0.96$^{+0.15}_{-0.11}$ & 1.7$^{+0.3}_{-0.2}$ & 6.5$^{+0.4}_{-0.4}$ & Low & 370 \\
C-NGC\,5156 & 3.12 & 388.6$ \pm $4.7 & 0.15$^{+0.40}_{-0.05}$ & 0.96$^{+0.15}_{-0.11}$ & 1.7$^{+0.3}_{-0.2}$ & 14.3$^{+2.2}_{-2.2}$ & Medium & 822 \\
C-NGC\,5156 (NOV13) & 3.29 & 398.0$ \pm $6.7 & 0.11$^{+0.01}_{-0.01}$ & 0.82$^{+0.04}_{-0.04}$ & 1.5$^{+0.1}_{-0.1}$ & [6.5$^{+0.4}_{-0.4}$] & Low & 430 \\
\vspace{+1mm}
C-NGC\,5156 (NOV13) & 3.29 & 398.0$ \pm $6.7 & 0.11$^{+0.01}_{-0.01}$ & 0.82$^{+0.04}_{-0.04}$ & 1.5$^{+0.1}_{-0.1}$ & [14.3$^{+2.2}_{-2.2}$] & Medium & 954 \\
C-NGC\,7162A-1 & 2.43 & 34.2$ \pm $0.8 & $<$0.21 & $<$2.25 & $<$4.1 & 4.6$^{+0.2}_{-0.2}$ & Low & $>$113 \\
\vspace{+1mm}
C-NGC\,7162A-1 & 2.43 & 34.2$ \pm $0.8 & $<$0.21 & $<$2.25 & $<$4.1 & 4.4$^{+1.3}_{-1.3}$ & Medium & $>$106 \\
C-NGC\,7162A-2 & 1.72 & 13.6$ \pm $0.6 & $<$0.38 & $<$4.01 & $<$7.3 & - & - & - \\
\hline
\end{tabular}
\end{minipage}
\end{table*}

\section{Absorption-line detection in NGC\,5156}
\label{results_part3}

\subsection{Initial detection and high-spectral resolution follow-up}
\label{results_part3:ngc5156_follow_up}

In our initial observations we detected a deep, narrow \mbox{H\,{\sc i}} absorption-line in the galaxy NGC\,5156, at an impact parameter of 19 kpc. 
Since the line was spectrally unresolved we conducted higher spectral resolution observations using the CABB 1M-0.5k configuration (and the 6A array) in order to resolve the line-profile. 
The follow-up observations confirm the presence of a deep absorption-line, and provide a spectrally resolved profile of the absorption feature. 
The absorption-line has a peak optical depth of 11 per cent, and a width of just 7.6 km s$^{-1}$, giving it an integrated optical depth of 0.82 km s$^{-1}$.

\begin{figure*}
\includegraphics[width=0.8\linewidth]{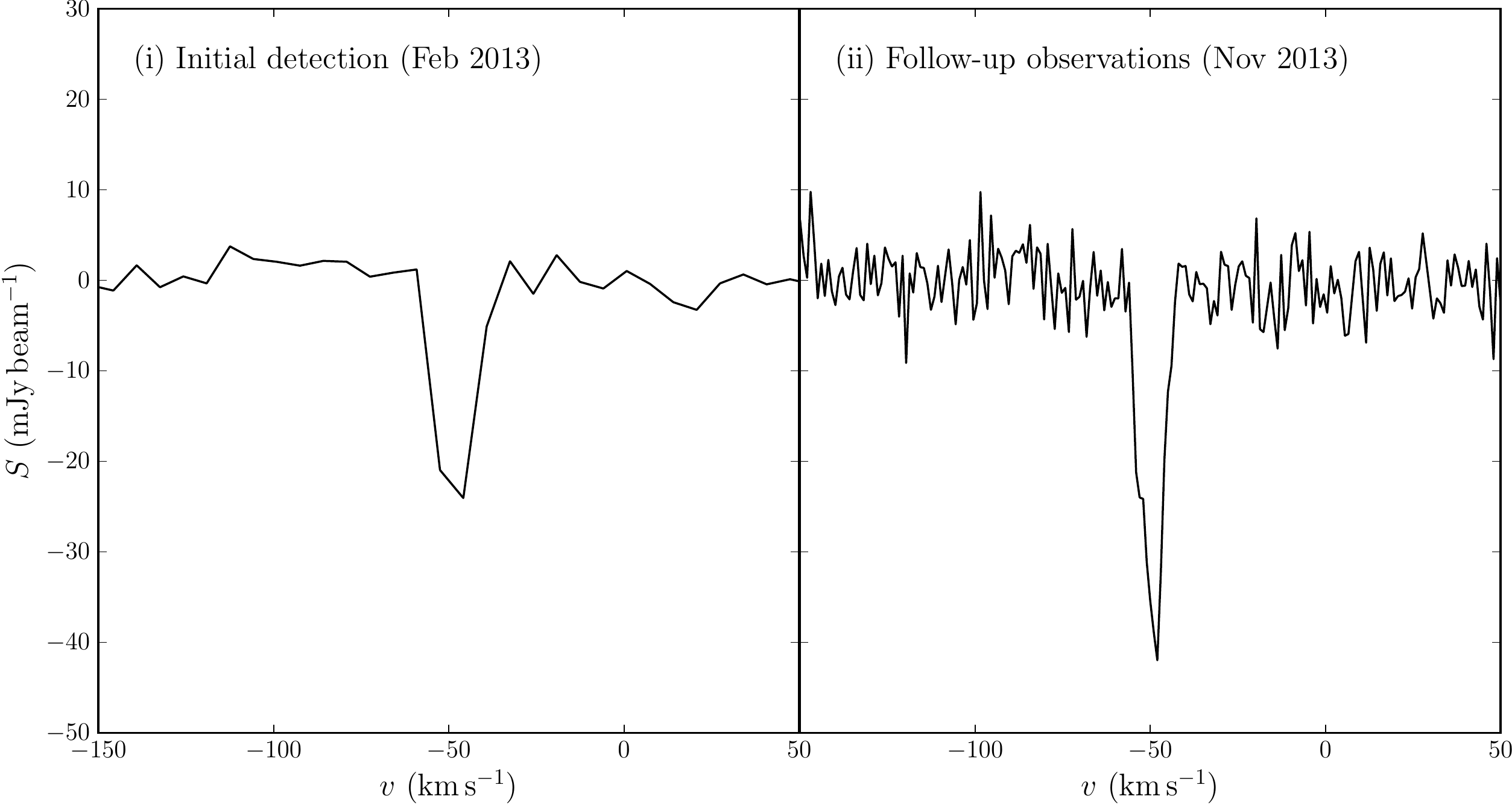}
\caption[]{Comparison of the low and high resolution \mbox{H\,{\sc i}} absorption spectra of NGC\,5156. While the two spectra appear to have very different line-depths, we have confirmed that this is merely an effect of the lower spectral resolution of the data in the initial spectrum, and that the data are completely consistent.}
\label{figure:ngc5156_comparison}
\end{figure*}

The two absorption-line spectra are presented in Figure \ref{figure:ngc5156_comparison}. 
We note that, while the two absorption-lines appear to have very different line-depths, this is merely the effect of the lower spectral resolution in the initial observation reducing the apparent depth of the line. 
We tested this by convolving the high resolution spectrum to the same resolution as the initial observation, and found the observations to be consistent.

\subsection{VLBI continuum observations}
\label{results_part3:vlbi}

To investigate the small-scale structure of the background radio source in the absorption-line system, a short Very Long Baseline Interferometry (VLBI) continuum observation of C-NGC\,5156 was conducted using the Long Baseline Array (LBA). 
The observations were made as part of a 1 $\times$ 18 h period on 2012 April 29-30 (along with a number of other sources which form part of a different project). 
We obtained two tracks at different hour angles, with a total on-source integration time of approximately 65 minutes. 
Ideally we would have obtained additional cuts, but the LST of the source, and scheduling constraints from other targets in our program made this impossible.

The array used consisted of 8 antennas -- the ATCA (6 dishes), Mopra, Parkes, Hobart, Ceduna, Katherine, Yaragadee, and Warkworth (New Zealand) -- with the longest baseline (Yaragadee-Warkworth, $\sim$5500 km) giving angular sensitivity down to $\sim$6 mas. 
Observations were conducted at 2.3 GHz which, although further from 1.4 GHz than we would have desired, offers the best sensitivity of all of the LBA bands, as well as additional stations for better \emph{uv}-coverage.

We observed with two polarisations (left- and right-hand circular polarisation), and a total bandwidth of 64 MHz (2240-2304 MHz). 
The ATCA, Parkes, Mopra, Hobart, and Ceduna observe with a single 64 MHz band, while Katherine, Yaragadee, and Warkworth observe with 4 contiguous 16 MHz bands. 
This means that band edges, where there is very poor sensitivity (and which are therefore normally flagged out), differ between antennas and these must therefore be treated separately when reducing the data.

Several observations of bright `fringe finders' (including 1921-293, the `main' fringe finder) were made throughout the observation, to allow for calibration of station clock delay and rate offsets. 
Since we did not know how bright the source would be at this resolution, we interleaved the observations of the target source (3.5 mins) with observations of a nearby, bright phase calibrator (1.5 mins) to allow phase referencing to be performed, if necessary.

The data were correlated at Curtin University of Technology using the DiFX-2 software correlator \citep{2007PASP..119..318D,2011PASP..123..275D} and standard continuum data parameters. 
All calibration was completed using the AIPS (`Classic') package \citep{1985daa..conf..195W}, following standard continuum reduction procedures. 
We note that, for the LBA, the system temperatures ($T_{\mathrm{sys}}$) are either not recorded or not reliable for many of the antennas, so amplitude calibration was instead performed using the nominal sensitivities available on the LBA Wiki\footnote{\url{http://www.atnf.csiro.au/vlbi/wiki/index.php?n=LBACalibrationNotes.NominalSEFD}}.

Unfortunately, owing to the limited \emph{uv}-coverage and complex source structure (discussed below), we found it was not possible to successfully image the source. 
However, examination of the \emph{uv}-data revealed evidence that the source is resolved into two compact components (seen as a sinusoid in the amplitude over the course of the observation). 
From the period of this sinusoid we estimated the separation of the two components to be a few arcseconds, meaning that the two sources should be resolvable with a higher frequency ATCA observation.

\subsection{ATCA 5 and 8 GHz continuum images}
\label{results_part3:atpmn}

Existing ATCA 5 and 8 GHz images of C-NGC\,5156 were available from the ATPMN survey \citep{2012MNRAS.422.1527M}. 
ATPMN presents high resolution ATCA observations of 8385 sources from the Parkes-MIT-NRAO survey \citep[PMN,][]{1993AJ....105.1666G}, allowing accurate positions and fluxes to be determined. 
Observations were conducted at two frequencies, 4800 and 8640 MHz, using the 6 km arrays, which results in a synthesised beam of 2.2 $\times$ 1.6 and 1.3 $\times$ 0.9 arcsec$^{2}$, respectively.

\begin{figure*}
\includegraphics[width=\linewidth]{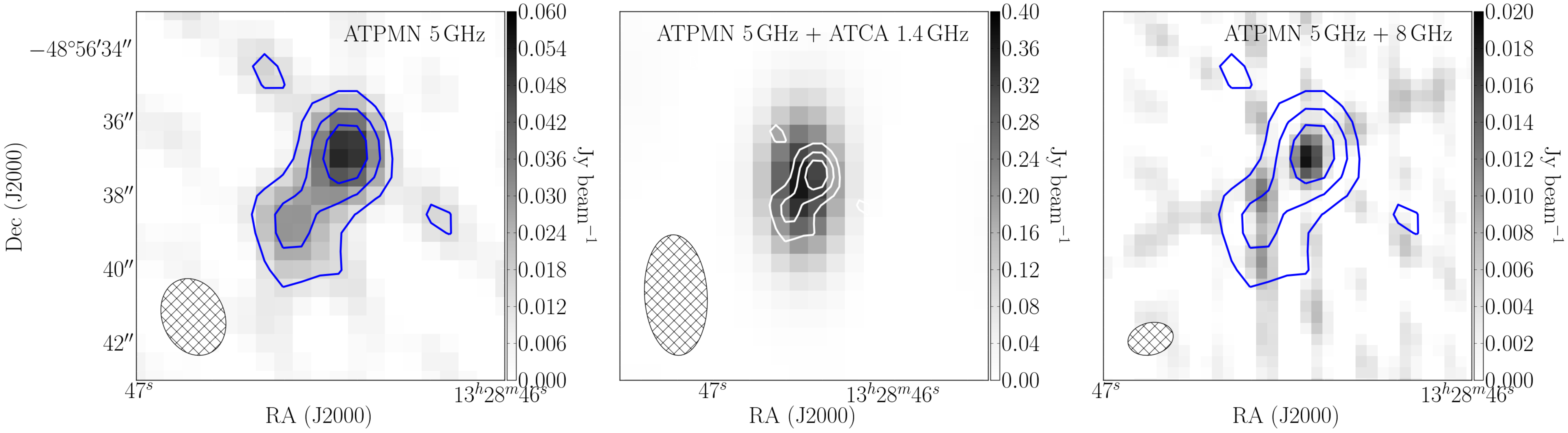}
\caption[]{Radio continuum images of the background radio source C-NGC\,5156, showing that the source is resolved into two components at higher resolution. 
From left to right we show: (i) the 5 GHz ATPMN image (with contours overlaid), (ii) the high resolution ATCA 1.4 GHz image with 5 GHz ATPMN contours overlaid, and (iii) the 8 GHz ATPMN image with 5 GHz ATPMN contours overlaid. The synthesised beam for the greyscale image is shown in the bottom left corner for each.}
\label{figure:ngc5156_atpmn}
\end{figure*}

As predicted from the VLBI data, the ATPMN images reveal that C-NGC\,5156 is resolved into two separate components at high resolution.   
The 5 and 8 GHz images are shown in Figure \ref{figure:ngc5156_atpmn}. 
The two sources have peak fluxes of 53.2 and 27.6 mJy beam$^{-1}$ at 5 GHz and 18.0 and 8.4 mJy beam$^{-1}$ at 8 GHz. 
The 5 GHz fluxes were measured by fitting an elliptical gaussian to each of the sources, but for the 8 GHz image (where the image quality is much worse) we have simply measured the peak flux at the same position as in the 5 GHz image. 
Upper limits on the integrated flux at 8 GHz were estimated by measuring the total flux in a region the same size as the source in the 5 GHz image. 
We note that the two components are catalogued as a single source in the ATPMN catalogue (presumably as a result of the small angular separation) but the integrated flux at each frequency is in good agreement with the combined fluxes that we measure for the two components.

The 5 GHz integrated flux from PMN is around 130 mJy, at a resolution of 4.2 arcmin. 
Given that the combined 5 GHz flux of the two components measured from ATPMN is around 110 mJy, this suggests about 20 mJy (or more) of the total 5 GHz flux of the source exists in a more diffuse component. 
We also see that at 8 GHz the second (weaker) component is only a marginal detection, suggesting that the continuum emission in this component is also more diffuse, with most of the compact continuum emission being associated with the brighter component. 
The 5 and 8 GHz fluxes of the two components are given in Table \ref{table:atpmn_fluxes}.

We measured the spectral index of both sources (from the peak fluxes, since the 8 GHz integrated fluxes are not very reliable) and find spectral indices of $\alpha_{5}^{8}$ = $-1.8$ and $-2.0$, respectively. 
The fact that both sources have a steep spectral index suggests that what we are seeing is two hotspots in the lobes of a distant radio galaxy. 
The angular separation between the two sources is 2.6 arcsec. 
Assuming a typical redshift of $z = 1$ for the radio source (see \citealt{1998AJ....115.1693C}) this would correspond to a physical separation of about 20 kpc, which is consistent with the hotspot scenario.

\begin{figure}
\includegraphics[width=\linewidth]{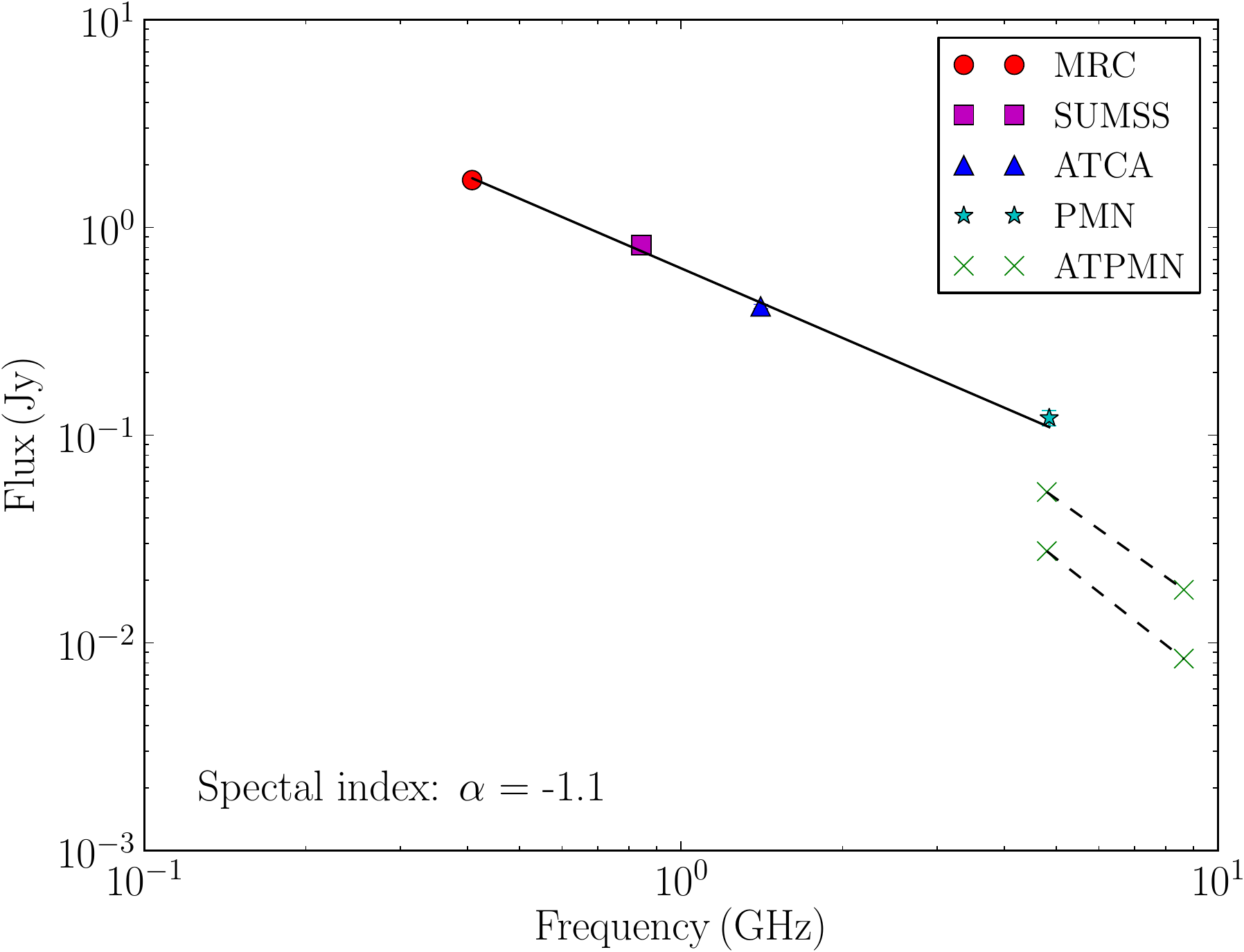}
\caption[]{The spectral energy distribution (SED) plot of the background radio source C-NGC\,5156 from 408 MHz to 4.85 GHz. The spectral index over these frequencies is $\alpha = -1.1$. We have also plotted the fluxes at 5 and 8 GHz, as measured from the ATPMN data, which show a steepening of the spectral index (likely due to more diffuse flux being resolving out).}
\label{figure:ngc5156_spectral_index}
\end{figure}

In Figure \ref{figure:ngc5156_spectral_index} we show the spectral energy distribution (SED) from the available radio continuum data. 
The spectral index between 408 MHz and 4.85 GHz is $\alpha = -1.1$, which is less steep than we found from the ATPMN data, and we suggest that this is likely due to the more diffuse continuum emission being resolved out as the resolution increases, as previously mentioned.

\begin{table*}
\begin{minipage}{\linewidth}
\centering
\caption{5 and 8 GHz fluxes for C-NGC\,5156 from the ATPMN and PMN data. 
Column (1) gives the source number, or the catalogue from which the measurement was taken. The first two rows are our measurements from the ATPMN images. The second two rows are the fluxes as published in the ATPMN \citep{2012MNRAS.422.1527M} and PMN \citep{1993AJ....105.1666G} catalogues. 
Columns (2) and (3) are the 5 GHz peak and integrated fluxes (where available). 
Columns (4) and (5) are the 8 GHz peak and integrated fluxes (where available). 
Column (6) is the spectral index between 5 and 8 GHz for each of the sources calculated from the ATPMN images.}
\label{table:atpmn_fluxes}
\begin{tabular}{@{} lrrrrrrr @{}} 
\hline
& \multicolumn{2}{c}{5 GHz Flux} & \multicolumn{2}{c}{8 GHz Flux}  \\
Source & S$_\mathrm{peak}$ & S$_\mathrm{int}$ & S$_\mathrm{peak}$ & S$_\mathrm{int}$ & $\alpha_{5}^{8}$ & \\
& (mJy beam$^{-1}$) & (mJy) & (mJy beam$^{-1}$) & (mJy) & \\
\hline
1 & 53.2 $\pm$ 6.7 & 74.3 $\pm$ 16.5 & 18.0 & $\lesssim$26 & $-$1.8  \\
2 & 27.6 $\pm$ 5.8 & 35.2 $\pm$ 11.7 & 8.4 & $\lesssim$29 & $-$2.0 &  \\
\hline
ATPMN & N/A & 105 $\pm$ 7 & N/A & 31 $\pm$ 10 & N/A \\ 
\hline
PMN & N/A & 129 $\pm$ 11 & N/A & N/A & N/A \\
\hline
\end{tabular}
\end{minipage}
\end{table*}

\subsection{Nature of the absorption-line system}
\label{discussion:ngc5156}

\begin{figure*}
\includegraphics[width=0.7\linewidth]{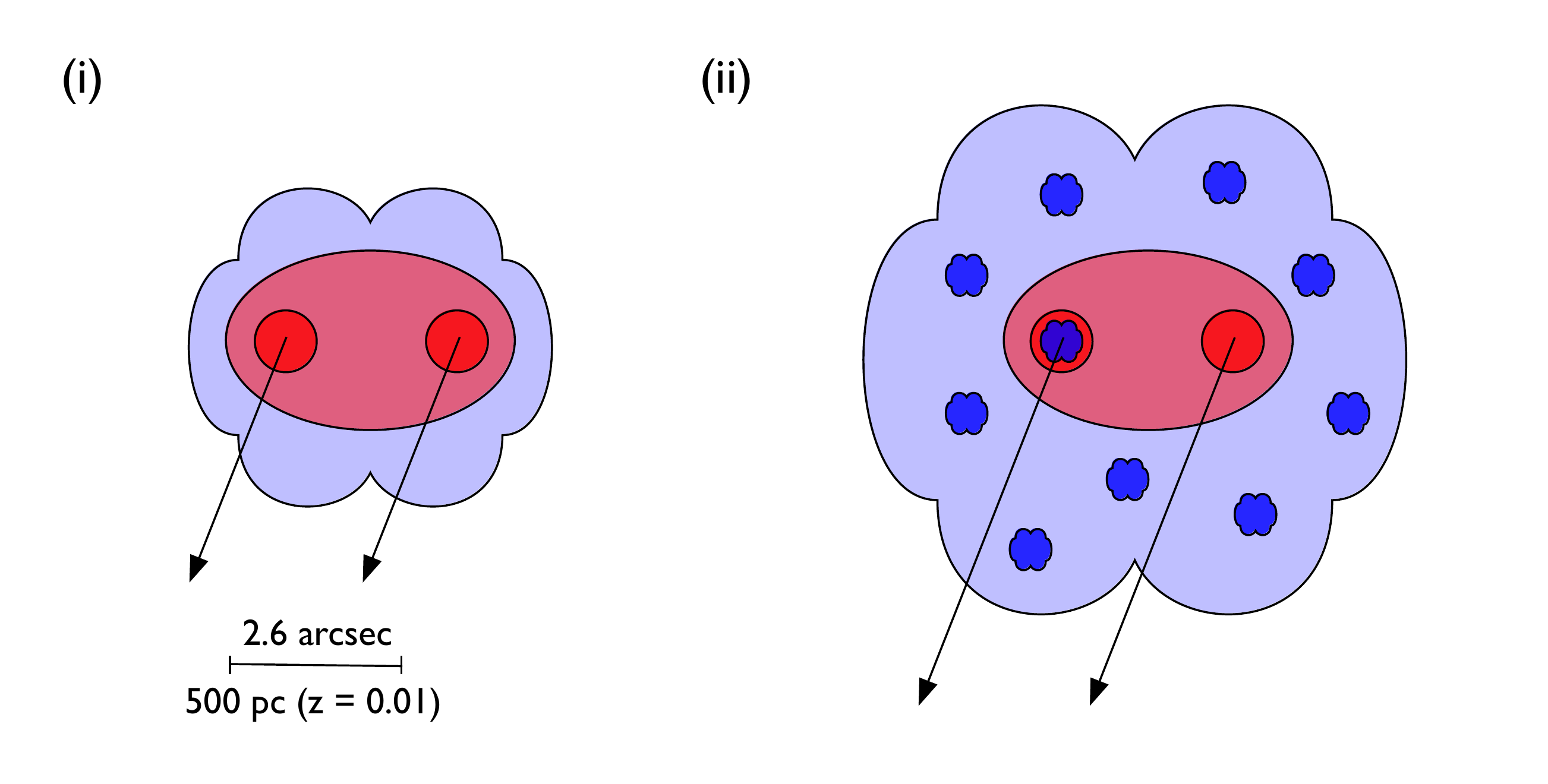}
\caption[]{Schematics showing the two different scenarios which may be giving rise to the absorption-line feature detected in the disc of NGC\,5156. Blue (cloud shapes) represent the \mbox{H\,{\sc i}} gas (with darker blue indicating denser clouds) and red (circles/ellipses) represent radio continuum emission (with darker red indicating more compact emission). The two possible scenarios are as follows: (i) the absorption is occurring against both components of the background source (as well as the more diffuse radio continuum emission) which would imply a minimum cloud size of 500 pc or (ii) the absorption is occurring against a single component of the background source, implying a small dense \mbox{H\,{\sc i}} cloud is responsible for the absorption.}
\label{figure:ngc5156_absorption_scenarios}
\end{figure*}

Knowing something about the structure of the background source, it is now possible to consider where the absorption is actually occurring. 
There are two likely scenarios, which are illustrated in Figure \ref{figure:ngc5156_absorption_scenarios}. 
Firstly the absorption may be occurring against both of the components (as well as against the more diffuse continuum emission). 
This would mean that the absorbing cloud is at least as large as the separation between the two components. 
At the redshift of the galaxy ($z = 0.01$), the angular separation of 2.6 arcsec corresponds to a physical size of around 500 pc, making this the minimum size of the absorbing cloud in this scenario.  
Alternatively, the absorption may be occurring against just one of the compact components. 
This would probably suggest that the absorbing gas is contained within a compact, dense \mbox{H\,{\sc i}} cloud (i.e. the same size or smaller than that of the background source). 

Given that the line is only 8 km s$^{-1}$ wide, this suggests that the absorption is occurring along only a very narrow sightline, leading us to favour the second scenario. 
This is also in agreement with previous work, suggesting that the typical size of absorbing clouds is around 100 pc \citep{2012ApJ...749...87B,2013MNRAS.431.3408C,2013MNRAS.428.2198S,2010ApJ...713..131B,2011ApJ...727...52B,2014ApJ...795...98B}. 
VLBI spectroscopy at 1.4 GHz would allow us to pin-point where the absorption is occurring and confirm that this is the case.

\section{Discussion}
\label{discussion}

\subsection{Detection rate of intervening absorption}
\label{discussion:detection_rate}

In our full sample of 16 galaxies (samples A+B) -- and 23 individual sightlines -- we have detected only one intervening \mbox{H\,{\sc i}} absorption-line, which is a detection rate of 4.3 per cent. 
This is low compared to previous surveys ($\sim$50 per cent, overall), so in this section we wish to establish the reason(s) for the low detection rate in our sample. 
We already established in Paper~I that the majority of our non-detections in sample A were due to the background sources being too faint (with many of the sources becoming resolved or extended, reducing our absorption-line sensitivity), and we now wish to determine whether this explains the low detection rate seen in sample B as well. 
As discussed in Section \ref{results_part1:hi_maps}, examination of the \mbox{H\,{\sc i}} distribution in Figure \ref{figure:overlay_maps} shows that only five of the 17 sightlines in our sample intersect the \mbox{H\,{\sc i}} disc at column densities greater than 1-2 $\times$ 10$^{20}$ cm$^{-2}$ (where our observations are sensitive to absorption). 
Therefore we find that in sample B, the majority of the non-detections are because the sightlines miss the \mbox{H\,{\sc i}} disc (or only intersect the disc at column densities below what this survey was designed to detect i.e. sub-DLA column densities). 
Of the six sightlines that \emph{do} intersect the \mbox{H\,{\sc i}} disc we have one detection (in the galaxy NGC\,5156) and four non-detections (two sightlines intersecting NGC\,1566, and one each in ESO\,300-G\,014 and NGC\,7162A).

To examine the reasons for the non-detections in these cases, we have produced plots of the `absorption-line detectable region' for each sightline. 
These plots (presented in Figure \ref{figure:column_density_contours}) show the region of the galaxy that would be detectable in absorption given the flux of the background source and the \mbox{H\,{\sc i}} column density along the sightline (derived from the emission-line maps), assuming a 3-$\sigma$ detection limit. 
Two contours show the detectable region calculated from (i) the 1.4 GHz flux measured in the highest resolution image, and (ii) the SUMSS flux (assuming a typical spectral index of $\alpha$ = $-$0.7). 
Therefore if the background source falls inside the first of these regions (represented by the solid blue contour), it indicates that we would have expected to detect intervening absorption along this sightline (we can ignore the second region, represented by the dashed green contour, for now, but discuss the significance of this in the following section). 
We refer the reader to paper I for a more detailed description of the interpretation of these plots and how the regions were calculated.

\begin{figure*}
\includegraphics[width=0.3\linewidth]{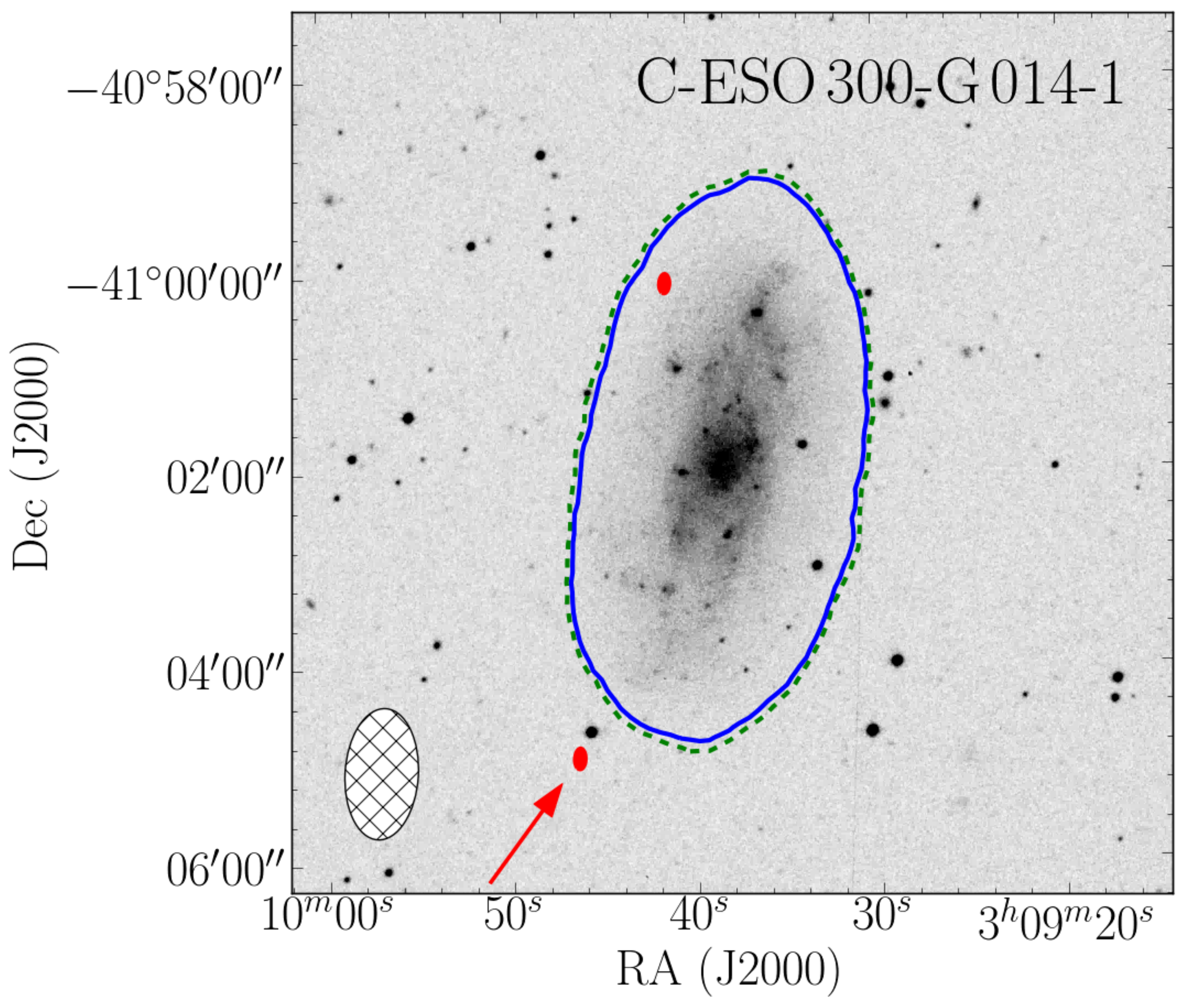}
\includegraphics[width=0.3\linewidth]{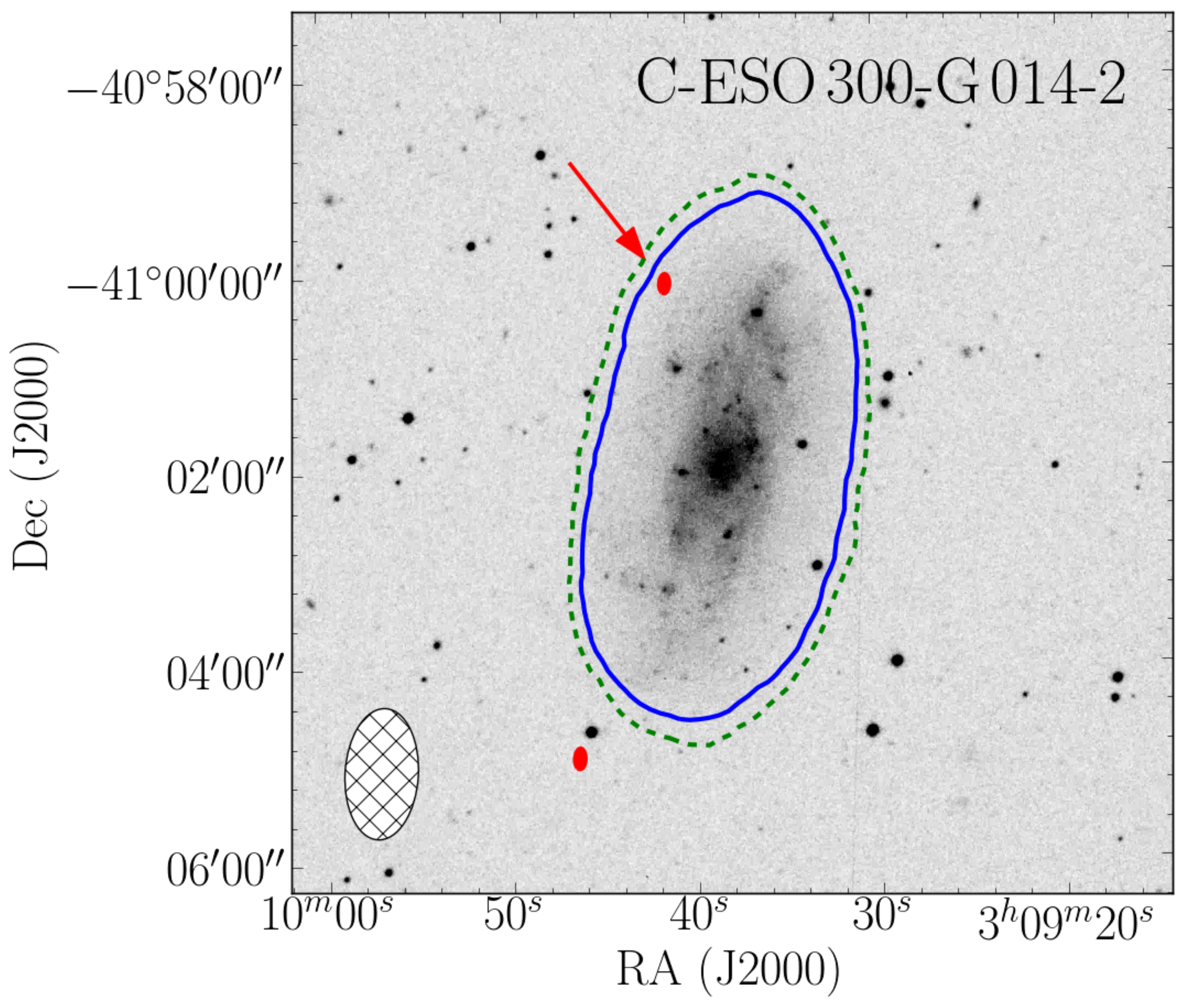}
\includegraphics[width=0.3\linewidth]{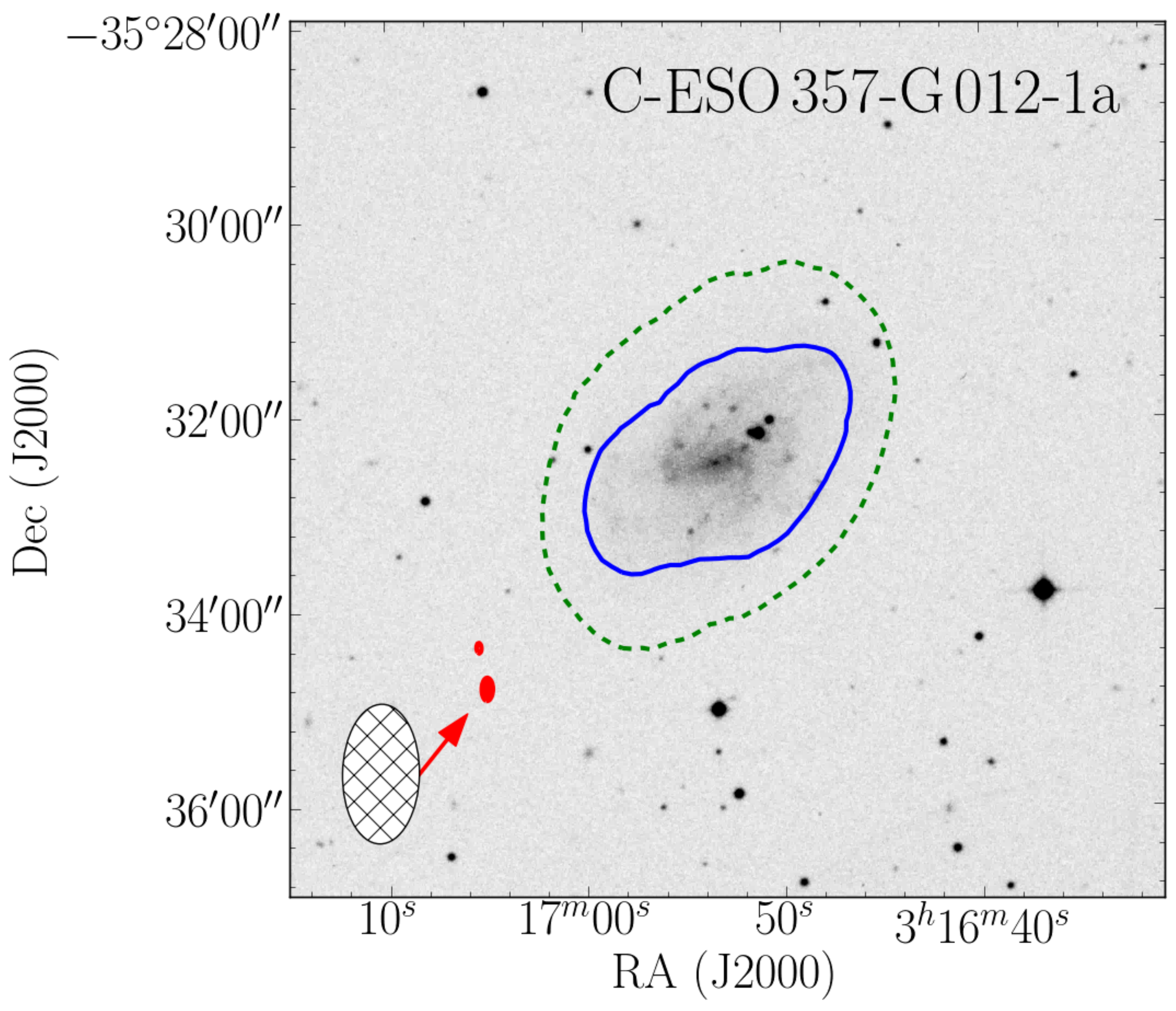}
\includegraphics[width=0.3\linewidth]{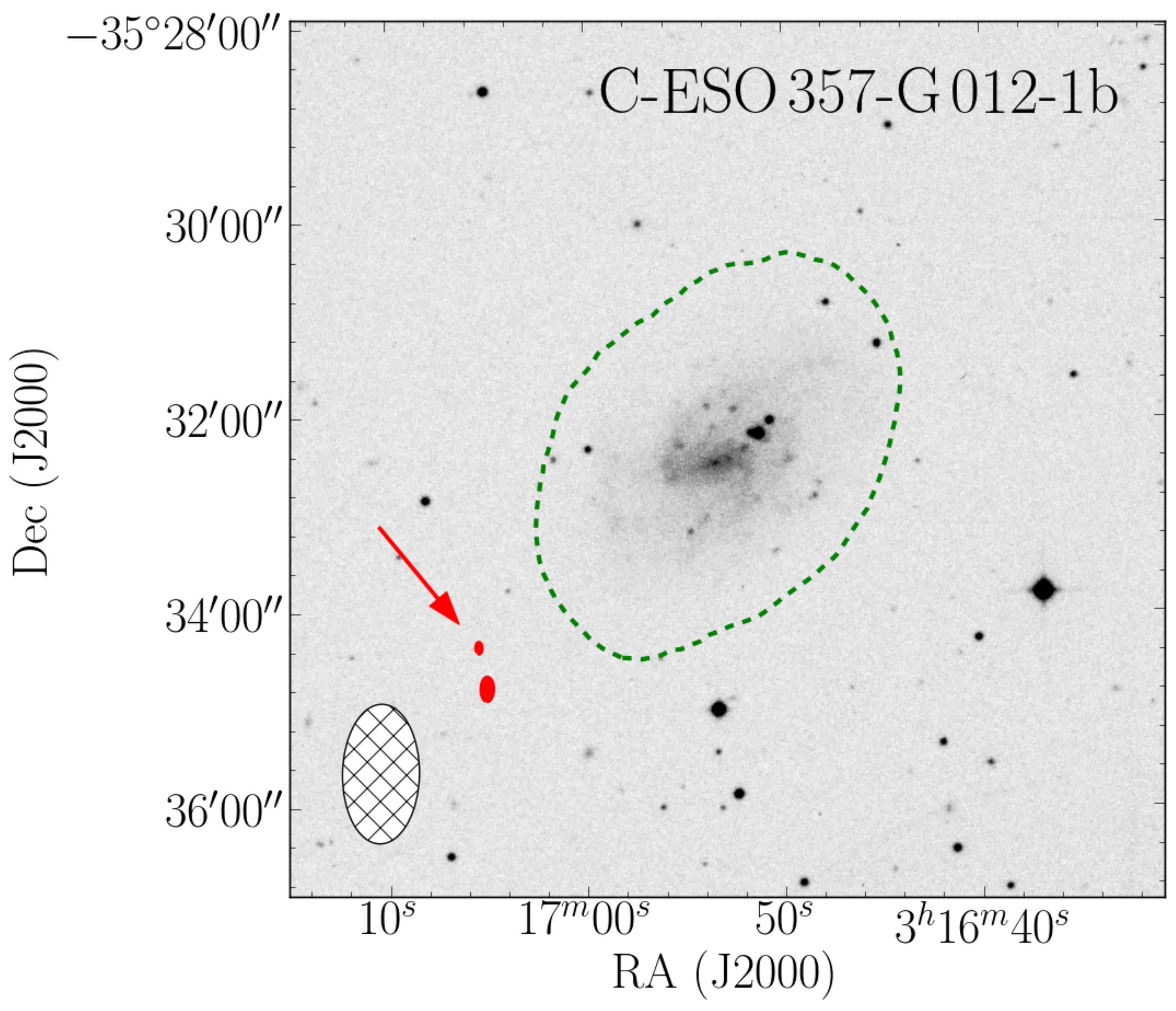}
\includegraphics[width=0.3\linewidth]{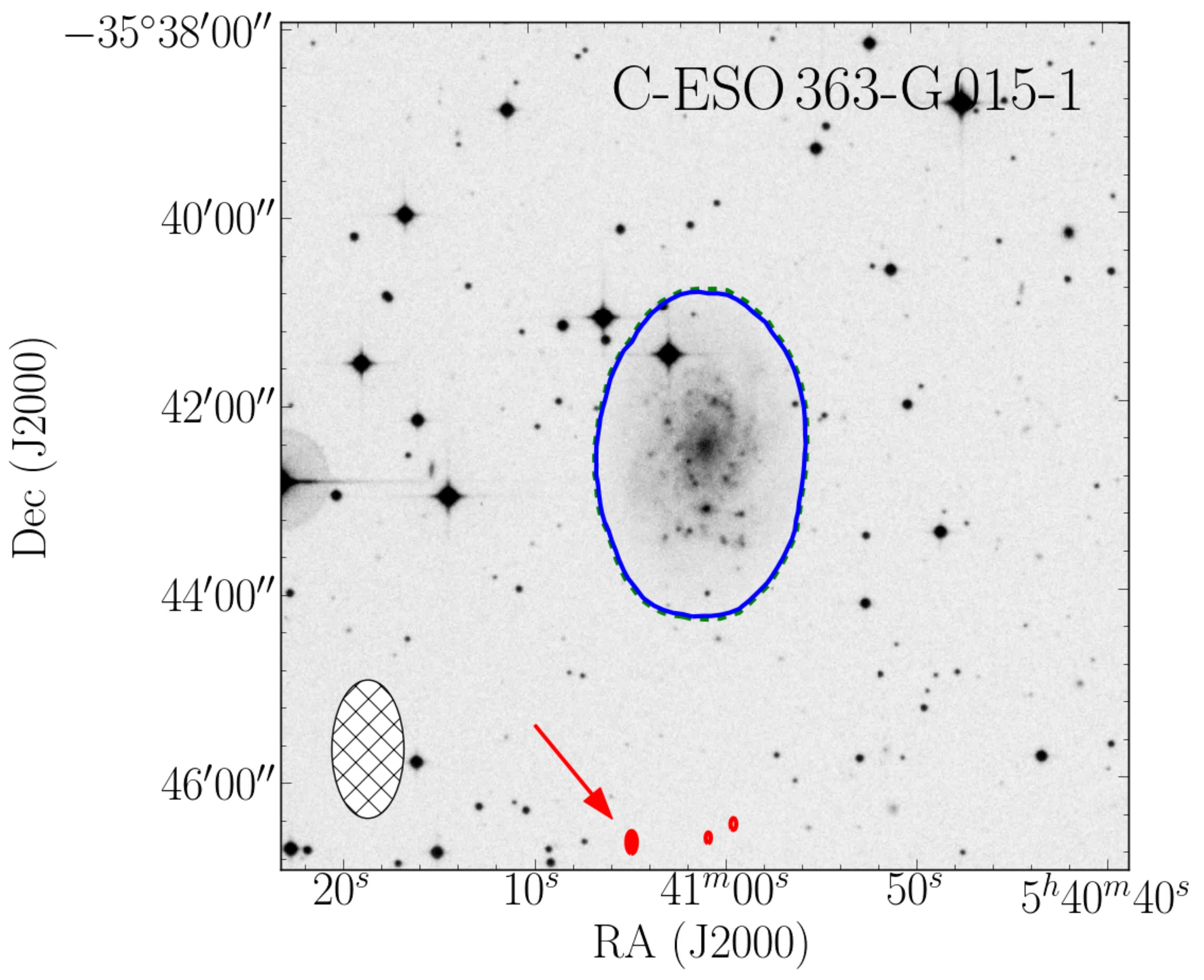}
\includegraphics[width=0.3\linewidth]{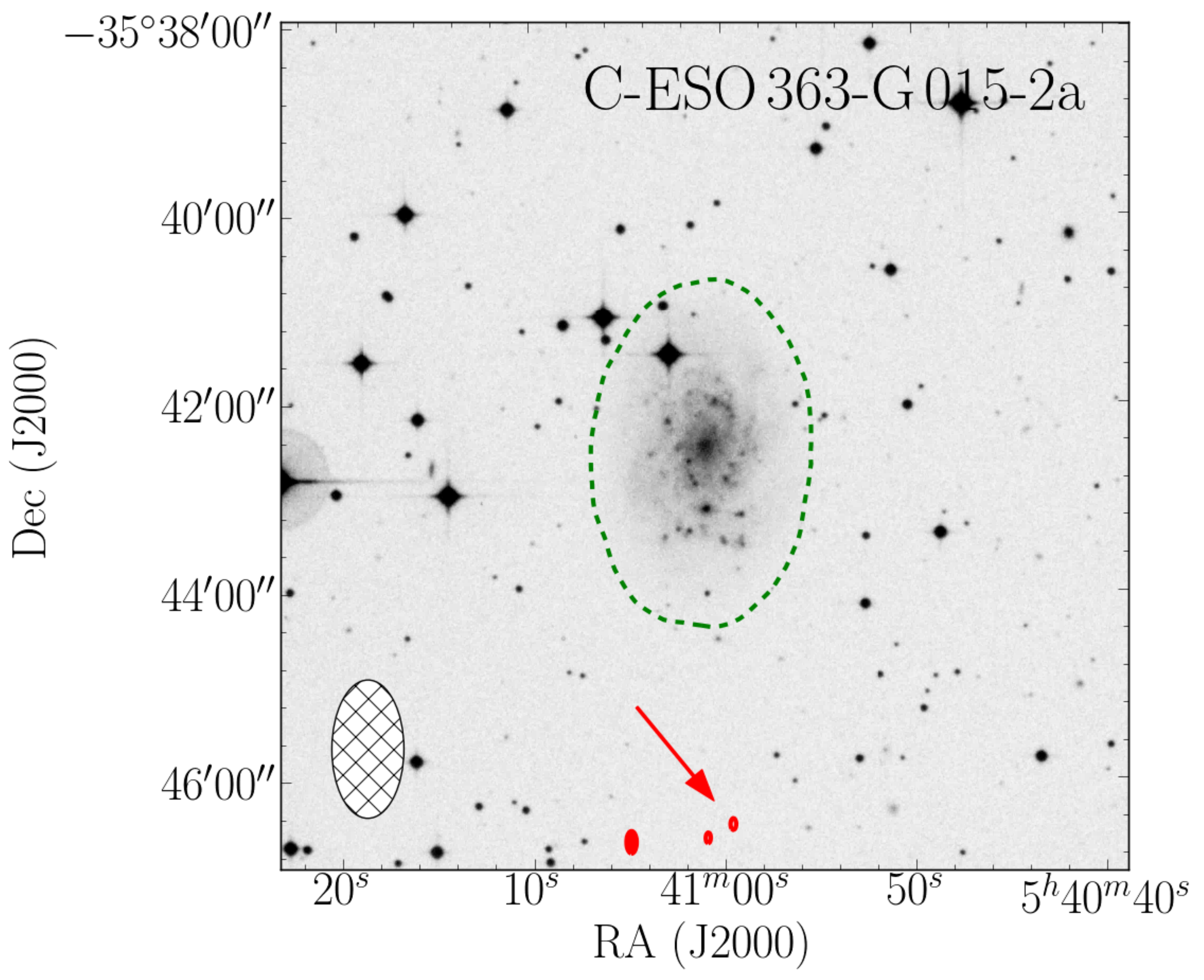}
\includegraphics[width=0.3\linewidth]{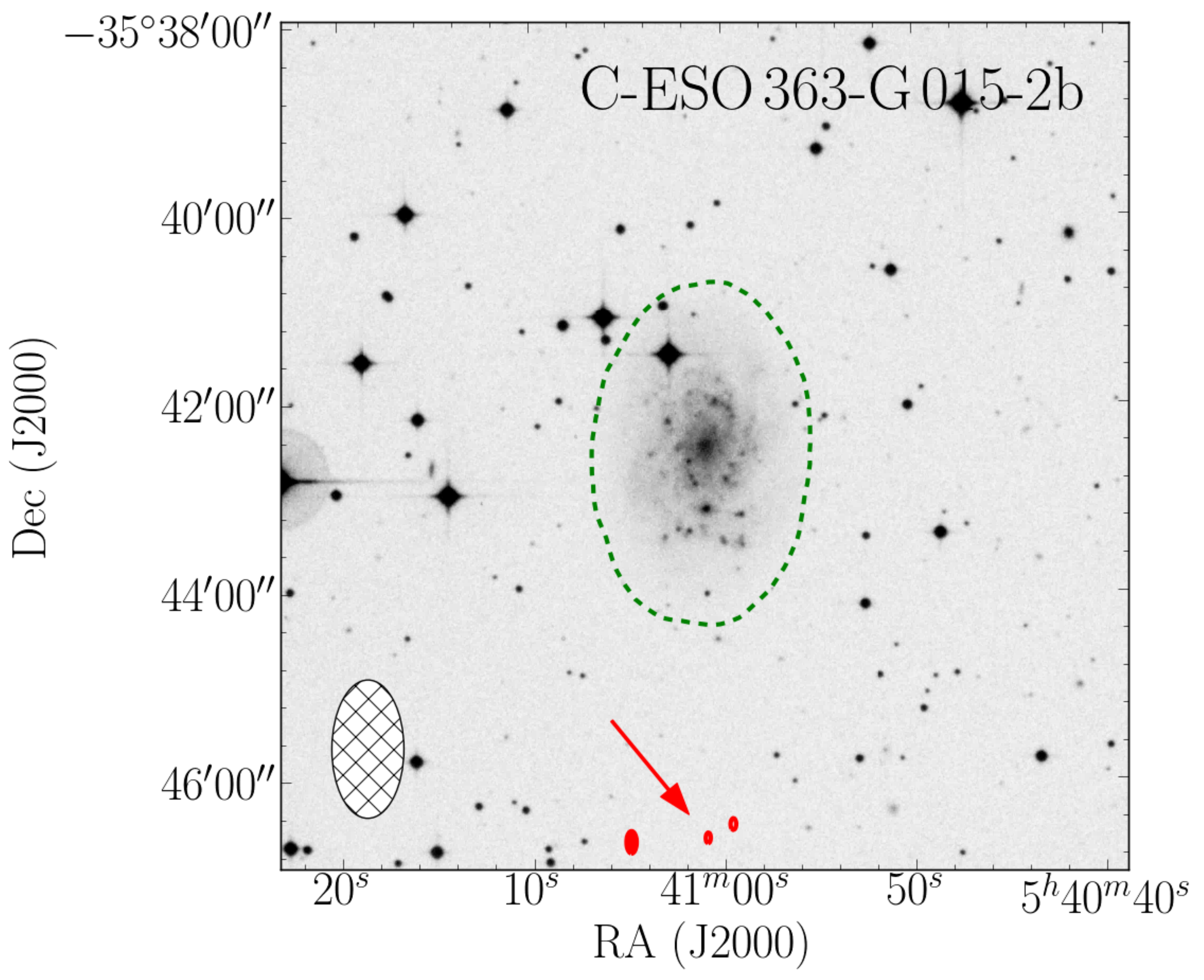}
\includegraphics[width=0.3\linewidth]{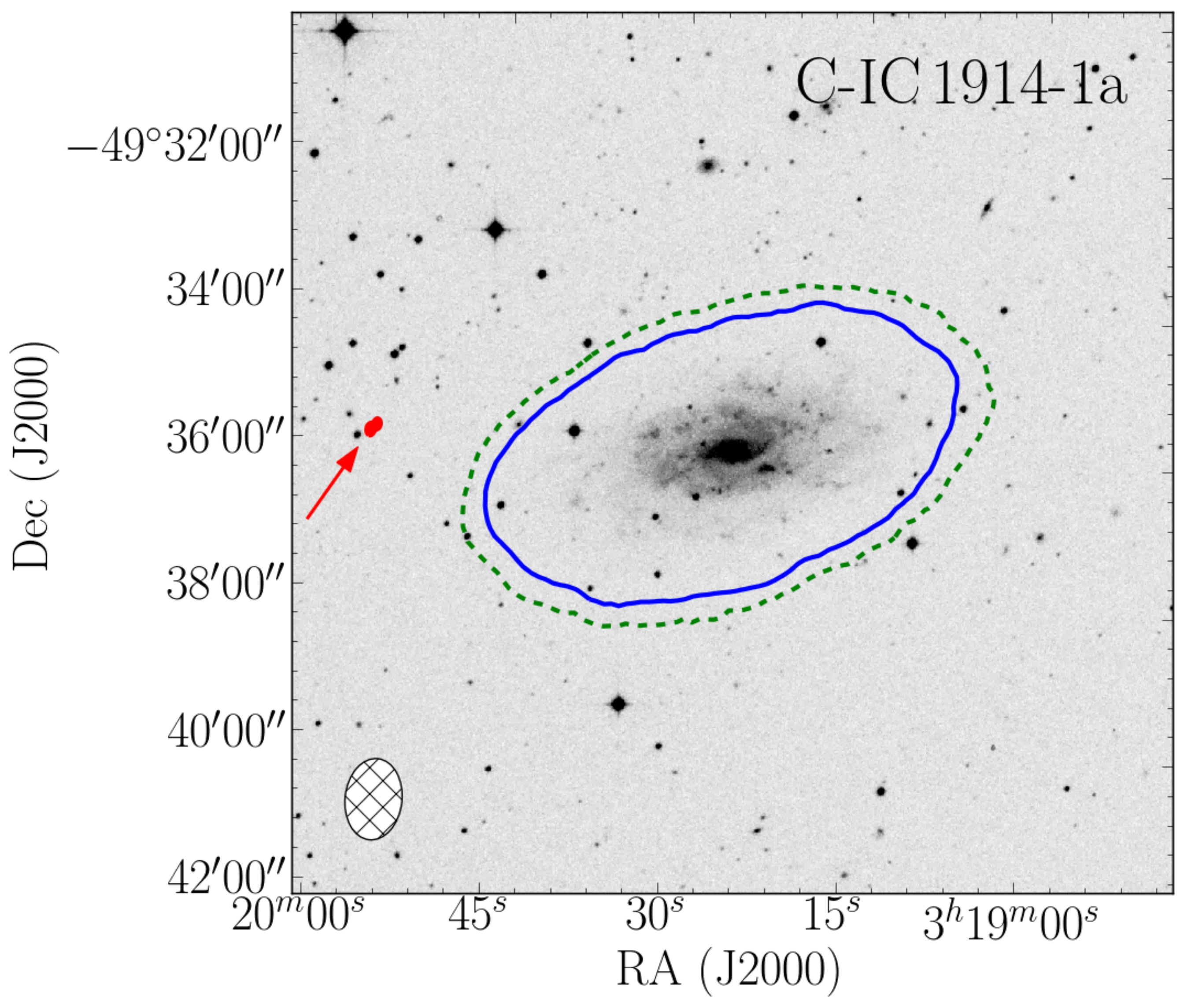}
\includegraphics[width=0.3\linewidth]{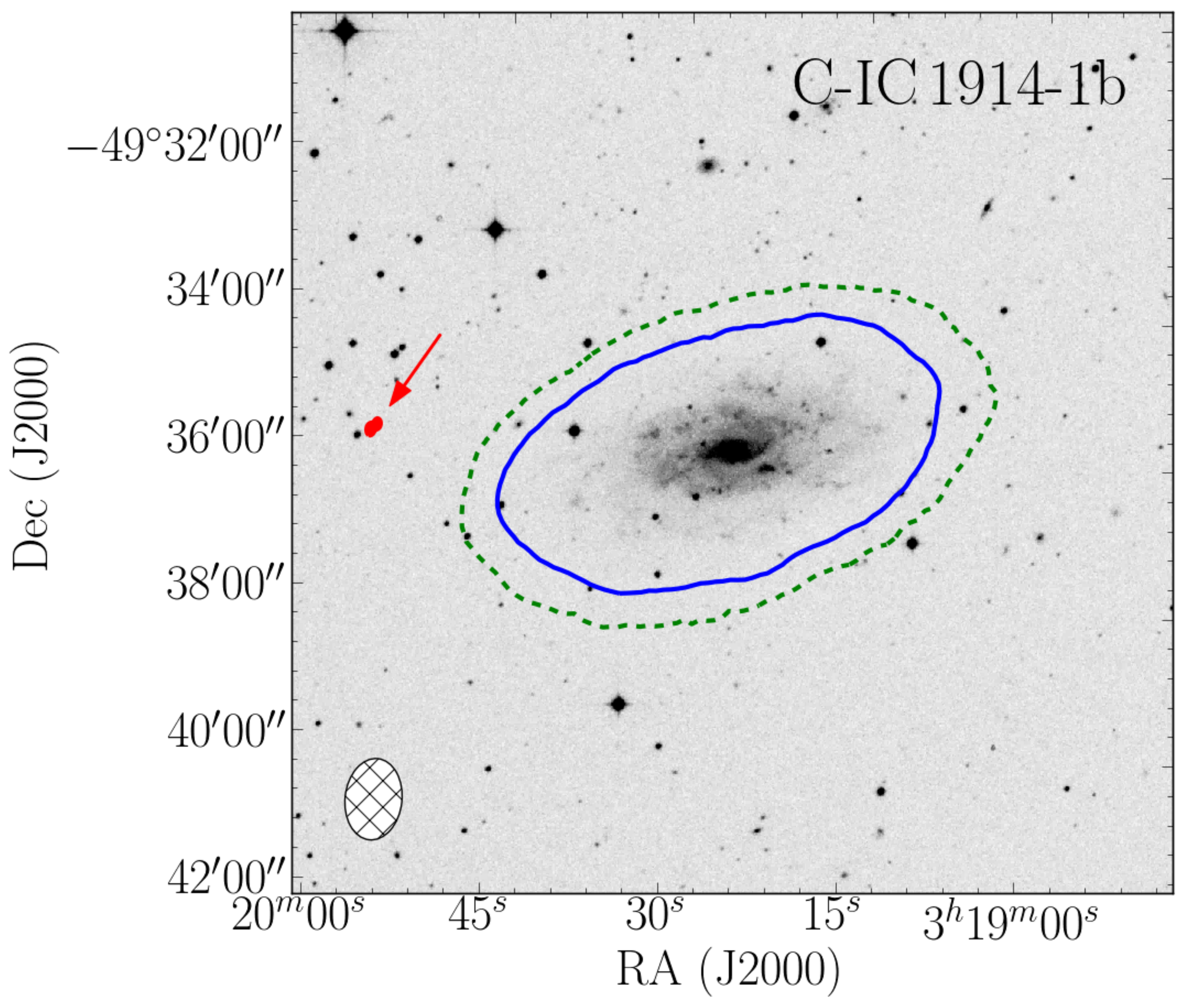}
\includegraphics[width=0.3\linewidth]{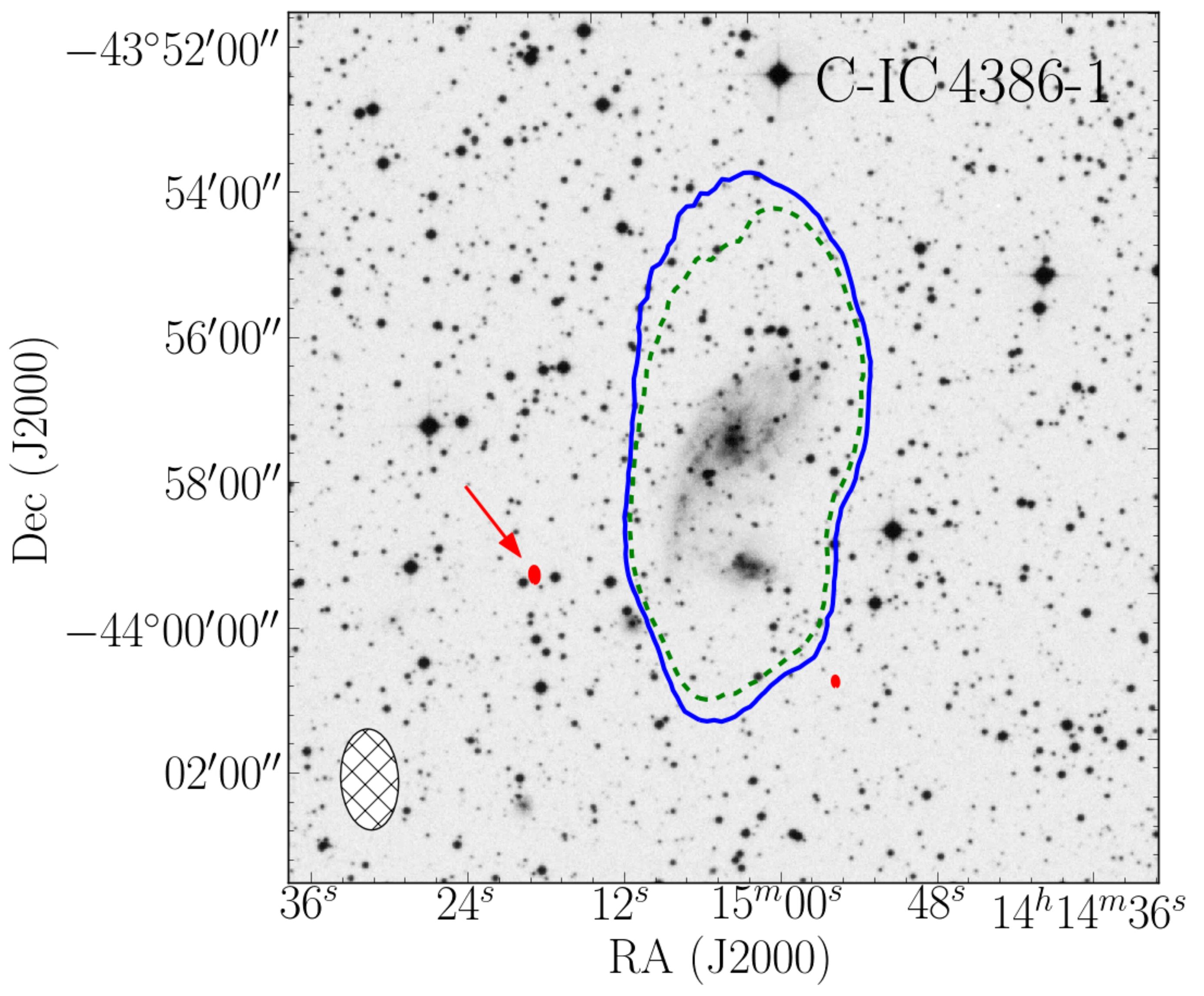}
\includegraphics[width=0.3\linewidth]{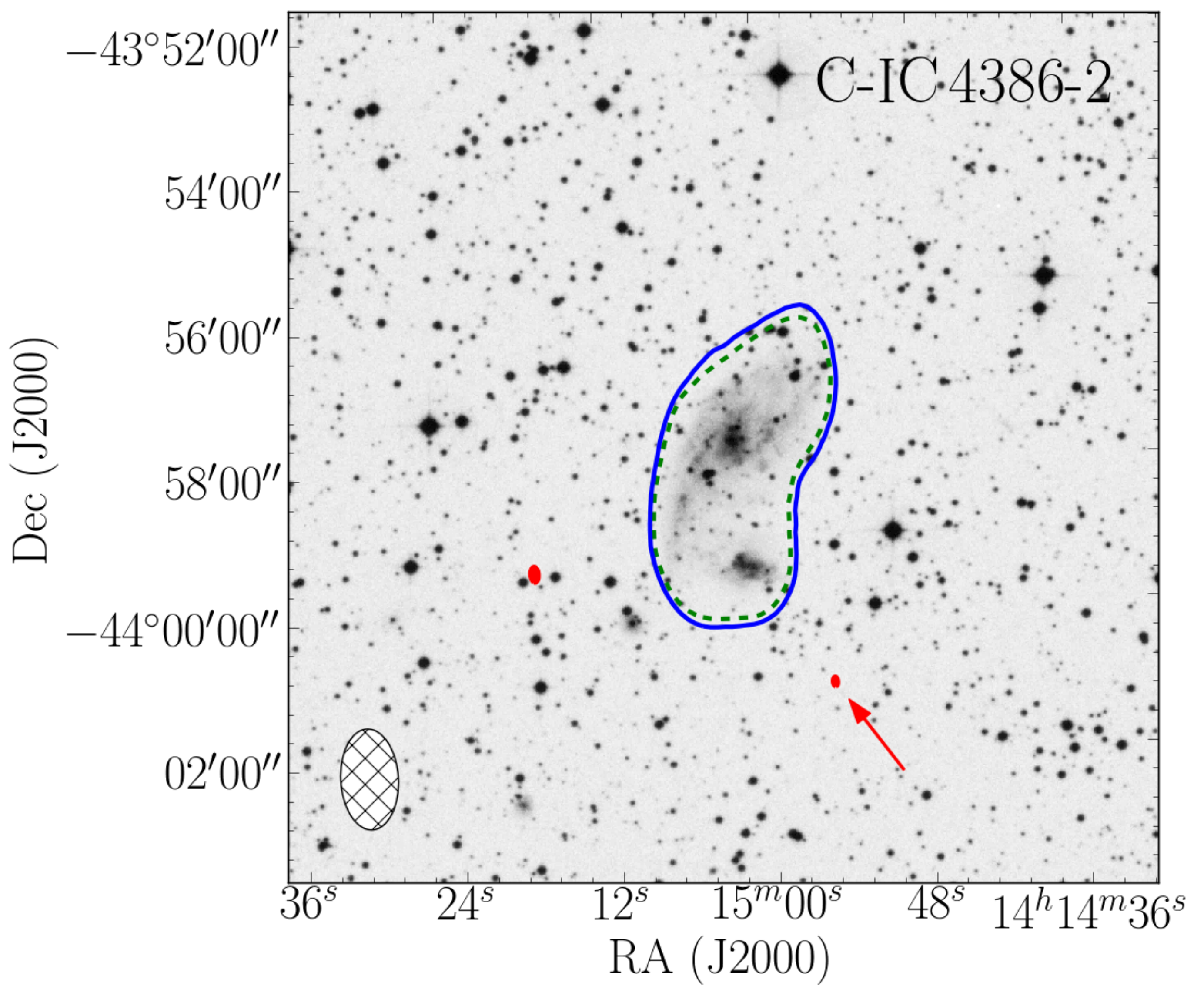}
\includegraphics[width=0.3\linewidth]{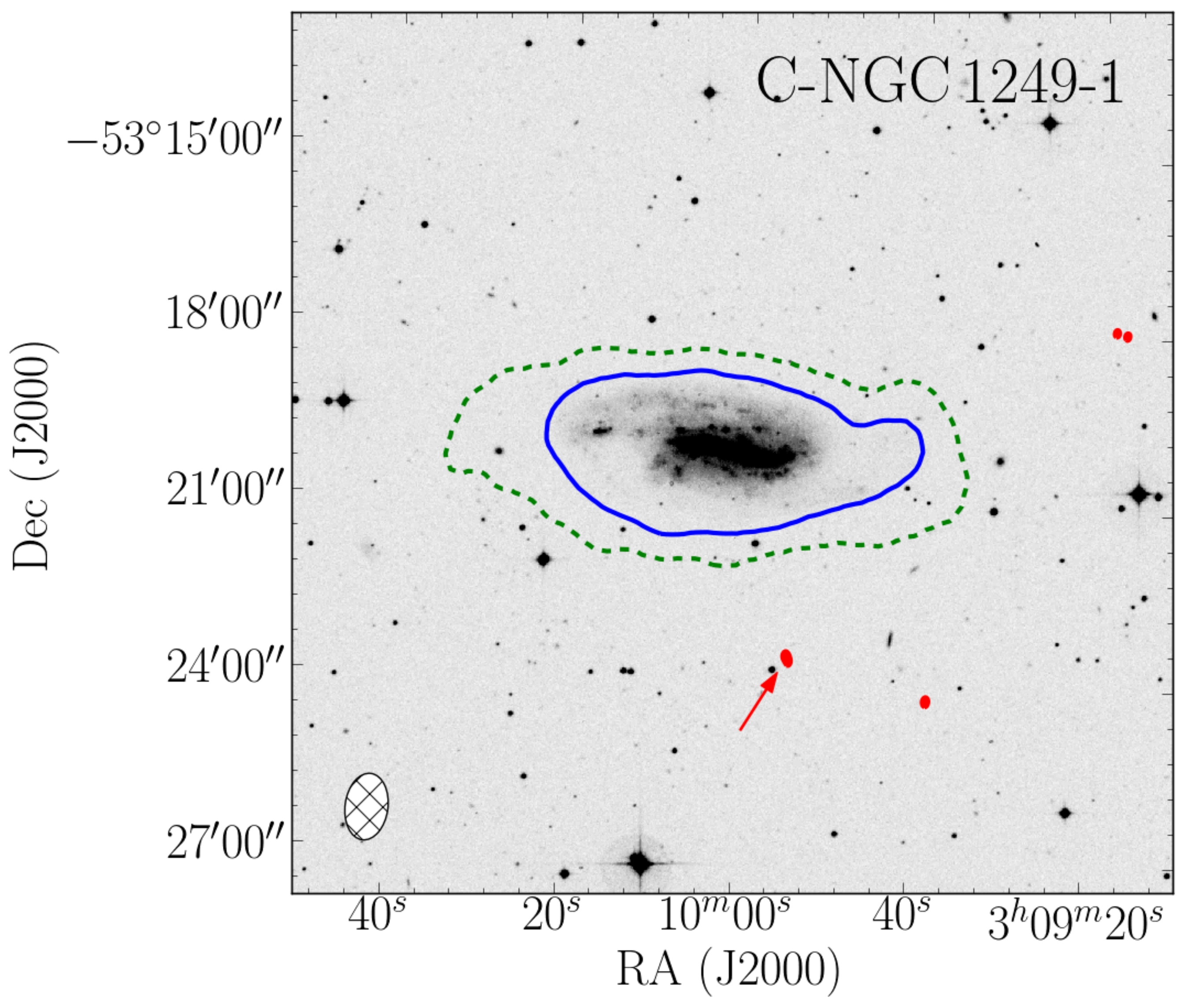}
\caption{Plots showing the absorption-line detectable region for each of the sightlines/galaxies in our sample. 
The blue (solid) contour is the `actual' detectable region, given the flux measured in the high resolution ATCA image. 
The green (dotted) contour is the `expected' detectable region based on the SUMSS 843 MHz flux (assuming $\alpha$ = $-$0.7). 
For sources that are resolved into multiple components we have produced a separate plot for each component. 
The 1.4 GHz continuum map is also overlaid (solid red contours), and the arrow indicates the position of the background continuum source (which is offset from the centre of the galaxy). 
If the continuum source lies inside the detectable region, we would have expected to detect an absorption-line (for normal values of $T_{\mathrm{S}}/f$), but if the continuum source lies outside the detectable region, it means the \mbox{H\,{\sc i}} distribution is consistent with an absorption-line non-detection. 
The synthesised beam for the \mbox{H\,{\sc i}} maps ($\sim$60 arcsec) is shown in the bottom left corner, and the synthesised beam for the continuum images (not shown) is $\sim$5 arcsec.}
\label{figure:column_density_contours}
\end{figure*}

\begin{figure*}
\includegraphics[width=0.3\linewidth]{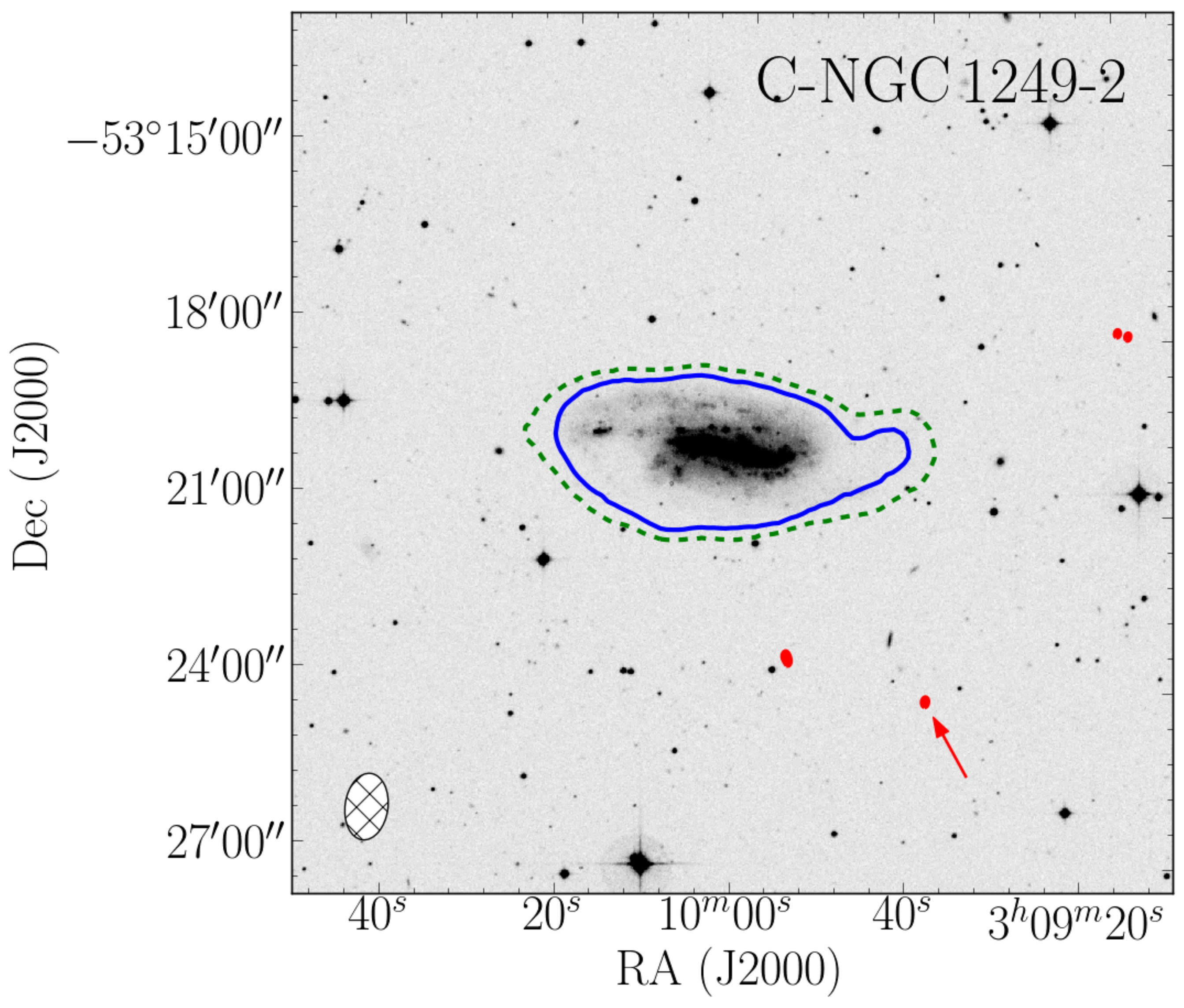}
\includegraphics[width=0.3\linewidth]{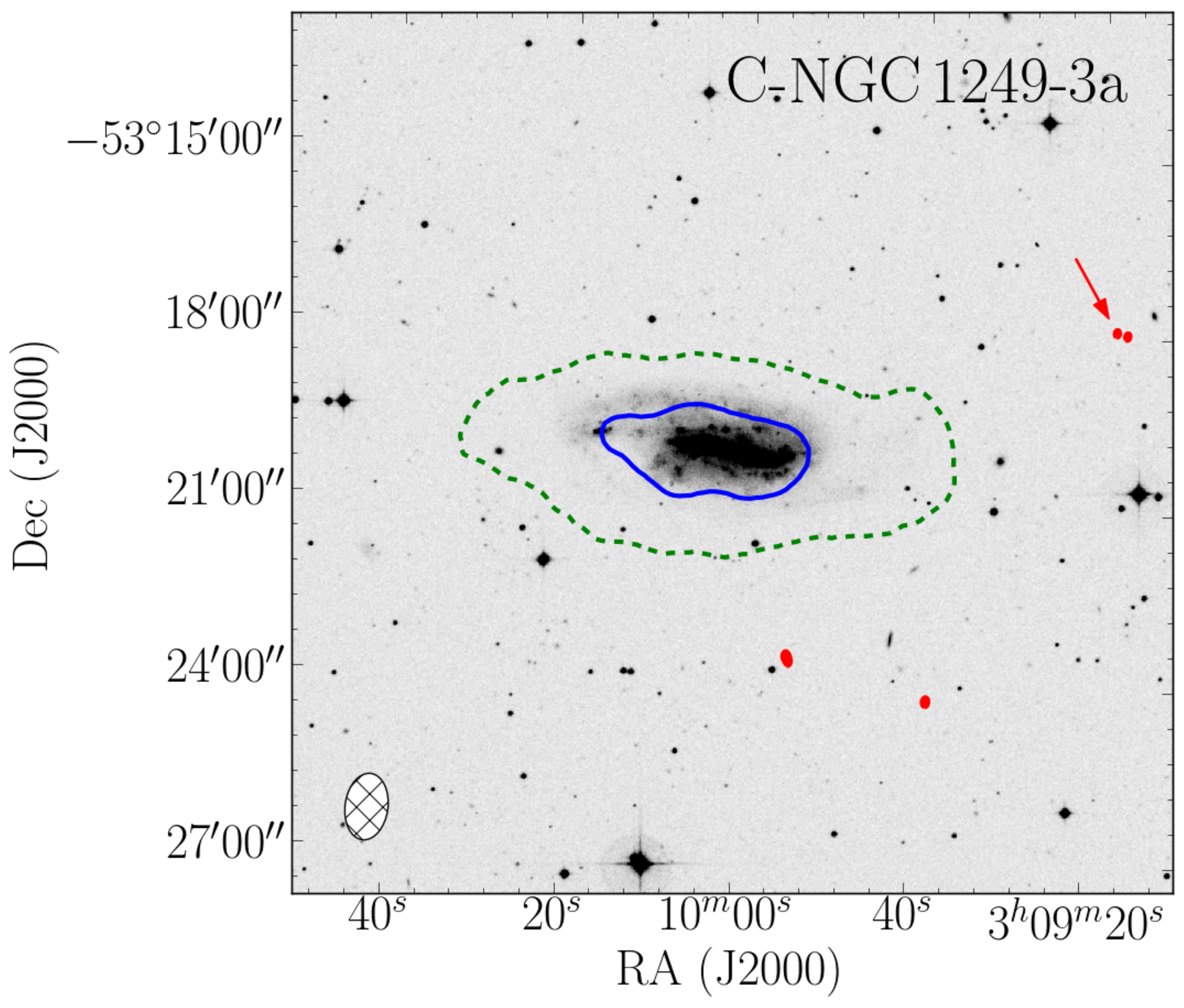}
\includegraphics[width=0.3\linewidth]{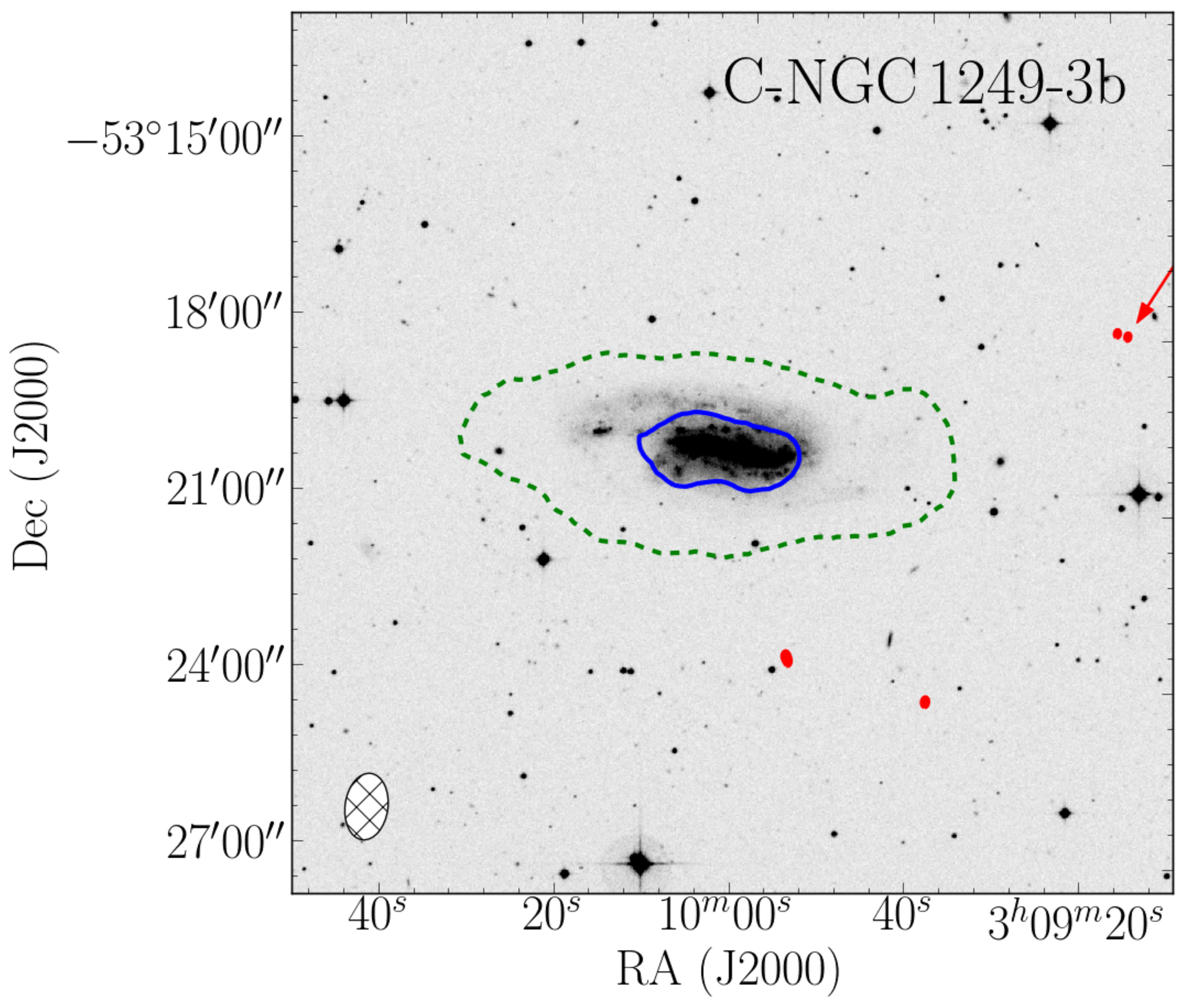}
\includegraphics[width=0.3\linewidth]{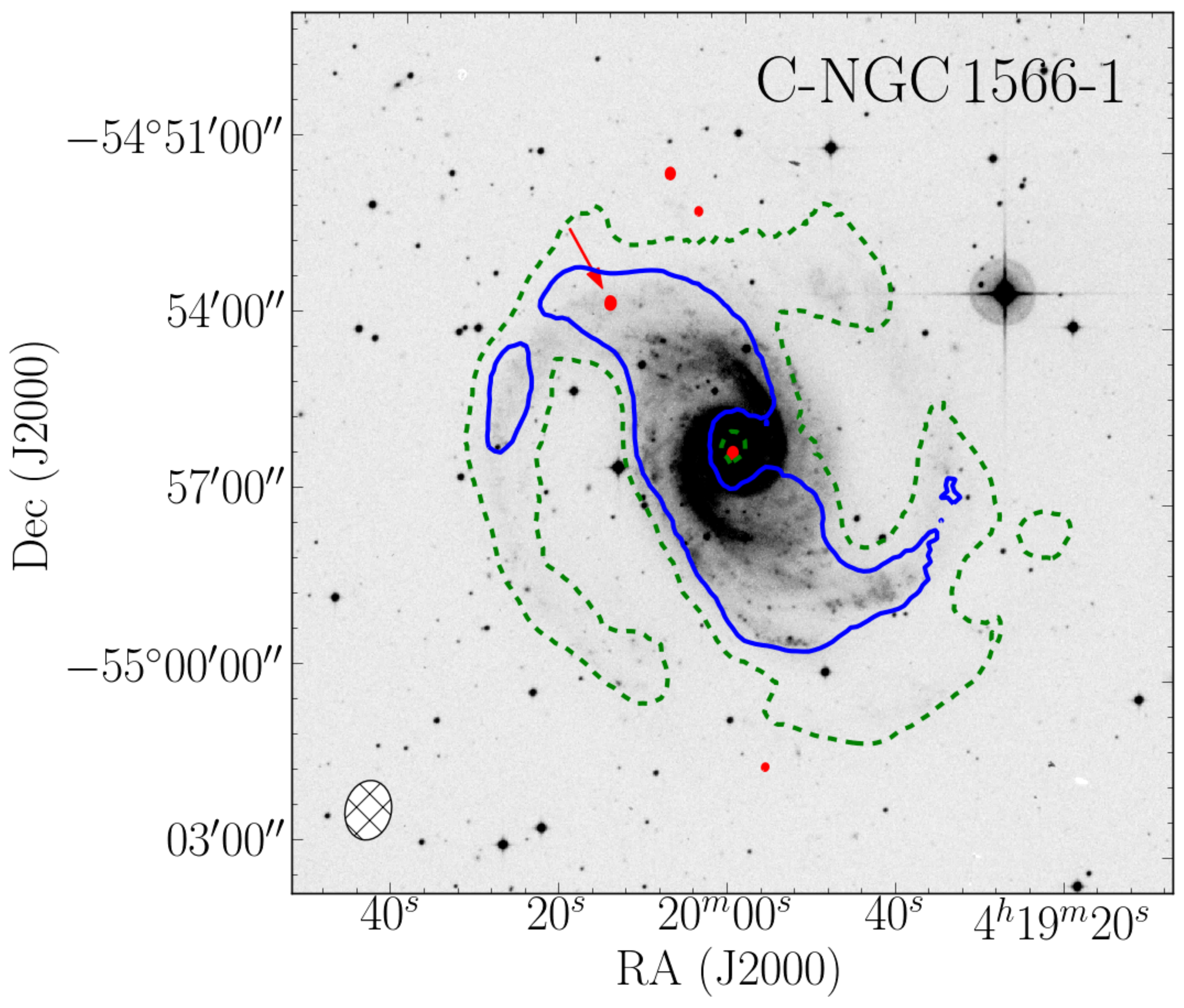}
\includegraphics[width=0.3\linewidth]{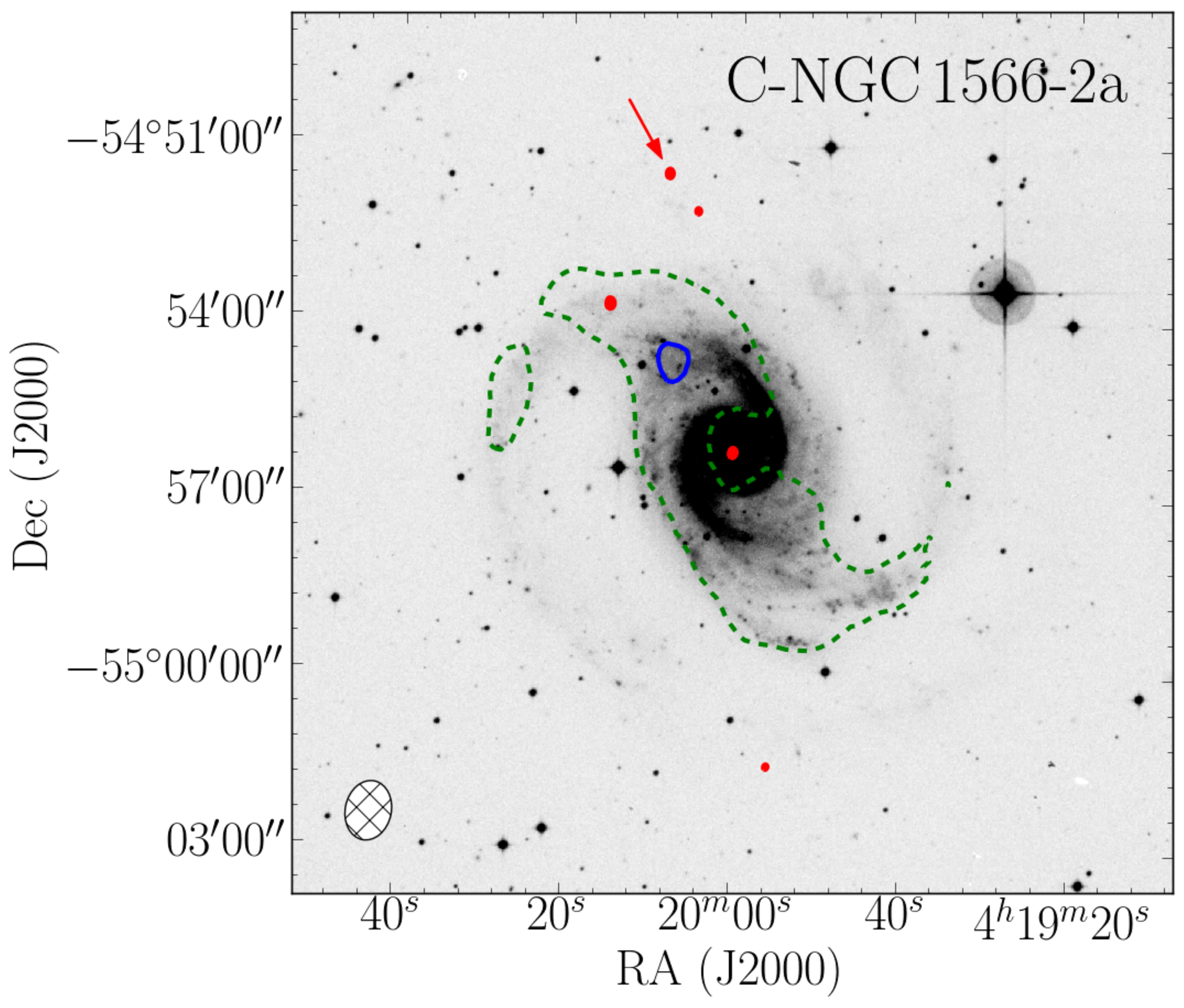}
\includegraphics[width=0.3\linewidth]{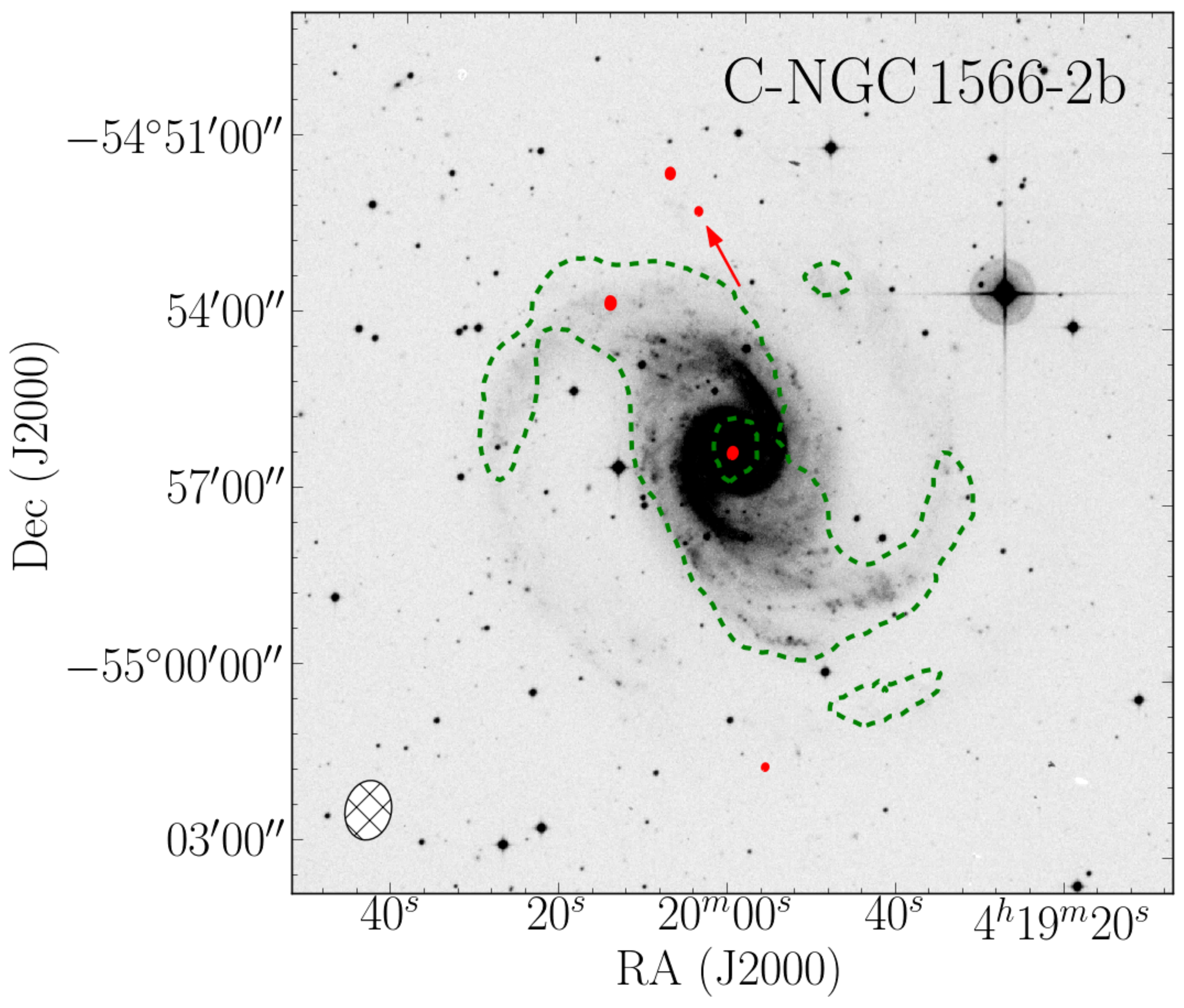}
\includegraphics[width=0.3\linewidth]{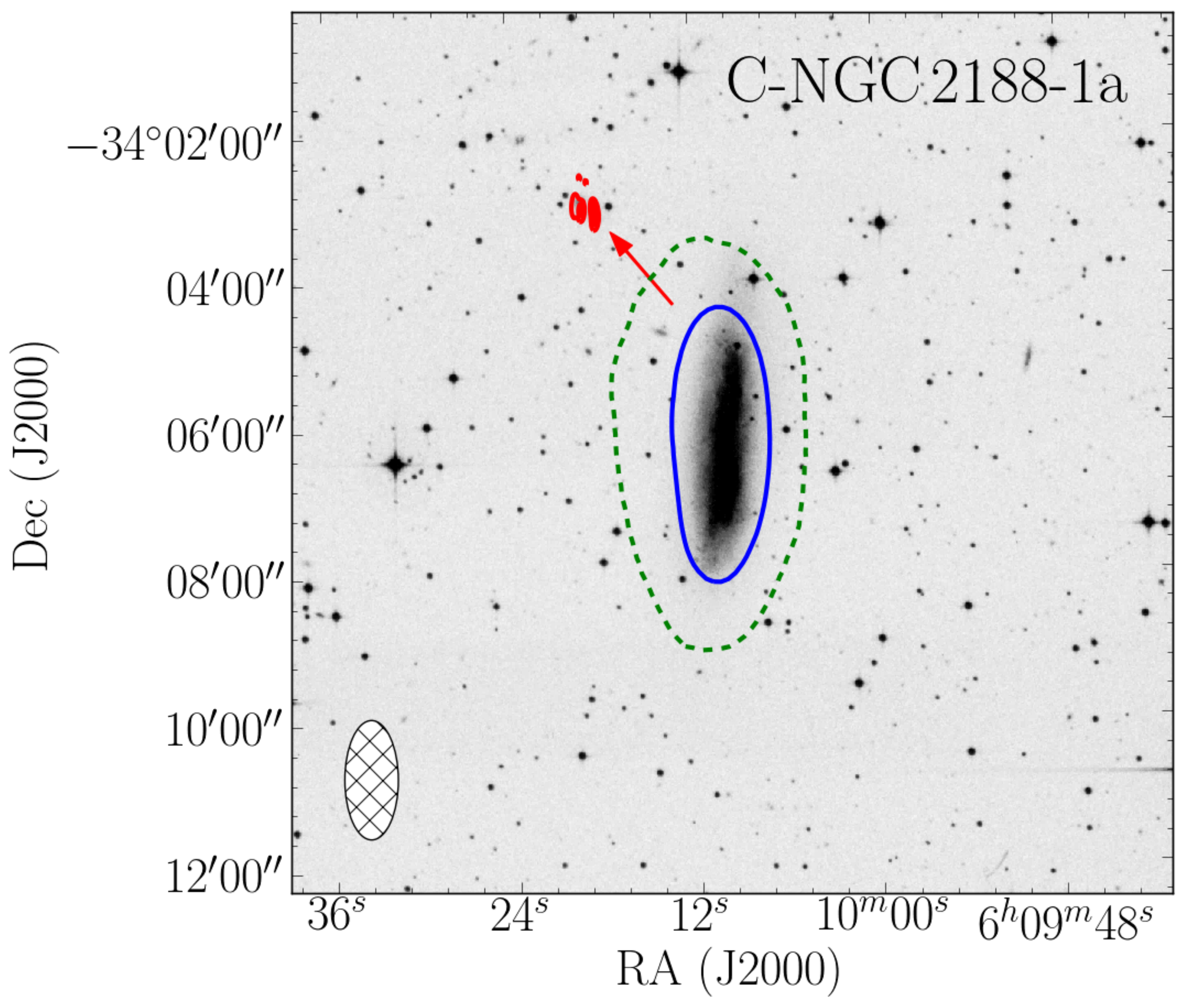}
\includegraphics[width=0.3\linewidth]{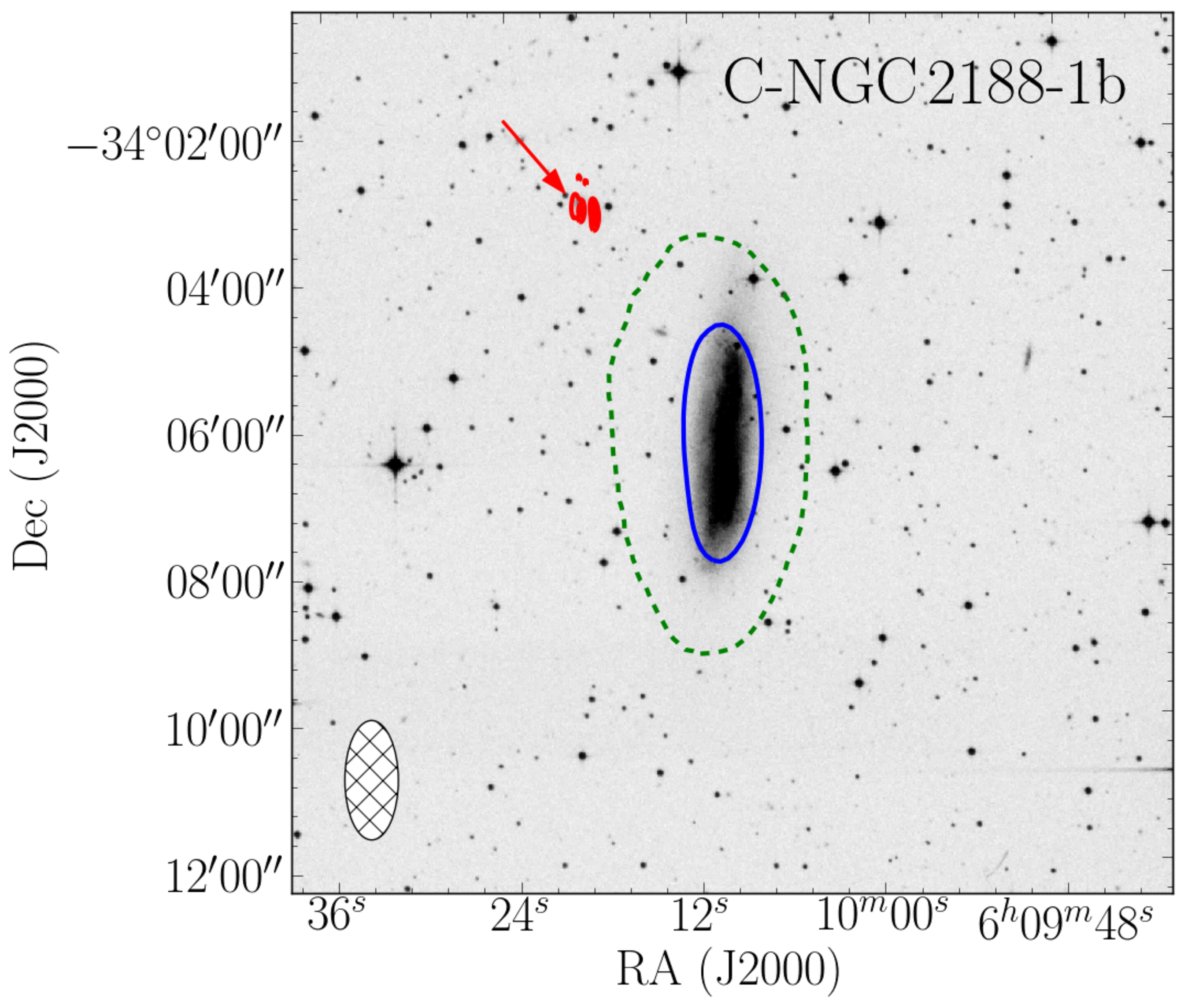}
\includegraphics[width=0.3\linewidth]{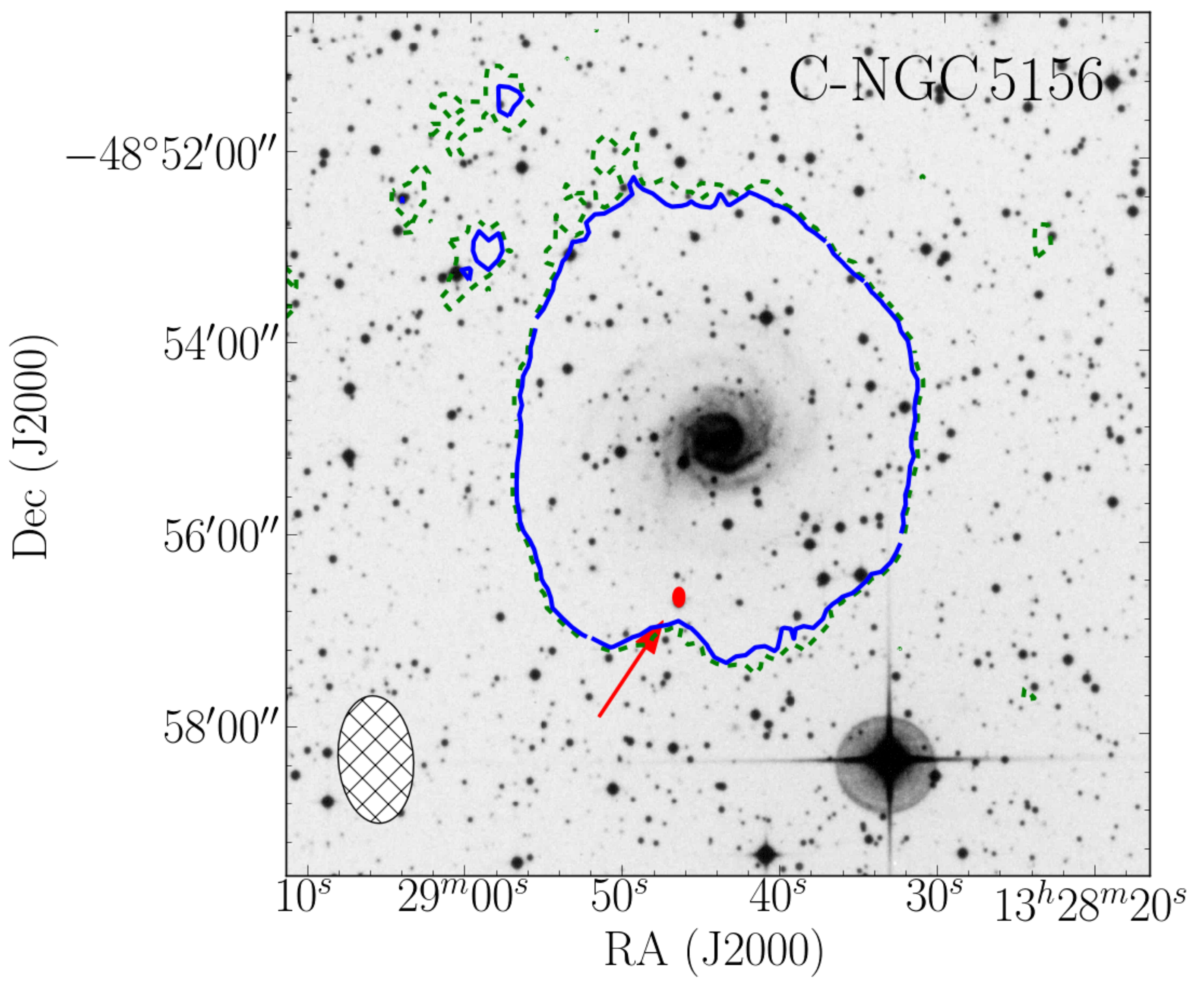}
\includegraphics[width=0.3\linewidth]{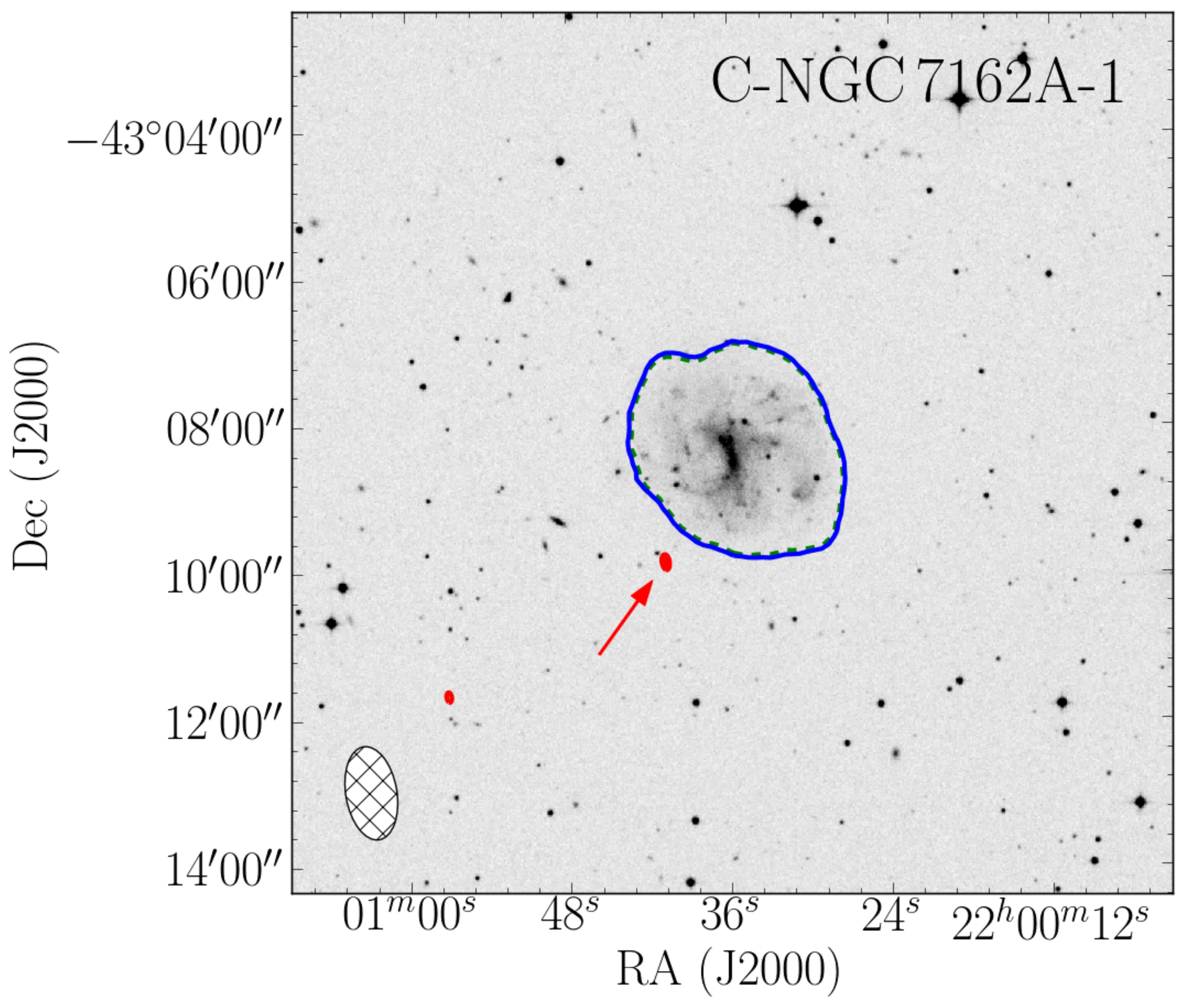}
\includegraphics[width=0.3\linewidth]{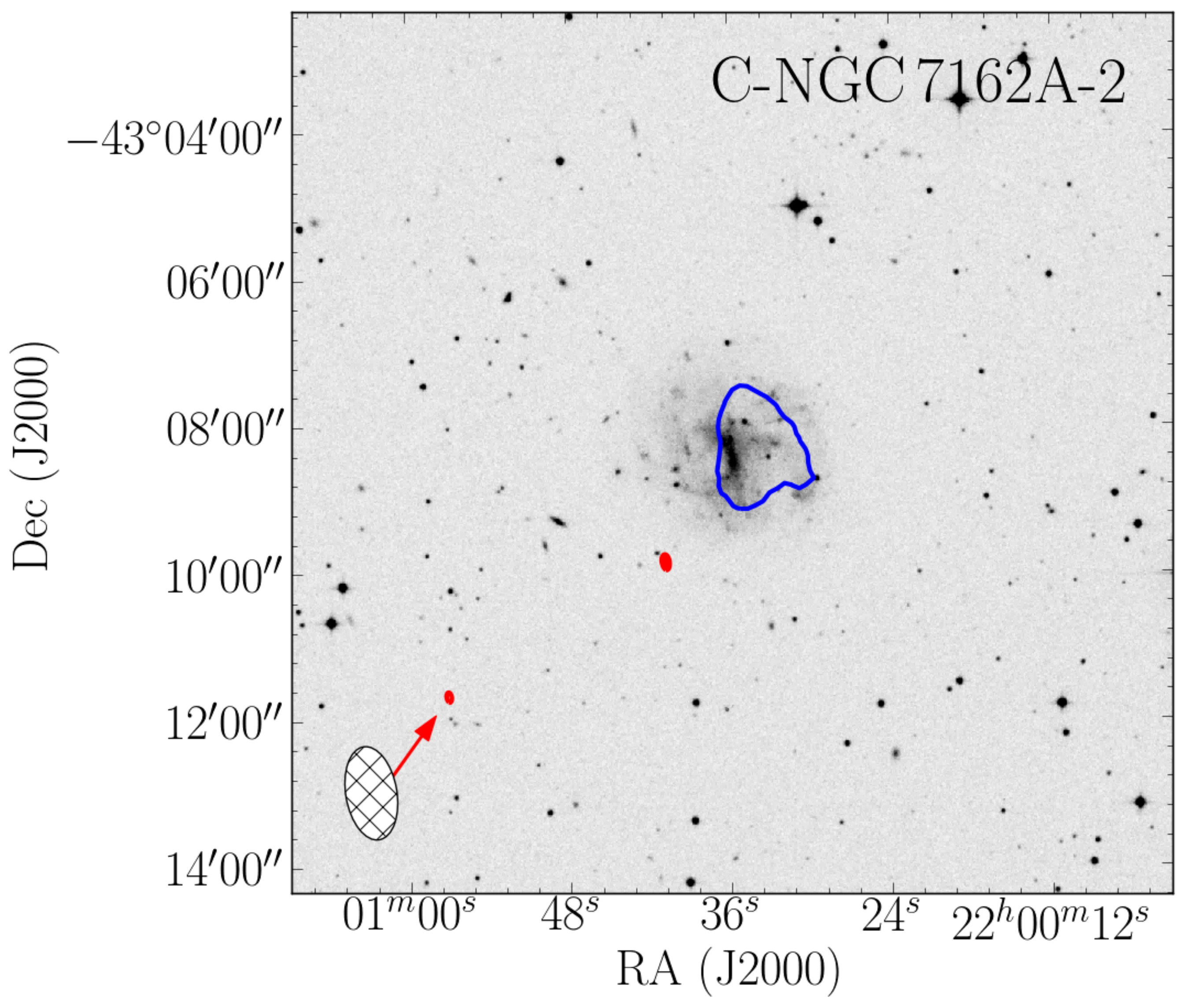}
\contcaption{}
\end{figure*}

Of the four non-detections mentioned above, two sources -- C-ESO\,300-G\,014-2 and C-NGC\,1566-1 -- fall inside the detectable region and two -- C-NGC\,1566-2 and C-NGC\,7162A-1 do not. 
Therefore, for C-NGC\,1566-2 and C-NGC\,7162A-1 the non-detections are consistent with the \mbox{H\,{\sc i}} distribution, given the flux of the background source, and are therefore simply due to the background source being too faint (meaning these sources are not as bright as we would have expected from SUMSS). 
For C-ESO\,300-G\,014-2 and C-NGC\,1566-1 the absorption-line non-detections are inconsistent with the emission-line maps for the assumed value of $T_{\mathrm{S}}/f$ = 100 K (i.e. the background sources are bright enough that we would have expected to detect an absorption-line). 
However, if $T_{\mathrm{S}}/f$ were much higher than expected ($\gg$100 K) this would decrease the size of the detectable regions, explaining this discrepancy, and we therefore suggest that this is the reason we haven't detected any \mbox{H\,{\sc i}} absorption along these sightlines.
A summary of the likely reasons for the non-detections in our sample are given in Table \ref{table:non-detections}. 
This has been done for both samples A and B separately, as well as for the combined sample (23 sightlines).

Although our results have helped to shed light on what factors influence the detection rate of intervening absorption, it is clear that there is still more work to be done. 
So far, our results indicate two or three main avenues of investigation. 
Firstly work on \mbox{H\,{\sc i}}/optical scaling relations (see e.g. \citealt{1997A&A...324..877B,2005A&A...442..137N,2009MNRAS.399.1447L}) would help us to better predict which sightlines are likely to intersect the \mbox{H\,{\sc i}} of a galaxy. 
Secondly, we need better methods of estimating source compactness from available lower resolution data, so that we know which radio sources are mostly likely  to yield a detection. 
VLBI observations would also help us to get a census of the covering factor in absorption-line systems detected to date, and thus a better estimation of the typical spin temperature in the discs of large spiral galaxies (in which a large proportion of future absorption-line detections will likely be made).

\begin{table}
\centering
\caption{Breakdown of detections and non-detections across the complete sample of 16 galaxies and 23 sightlines searched. 
The non-detections are divided into three categories: sightlines that do not intersect the \mbox{H\,{\sc i}} disc, sightlines where the continuum flux was too low, and sightlines where the ratio of $T_{\mathrm{S}}/f$ was much higher than expected ($\gg$100 K).}
\label{table:non-detections}
\begin{tabular}{@{} lrrr @{}} 
\hline
Reason for non-detections & Sample A & Sample B & Full sample \\
\hline
Does not intersect disc & 2 & 12 & 14 \\
Continuum flux too low & 3 & 2 & 5 \\
$T_{\mathrm{S}}/f$ $\gg$ 100 K & 1 & 2 & 3 \\
\hline
Detections & 0 & 1 & 1 \\
\hline
\hline
Total & 6 & 17 & 23 \\
\hline
\end{tabular}
\end{table}

\subsection{Influence of background source structure}
\label{discussion:background_source_structure}

In Paper I we found that the structure of the background sources had likely had a significant effect on the detection rate in sample A. 
Although this does not constitute as many of the non-detections in sample B, we still wish to investigate how the structure of the background sources in our sample might have affected the detection rate if more of the sightlines had intersected the \mbox{H\,{\sc i}} disc.

Examining the continuum images at different resolutions, we find that seven of the 17 SUMSS sources in sample B become resolved or extended at higher resolution, frequently splitting into two or more components. 
We also calculated the `compactness' factor (defined as the ratio of the flux on the longest and shortest baselines) and found this to be less then 10 per cent for four of the sources -- or almost a quarter -- of our sample. 

Examining the plots of the absorption-line detectable region (in Figure \ref{figure:column_density_contours}) allows us to determine the effect of source structure, even in galaxies where the continuum source sightline(s) do not intersect the \mbox{H\,{\sc i}} disc. 
We find that the `actual' detectable region (calculated from the ATCA 1.4 GHz flux) is significantly smaller than the `expected' detectable region (based on the SUMSS flux) for eight of the 17 sightlines -- or more than half our sample. 
As expected, the worst affected sightlines are those where the background source is extended or resolved into multiple components, showing that background source structure is an important consideration in the preparation for future absorption-line surveys.

Given this, it is important to further investigate the structure of the overall radio source population. 
Comparison of the structure of radio continuum sources in existing low and high resolution surveys -- such as SUMSS/the NRAO VLA Sky Survey \citep[NVSS,][]{1998AJ....115.1693C} and the Faint Images of the Radio Sky at Twenty-centimeters \citep[FIRST,][]{1995ApJ...450..559B}, respectively --  would allow us to better quantify what fraction of sources are likely to be compact versus extended on scales relevant for detecting \mbox{H\,{\sc i}} absorption. 
In addition, proxies for estimating source compactness, such as the spectral index (see e.g. \citealt{2011MNRAS.412..318M}) should be investigated to help with sample selection and analysis.

\subsection{Implications for future blind absorption-line surveys}
\label{discussion:detection_rate_impact_parameter}

\begin{figure}
\includegraphics[width=0.475\textwidth]{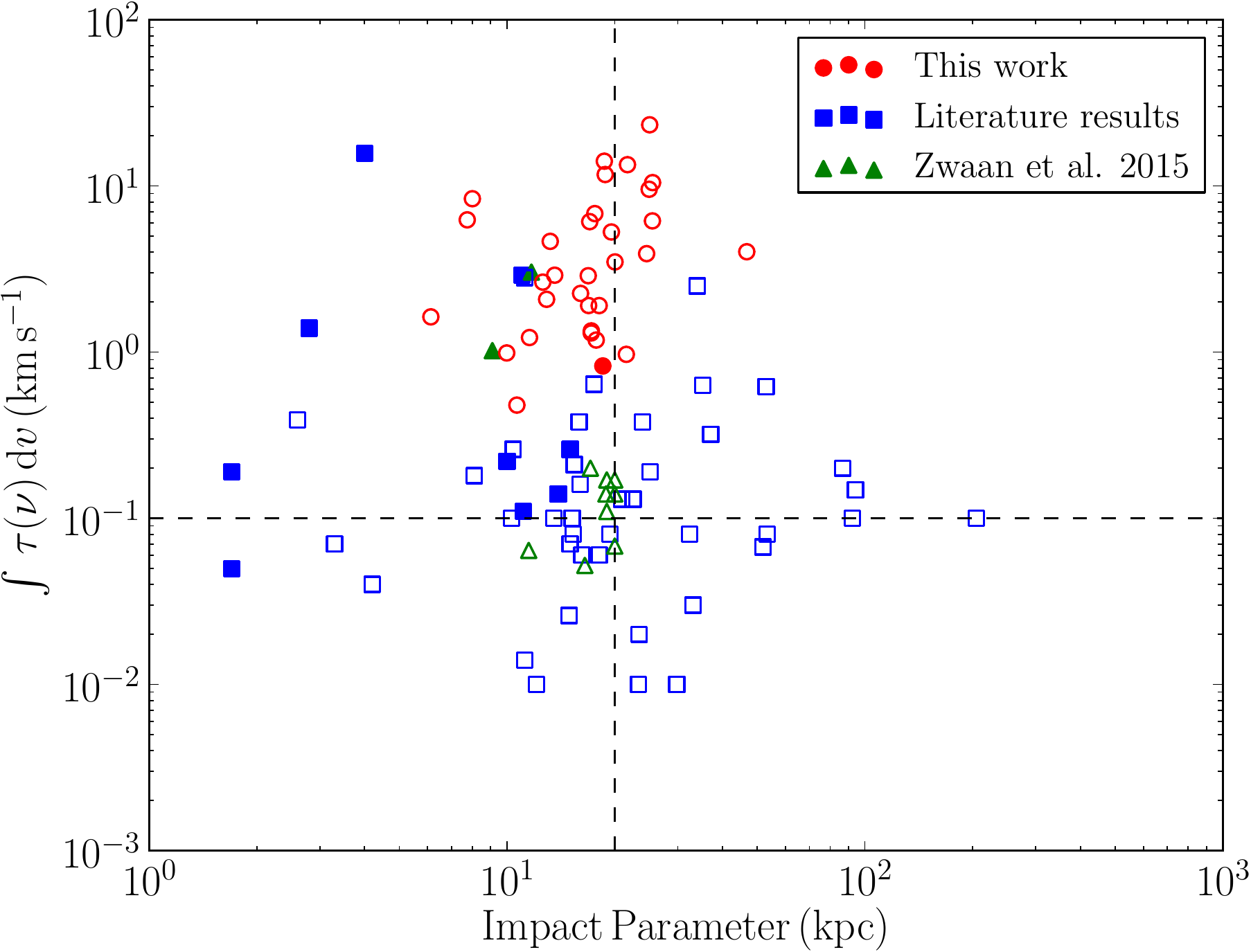}
\caption{Plot of integrated optical depth vs. impact parameter for different intervening absorption-line searches. 
The results of our survey (this work + \citealt{2015MNRAS.450..926R}) are indicated by circles (red), and the literature results (quasar sightlines only) by squares (blue). 
We also include the results of \citet{2015MNRAS.453.1268Z}, which comprises a mix of radio galaxies and quasars (but predominantly quasars) which are indicated by the triangular markers (green).}
Detections are shown as filled markers, and non-detections as open markers. 
Non-detections are plotted as the 3-$\sigma$ upper limit on the integrated optical depth.
\label{figure:gupta_comparison}
\end{figure}

To investigate the detection rate as a function of impact parameter, we have plotted in Figure \ref{figure:gupta_comparison} the integrated optical depth against the impact parameter for each of the sightlines in our sample. 
We have also included relevant literature results -- the list compiled by \citealt{2010MNRAS.408..849G} (\citealt{1975ApJ...200L.137H,1988A&A...191..193B,1990ApJ...356...14C,1992ApJ...399..373C,2004ApJ...600...52H,2010ApJ...713..131B}) as well as more recent results (\citealt{2013MNRAS.428.2198S,2011ApJ...727...52B,2014ApJ...795...98B,2015MNRAS.453.1268Z}).

We note that this is not a complete list of all previous absorption-line surveys -- in keeping with the similar plot presented in \citet{2010MNRAS.408..849G} we include only the quasar sightlines from the above literature results (apart from the \citet{2015MNRAS.453.1268Z} sample where we have included all sightlines in order to show the detections made in this survey). 
As we wish to compare our results to those of \citet{2010MNRAS.408..849G} (who conducted the only other recent systematic study of detection rate as a function of impact parameter) it is sensible to maintain this selection in compiling the literature results.

\citet{2010MNRAS.408..849G} estimate the detection rate of intervening \mbox{H\,{\sc i}} absorption to be approximately 50 per cent, for impact parameters less than 20 kpc and integrated optical depths $>$ 0.1 km s$^{-1}$. 
The addition of more recent surveys yields a similar overall detection rate. 
In the same parameter space our detection rate is less than 6 per cent (slightly higher than the detection rate for our full sample as some of the sightlines searched fall outside this region of the plot). 
The inclusion of our results thus brings the overall detection rate down to approximately 25 per cent. 
Given the large difference in detection rate between our work and previous surveys, it is clearly important to determine the reasons behind this, in order to better estimate the expected detection rate for future blind absorption-line surveys.

We suggest two possible explanations for differences in detection rate. 
Firstly, that the background radio sources in our sample are in general much fainter than those in other surveys, meaning we are not as sensitive to \mbox{H\,{\sc i}} absorption. 
The limiting flux for our sample is $S_{\mathrm{843}}$ $>$ 50 mJy (expected 1.4 GHz fluxes $\gtrsim$35 mJy), while many other surveys comprise sources $>$300-400 mJy (and often as bright as a few Jy). 
There are a number of cases in our sample where the sightline does intersect the \mbox{H\,{\sc i}} disc of the galaxy, but at sub-DLA column densities, which suggests that if we had conducted the same survey with a brighter sample of radio sources we would have expected to obtain a higher detection rate. 
The only other survey with a similar range of continuum fluxes is \citet{2011ApJ...727...52B}, which shows a detection rate of 2 in 23 (9 per cent), very similar to what we find.

A second, related reason is to do with the structure of the radio sources. 
As discussed in Paper I, the literature results shown are all surveys of quasar sightlines, whereas our own is an unbiased sample of radio sources (a mixture of radio galaxies and quasars, but predominantly radio galaxies). 
Since quasars provide very bright, compact radio sources, ideal for detecting \mbox{H\,{\sc i}} absorption against, while radio galaxies are more likely to be extended (reducing the absorption-line sensitivity, as we have seen), we suggest that differences in source type, and not just the intrinsic flux of the radio sources, may also be partially responsible for the observed differences in detection rate.

Our selection criteria much more closely match those of planned blind \mbox{H\,{\sc i}} absorption-line surveys (for example FLASH, which will search for absorption \emph{all} sightlines $S_{1.4}$ $>$ 50 mJy, regardless of source type). 
Therefore, while both of the above reasons essentially come down to the fact that the background source is not sufficiently bright, it is important to note that (because of the influence of source type) simply increasing the flux limit would likely not be sufficient to reproduce the $\sim$50 per cent detection rate seen in previous surveys. 
Similarly, if we wish to estimate the expected detection rate for future absorption-line surveys, basing this purely on studies of quasar sightlines (even if the flux limit is similar) is likely to over-estimate the true detection rate. 
Given this, we suggest our results are likely more representative of the expected detection rate for future blind surveys such as FLASH. 
From our results, we estimate that the expected detection rate (for an unbiased sample of radio sources) probably does not exceed about 20 per cent, for radio sources $\gtrsim$ 50 mJy and impact parameters less than 20 kpc, and may even be significantly lower than this ($<$5-10 per cent).

\section{Conclusions}
\label{conclusions}

We have conducted a survey of \mbox{H\,{\sc i}} absorption along 17 sightlines in 10 nearby galaxies, having impact parameters of between 6 and 46 kpc. 
These results add to our previous work \citep{2015MNRAS.450..926R} which searched 6 sightlines, giving a full sample of 23 sightlines.

In our sample we detected one intervening \mbox{H\,{\sc i}} absorption-line in the galaxy NGC\,5156 at an impact parameter of 19 kpc. 
This is the highest impact parameter detection of an intervening \mbox{H\,{\sc i}} absorption-line to date. 
The absorption-line line is deep and narrow, having a width of just 7.6 km s$^{-1}$ and an integrated optical depth of 0.82 km s$^{-1}$. 
High resolution ATCA images at 5 and 8 GHz from the ATPMN survey reveal that the background source is resolved into two separate components with a separation of 2.6 arcsec, which corresponds to a physical size of 500 pc at the redshift of the galaxy ($z = 0.01$). 
The narrowness of the line suggests that the absorption is mostly likely occurring against a single component of the background source. 
VLBI spectroscopy would allow us to confirm this, as well as providing an accurate measurement of the filling factor and spin temperature of the gas.

By targeting nearby galaxies we have also been able to map the extended \mbox{H\,{\sc i}} emission of all the galaxies in our sample, allowing us to directly relate the gas distribution to the absorption-line detection rate. 
In our sample we found that sightlines not intersecting the \mbox{H\,{\sc i}} disc accounted for the majority of the non-detections (14/23). 
The remaining non-detections were found to be because either the continuum source was too faint (5/23) or $T_{\mathrm{S}}/f$ was much higher than expected (3/23). 
In the cases where the continuum flux was too low, this was almost always due to the background sources becoming extended or resolved at higher resolution, reducing the absorption-line sensitivity.

Our detection rate (4.3 per cent) is much lower than previous surveys and we suggest that this is either the result of (i) the lower continuum fluxes in our sample or (ii)  differences in source type (radio galaxies versus quasars) meaning the sources in our sample are far more likely to be extended, reducing the absorption-line sensitivity (or some combination of these effects). 
Since our selection criteria (both in flux limit and source type) much more closely match those of large planned absorption-line surveys, such as FLASH, we suggest that our results may be more representative of the expected detection rate for such surveys. 
Based on our results, we estimate that, for an unbiased sample of radio sources, the expected detection rate is probably not greater than about 10-20 per cent for sources $\gtrsim$ 50 mJy and impact parameters less than 20 kpc. 

Given the above results, we suggest that further investigations should be conducted into (i) optical/\mbox{H\,{\sc i}} scaling relations (to help predict which sightlines will intersect the \mbox{H\,{\sc i}} disc) and (ii) proxies for estimating source compactness (to help identify the sources most likely to detect absorption against). 
\mbox{H\,{\sc i}} emission and absorption-line observations of nearby galaxies (as well as VLBI observations) to help estimate the typical spin temperature and covering factor in galaxy discs would also be worthwhile. 
These measurements will help us to better prepare for, and anticipate the results of future large-scale \mbox{H\,{\sc i}} absorption-line surveys with next-generation radio telescopes. 
However, the detection of a deep absorption-line in our sample highlights the discovery potential for future, absorption-line surveys like FLASH, which will target large numbers of Jy-level sources. 
Surveys like FLASH can be expected to make many new absorption-line detections, and provide an exciting new insight into the evolution of neutral gas in the distant Universe.

\section{Acknowledgements}
\label{acknowledgements}

JRA acknowledges support from a CSIRO Bolton Fellowship. 
The Australia Telescope and Parkes radio telescope are funded by the Commonwealth of Australia for operation as a 
National Facility managed by CSIRO.
The Centre for All-sky Astrophysics is an Australian Research Council Centre of Excellence, 
funded by grant CE110001020. 
This research has made use of the following resources: the NASA/IPAC Extragalactic Database (NED) which is operated by the Jet Propulsion Laboratory, California Institute of Technology, under contract with the National Aeronautics and Space Administration;  the SIMBAD database, operated at CDS, Strasbourg, France; NASAs Astrophysics Data System Bibliographic Services; APLpy, an open-source plotting package for Python hosted at http://aplpy.github.com, and Astropy, a community-developed core Python package for Astronomy (Astropy Collaboration, 2013). 

\bibliographystyle{mn2e_edit} 
\bibliography{bibliography}

\appendix

\section[]{Figures}
\label{appendix:figures}

In this appendix we present \mbox{H\,{\sc i}} moment maps (\mbox{H\,{\sc i}} total intensity, mean velocity and velocity dispersion) of the target galaxies (Figure \ref{figure:moment_maps}) and the 1.4 GHz continuum images of the background radio sources and the spectra extracted towards each source (Figure \ref{figure:continuum_and_spectra}). 
As these figures span multiple pages, placing them in the main text reduced the readability of the paper, so for convenience we have collated them in an appendix instead.

\begin{figure*}
\includegraphics[width=\linewidth]{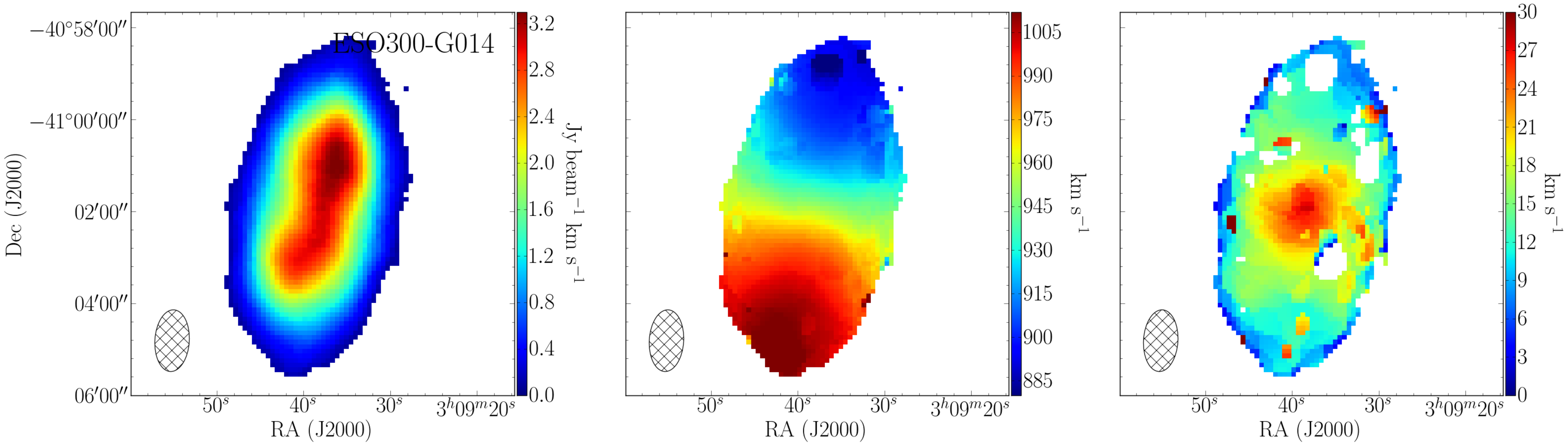}
\includegraphics[width=\linewidth]{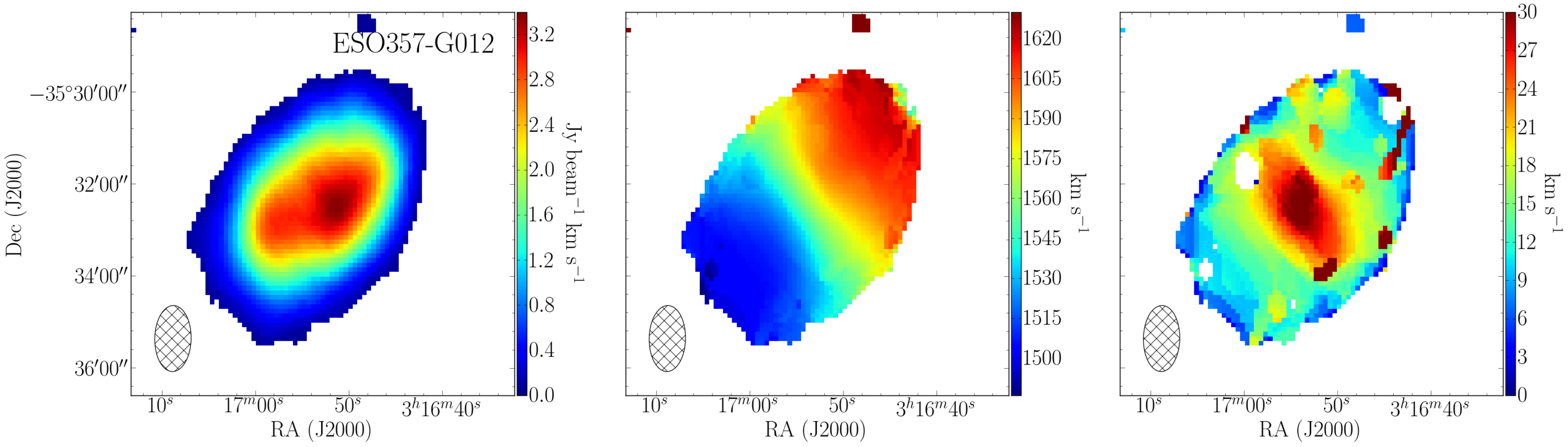}
\includegraphics[width=\linewidth]{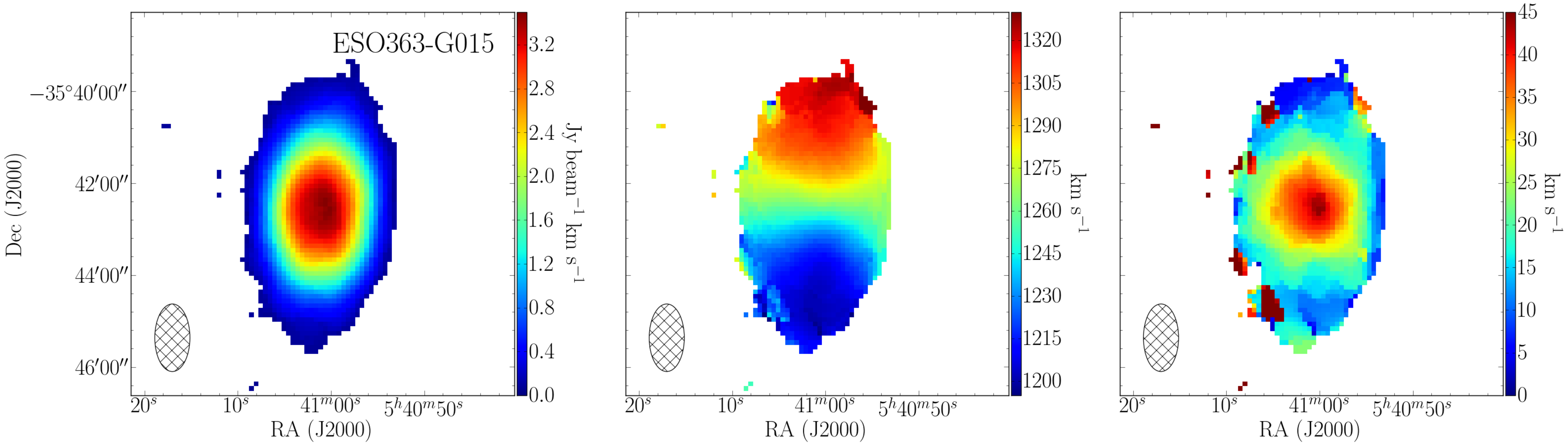}
\includegraphics[width=\linewidth]{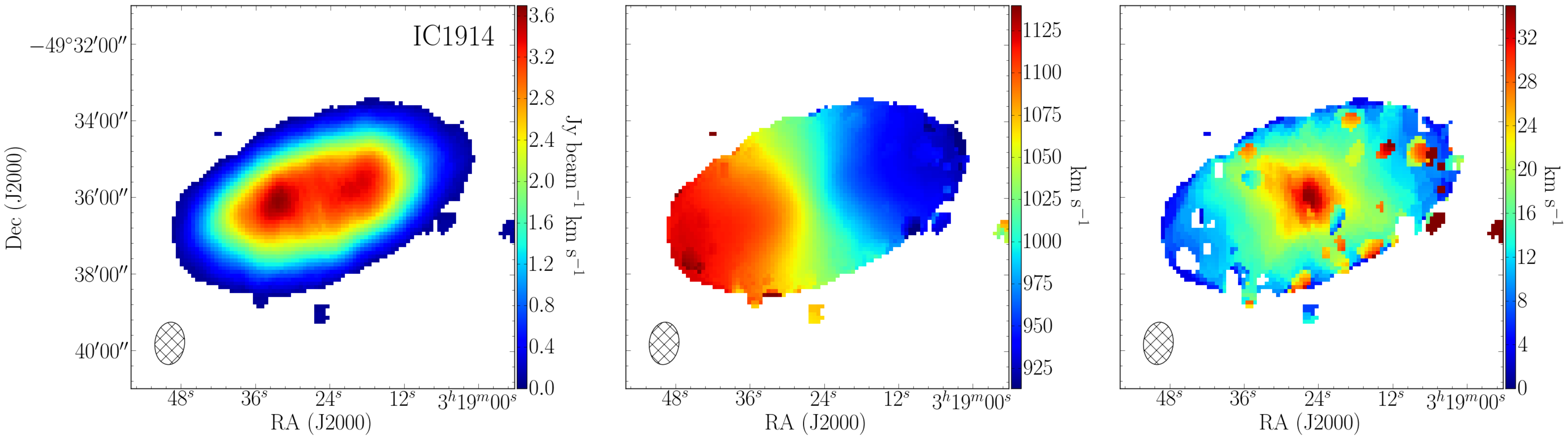}
\caption[]{\mbox{H\,{\sc i}} moment maps of the target galaxies (produced from the low resolution cubes). 
Left to right: \mbox{H\,{\sc i}} total intensity (zeroth moment), mean \mbox{H\,{\sc i}} velocity (first moment), and \mbox{H\,{\sc i}} velocity dispersion (second moment) maps.
The synthesised beam is shown in the bottom left corner of each image.}
\label{figure:moment_maps}
\end{figure*}

\begin{figure*}
\includegraphics[width=\linewidth]{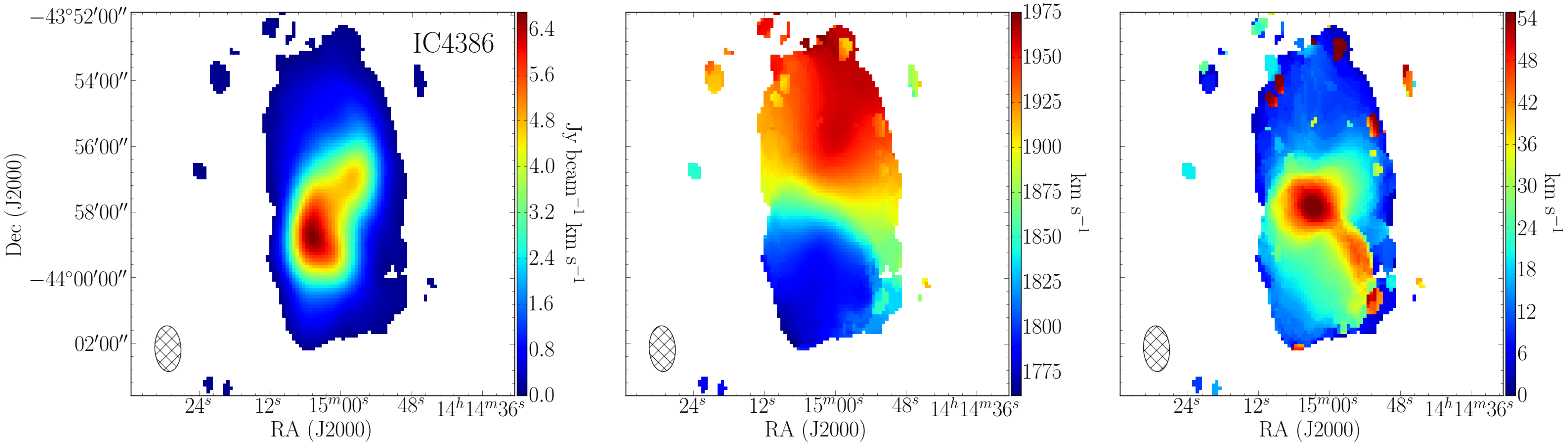}
\includegraphics[width=\linewidth]{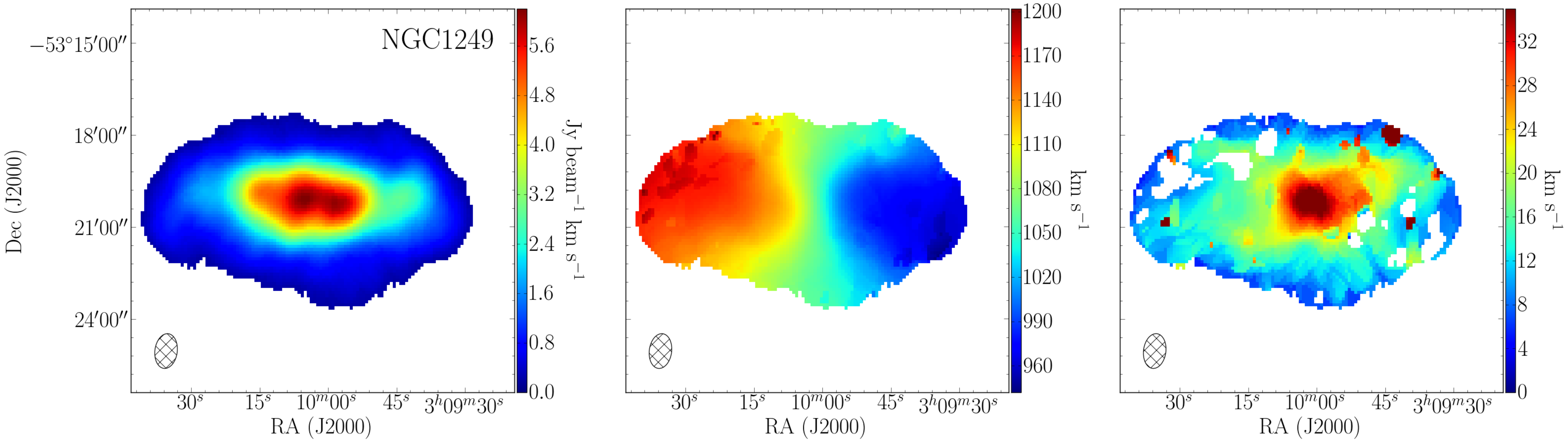}
\includegraphics[width=\linewidth]{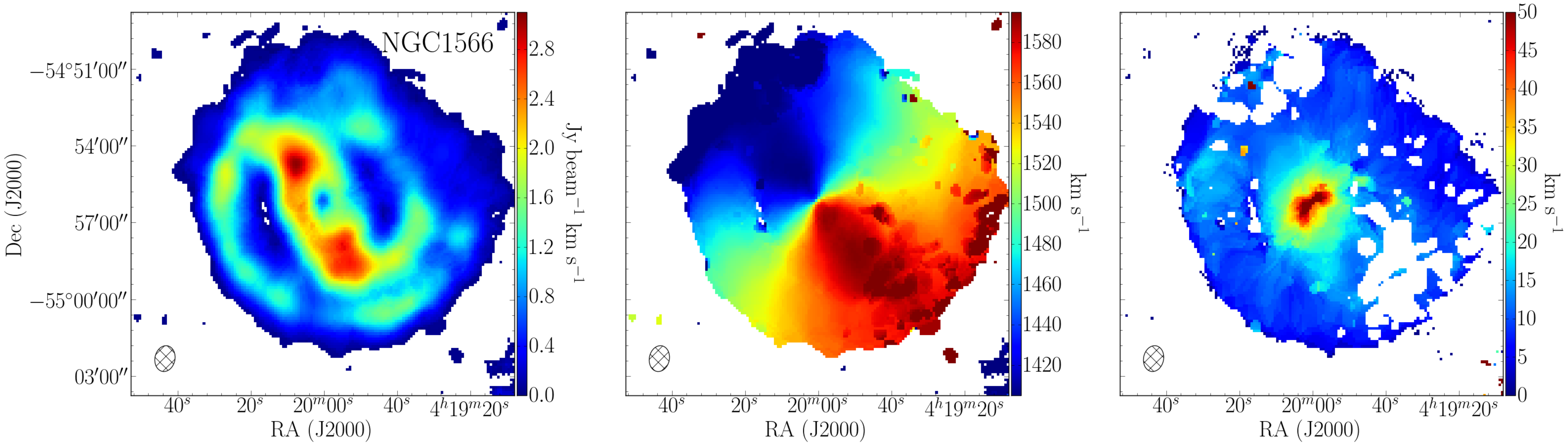}
\includegraphics[width=\linewidth]{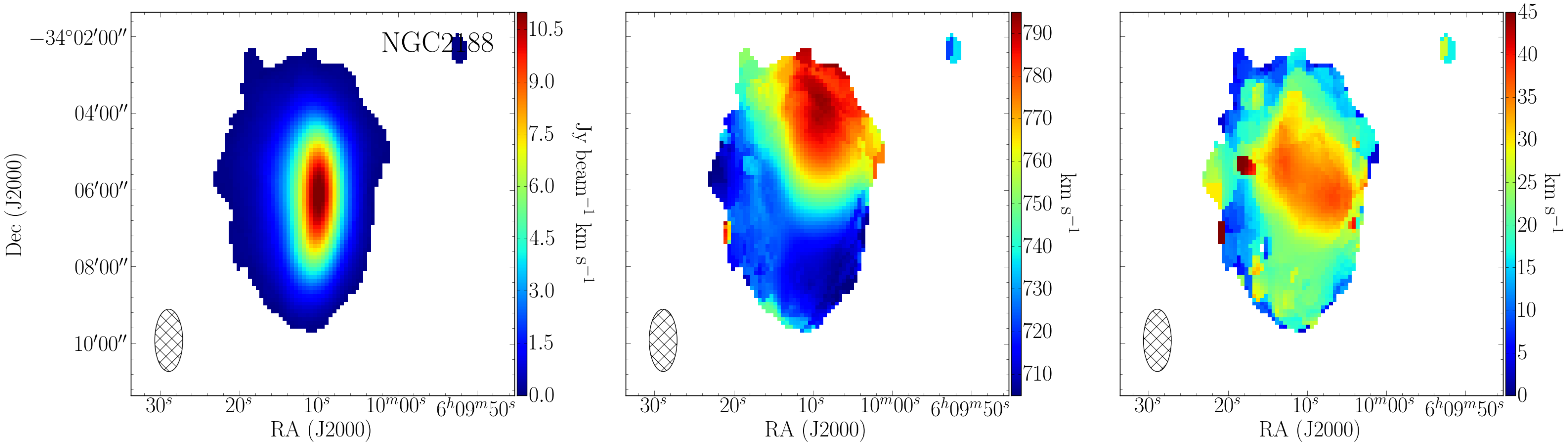}
\contcaption{}
\end{figure*}

\begin{figure*}
\includegraphics[width=\linewidth]{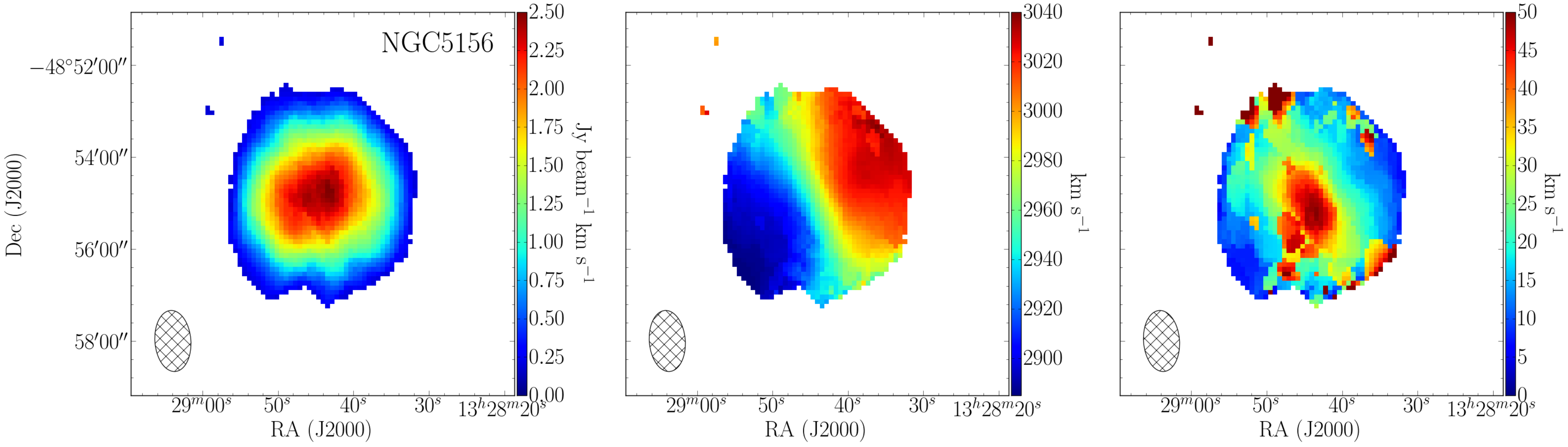}
\includegraphics[width=\linewidth]{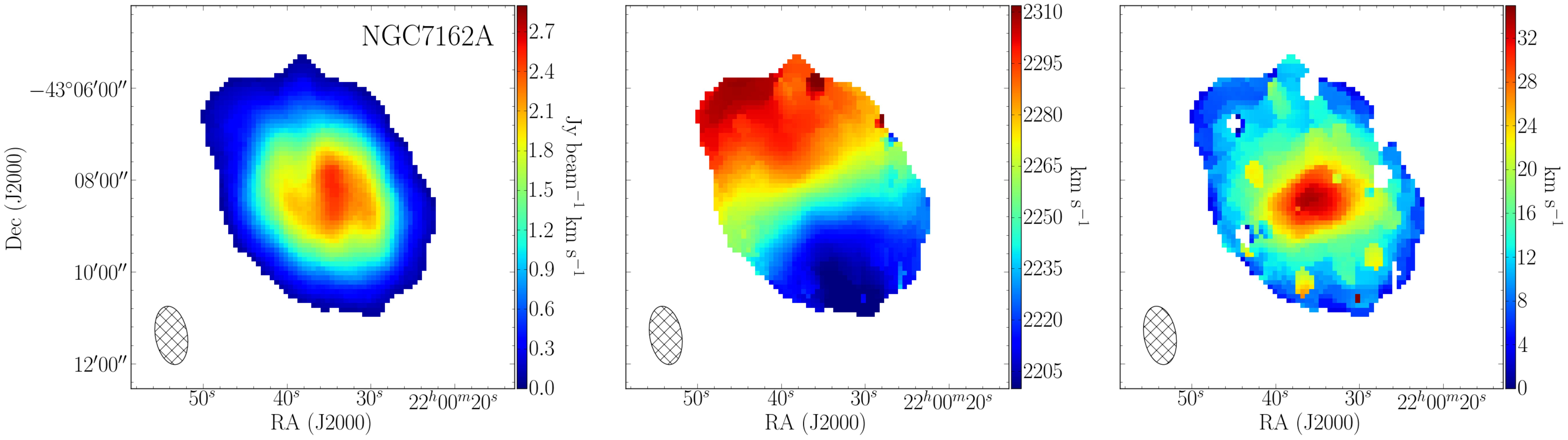}
\contcaption{}
\end{figure*}

\begin{figure*}
\includegraphics[width=\linewidth]{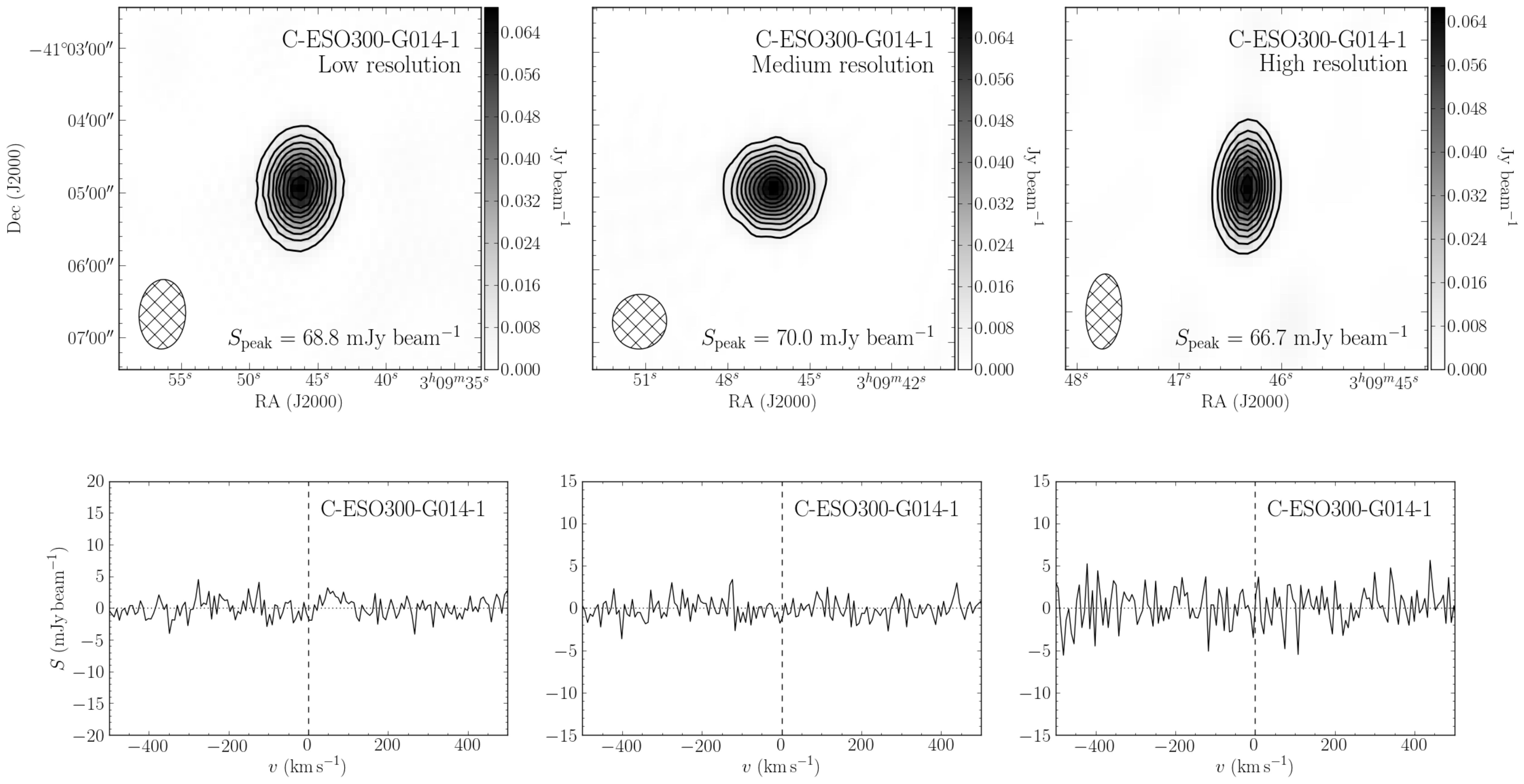}
\includegraphics[width=\linewidth]{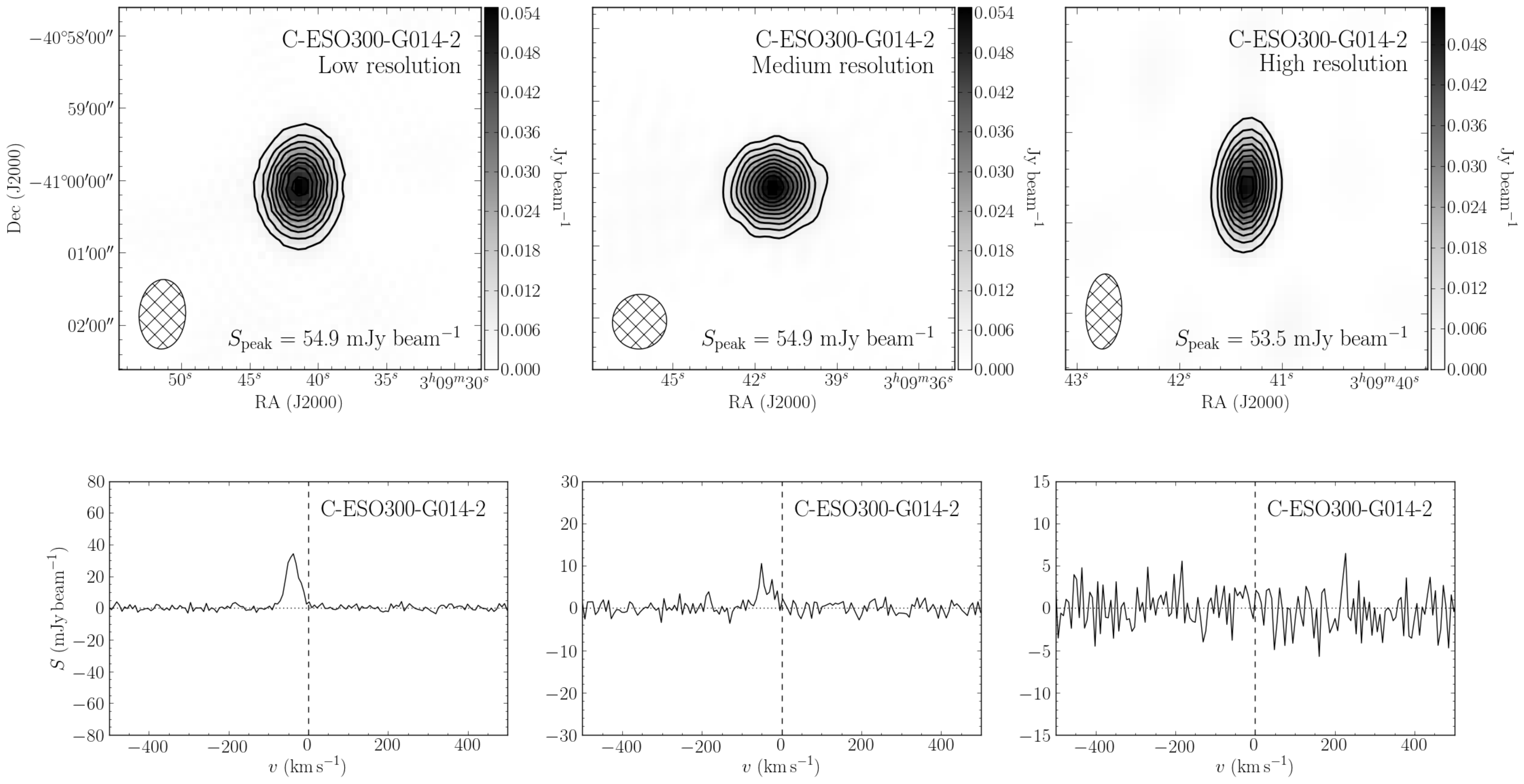}
\caption[]{ATCA 1.4 GHz continuum images of the background radio sources, with the spectrum along that sightline below it (left to right: low, medium, and high resolution). 
All spectra have been shifted to the rest-frame of the galaxy, with the velocity axis expressed with respect to the galaxy systemic velocity. 
The dotted line represents the approximate expected 21 cm line position. 
We note that, in general, the continuum images are not shown on the same scale, to allow the structure of the sources to be seen as the resolution increases. 
The radio contours start at 90 per cent of the peak flux and decrease in 10 per cent increments, and the peak flux is given in the bottom right corner. 
The synthesised beam is shown in the bottom left corner of each continuum image.}
\label{figure:continuum_and_spectra}
\end{figure*}

\begin{figure*}
\includegraphics[width=\linewidth]{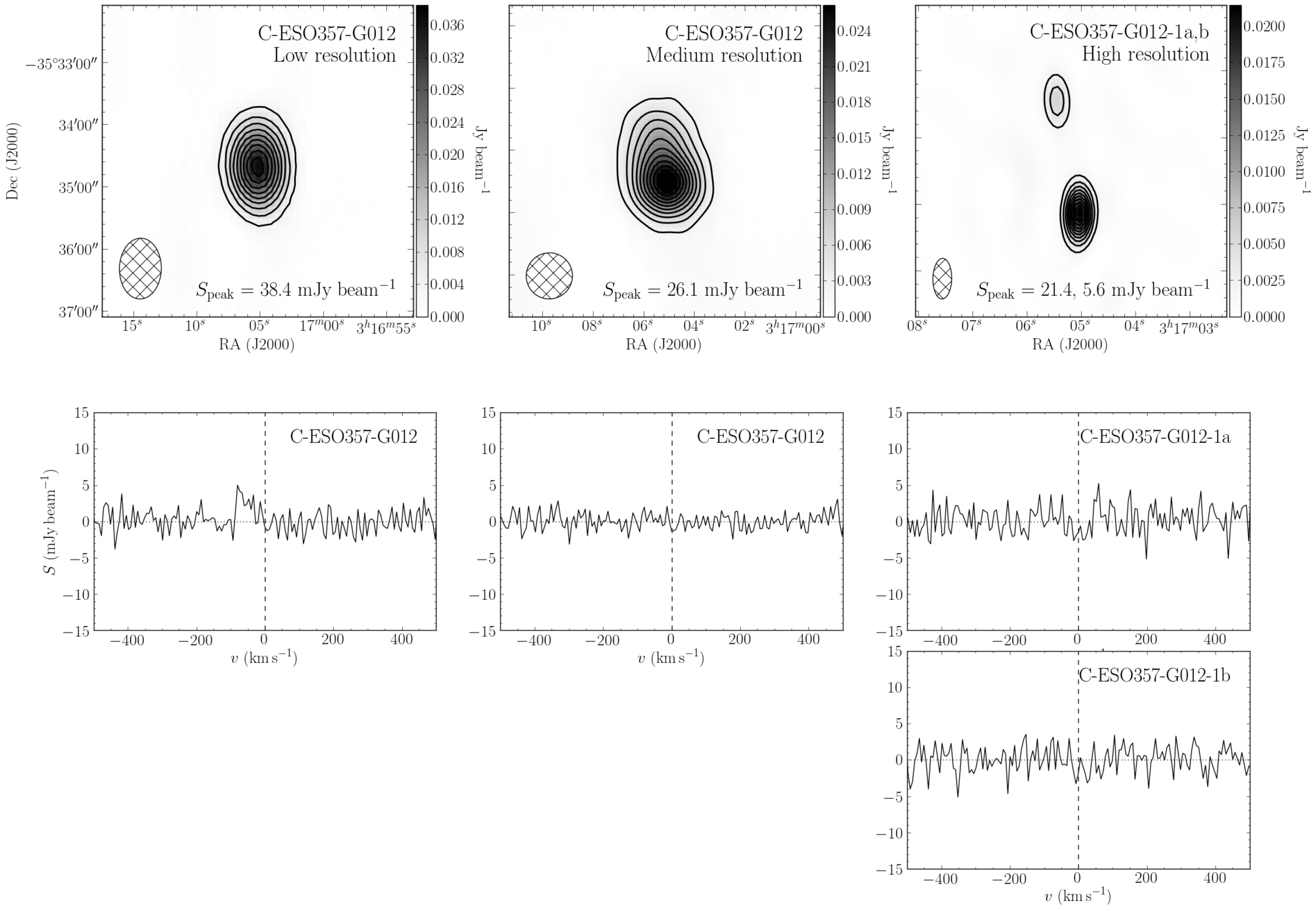}
\contcaption{}
\end{figure*}

\begin{figure*}
\includegraphics[width=\linewidth]{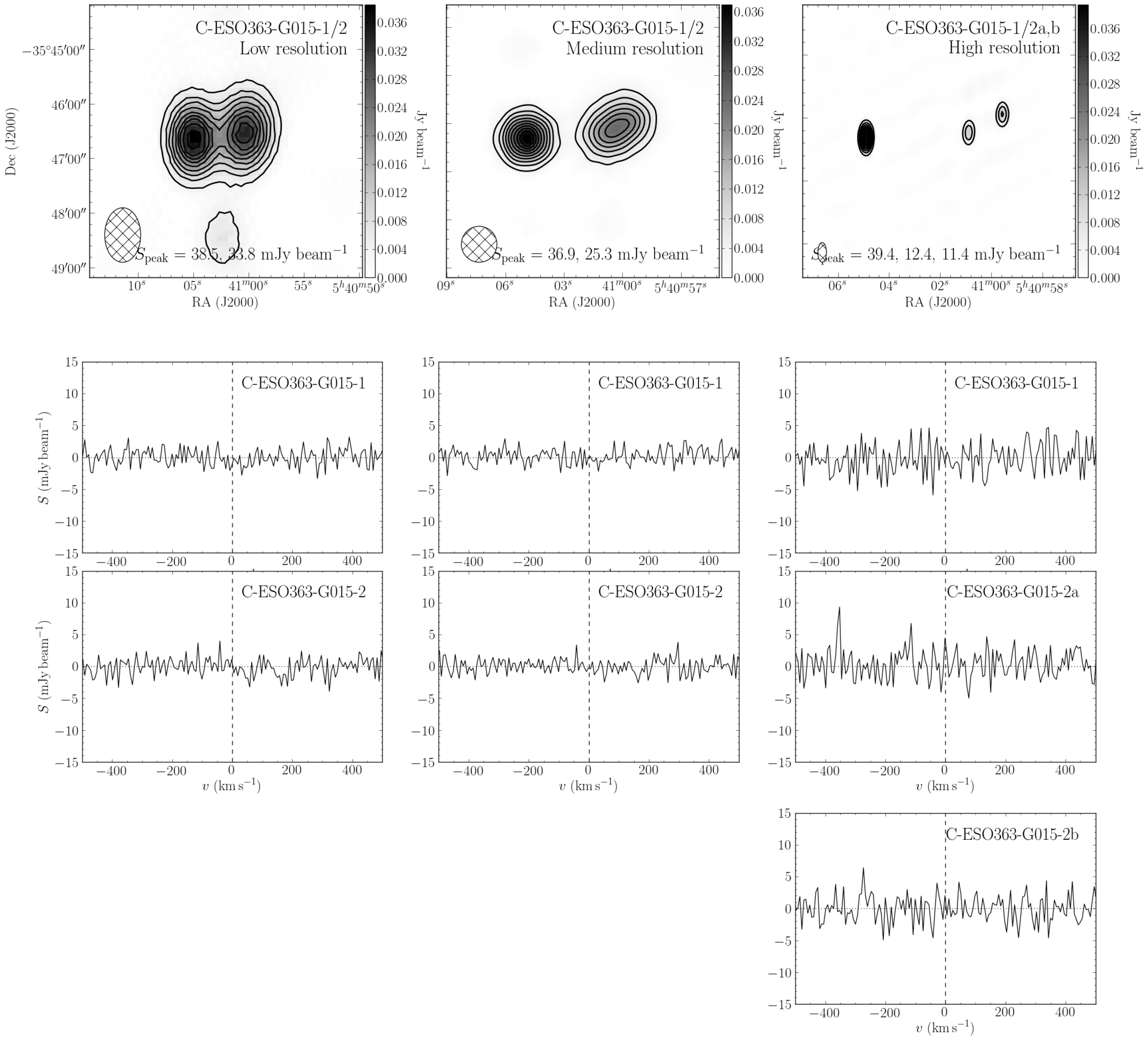}
\contcaption{}
\end{figure*}

\begin{figure*}
\includegraphics[width=\linewidth]{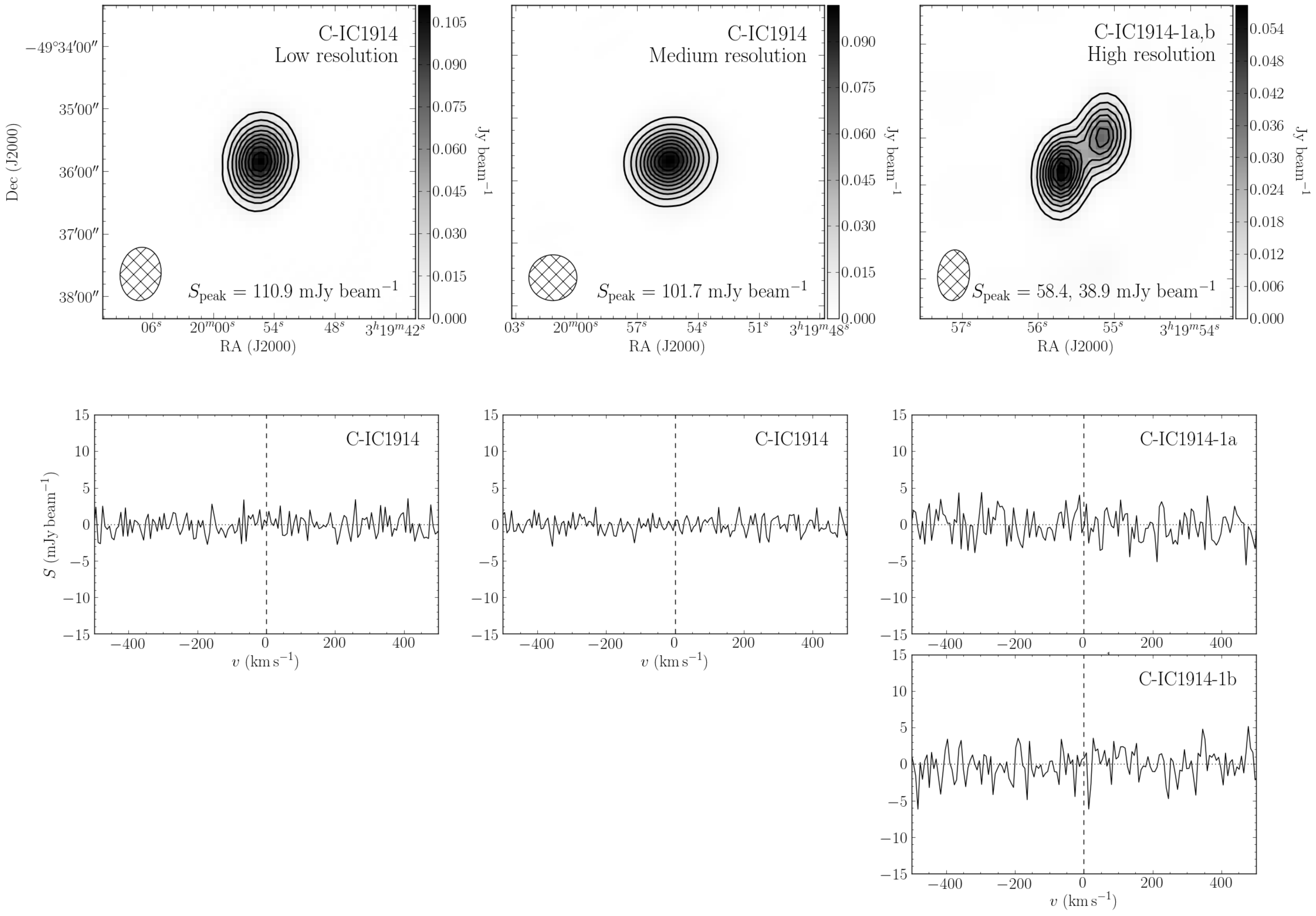}
\includegraphics[width=\linewidth]{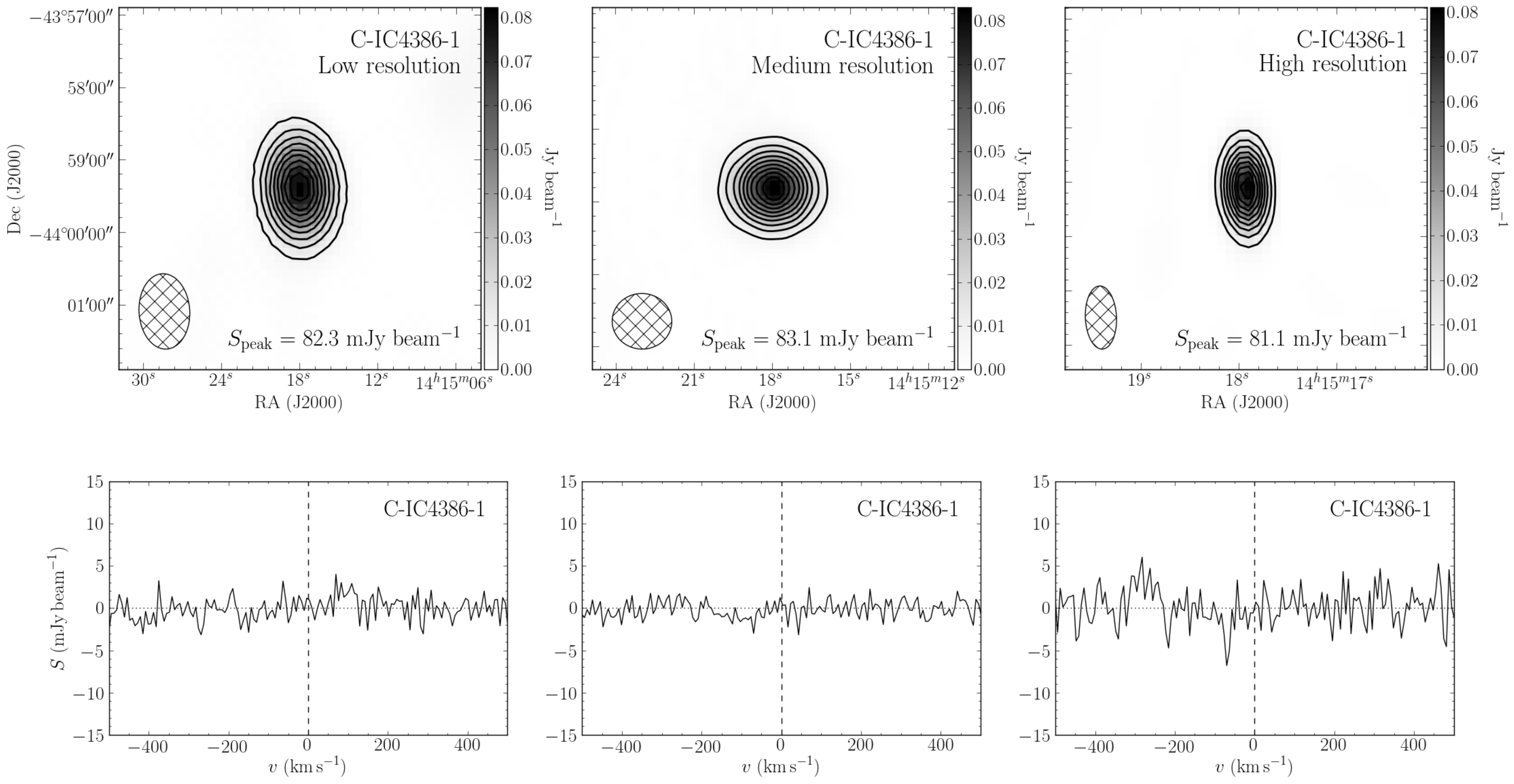}
\contcaption{}
\end{figure*}

\begin{figure*}
\includegraphics[width=\linewidth]{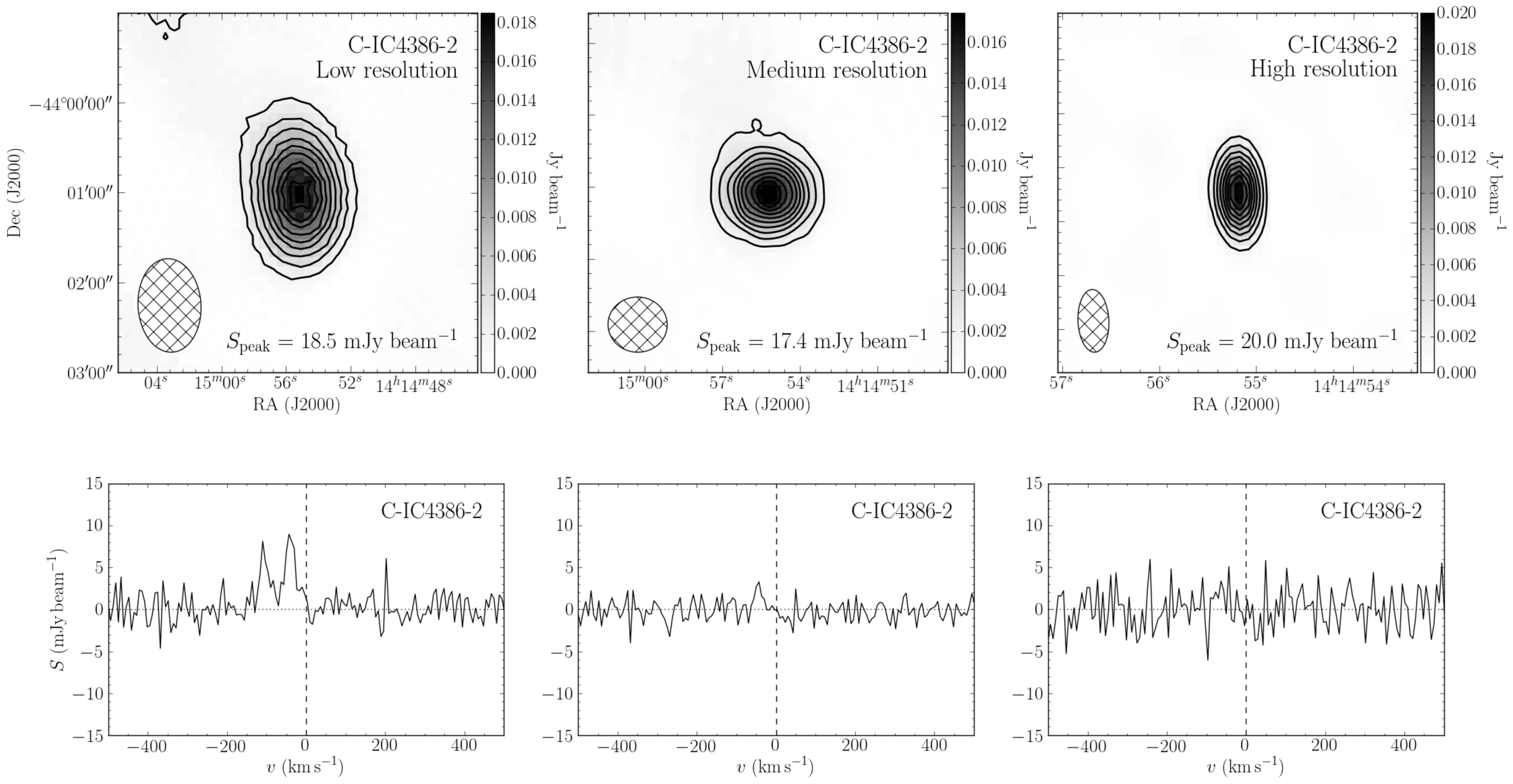}
\includegraphics[width=\linewidth]{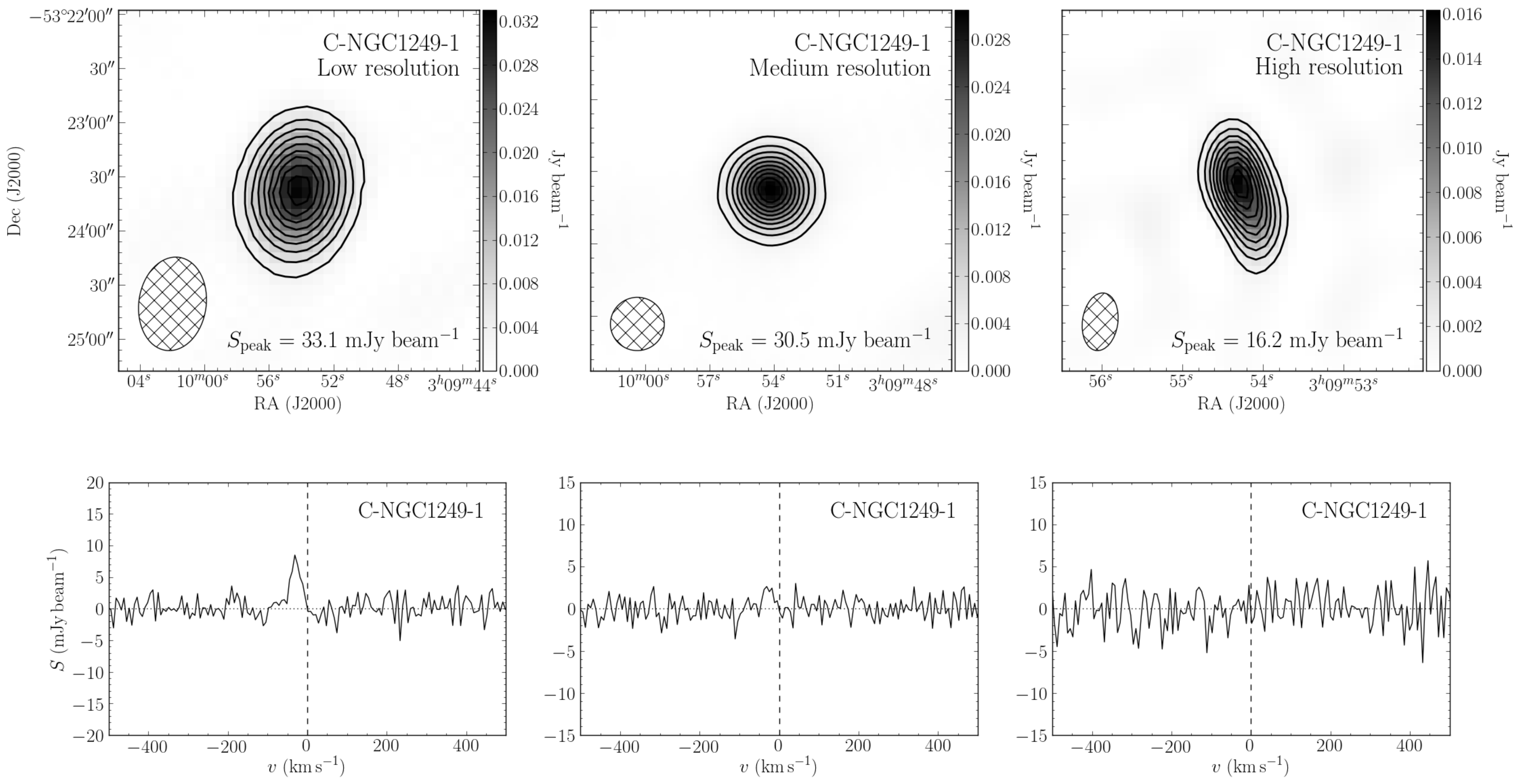}
\contcaption{}
\end{figure*}

\begin{figure*}
\includegraphics[width=\linewidth]{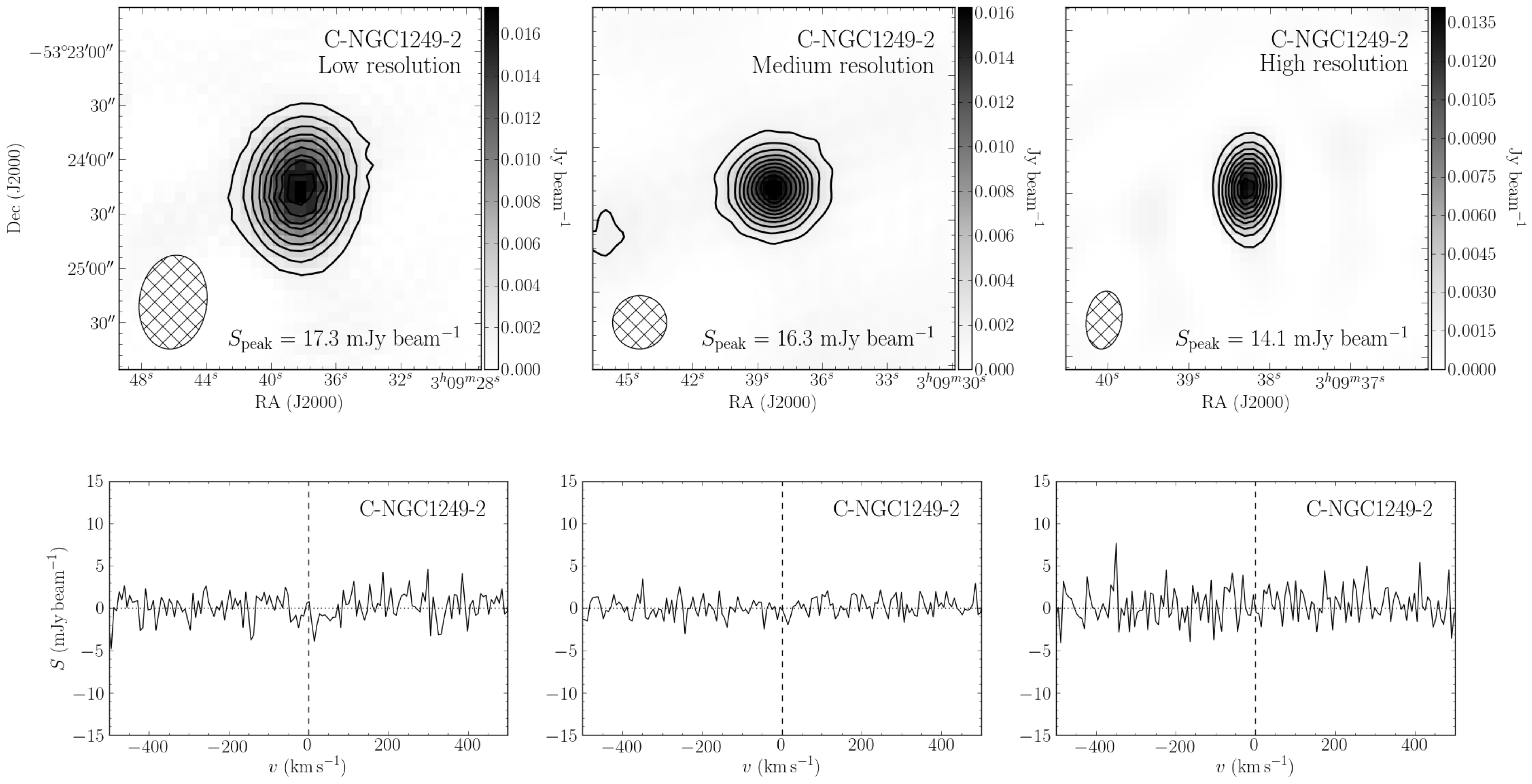}
\includegraphics[width=\linewidth]{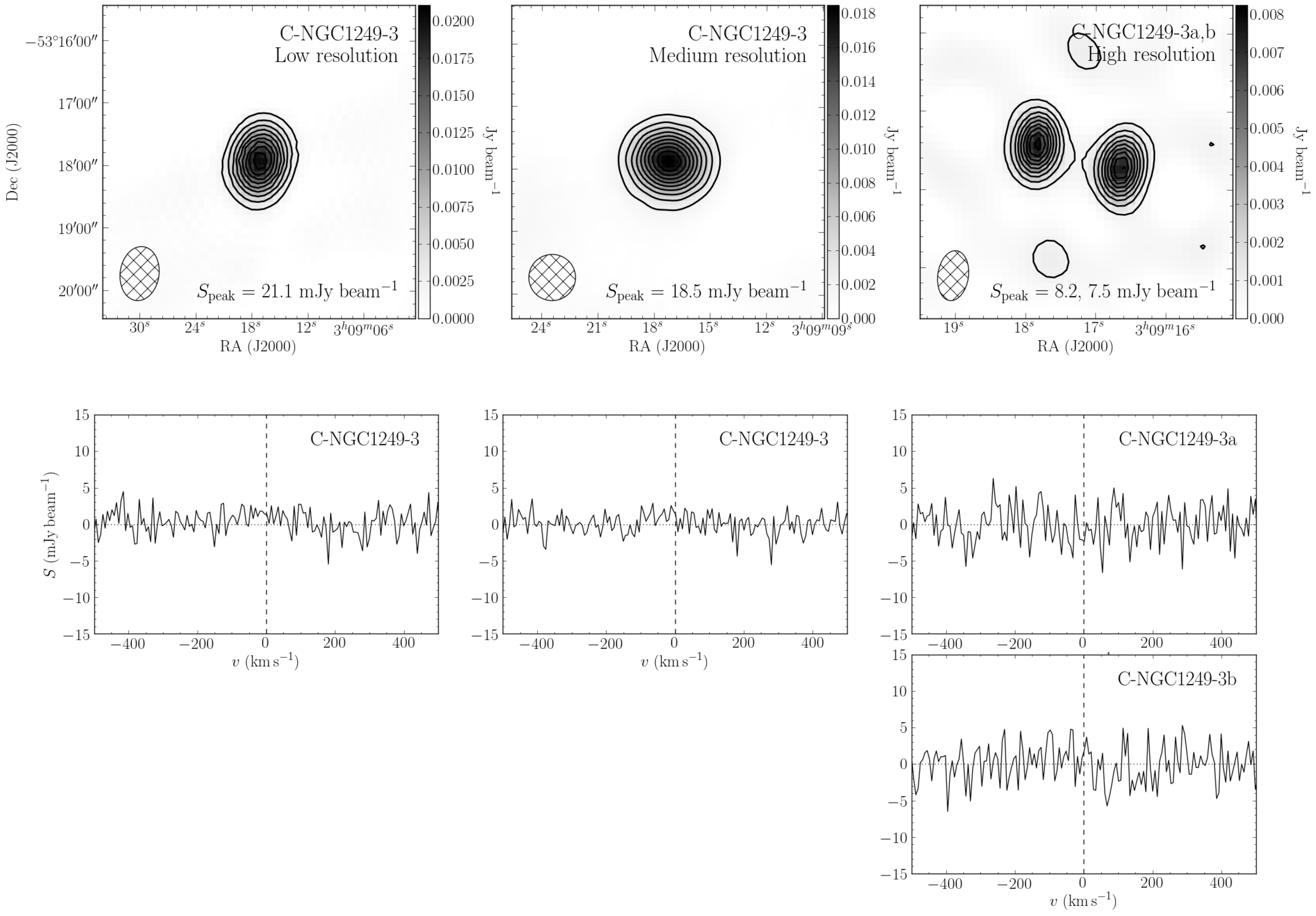}
\contcaption{}
\end{figure*}

\begin{figure*}
\includegraphics[width=\linewidth]{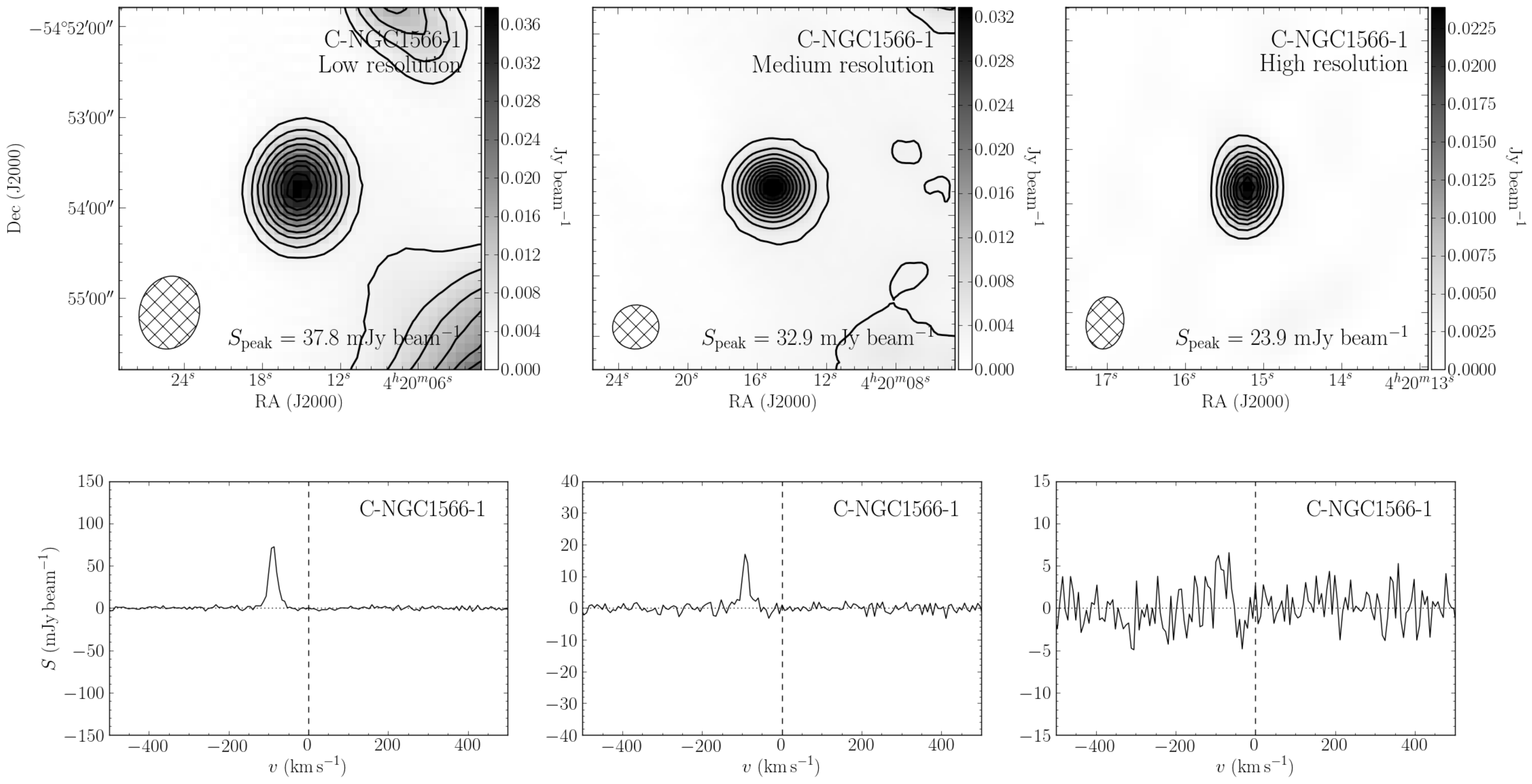}
\includegraphics[width=\linewidth]{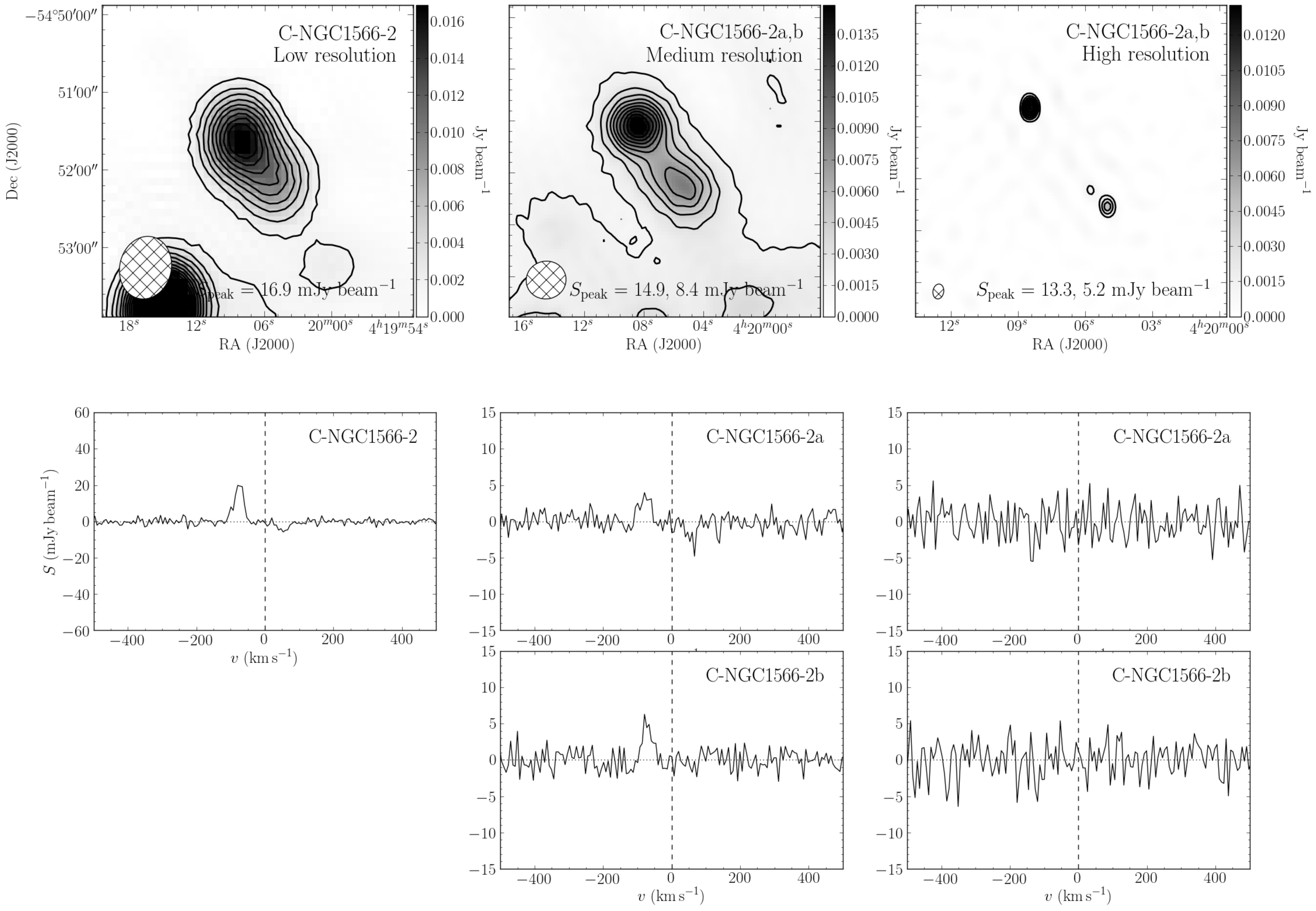}
\contcaption{}
\end{figure*}

\begin{figure*}
\includegraphics[width=\linewidth]{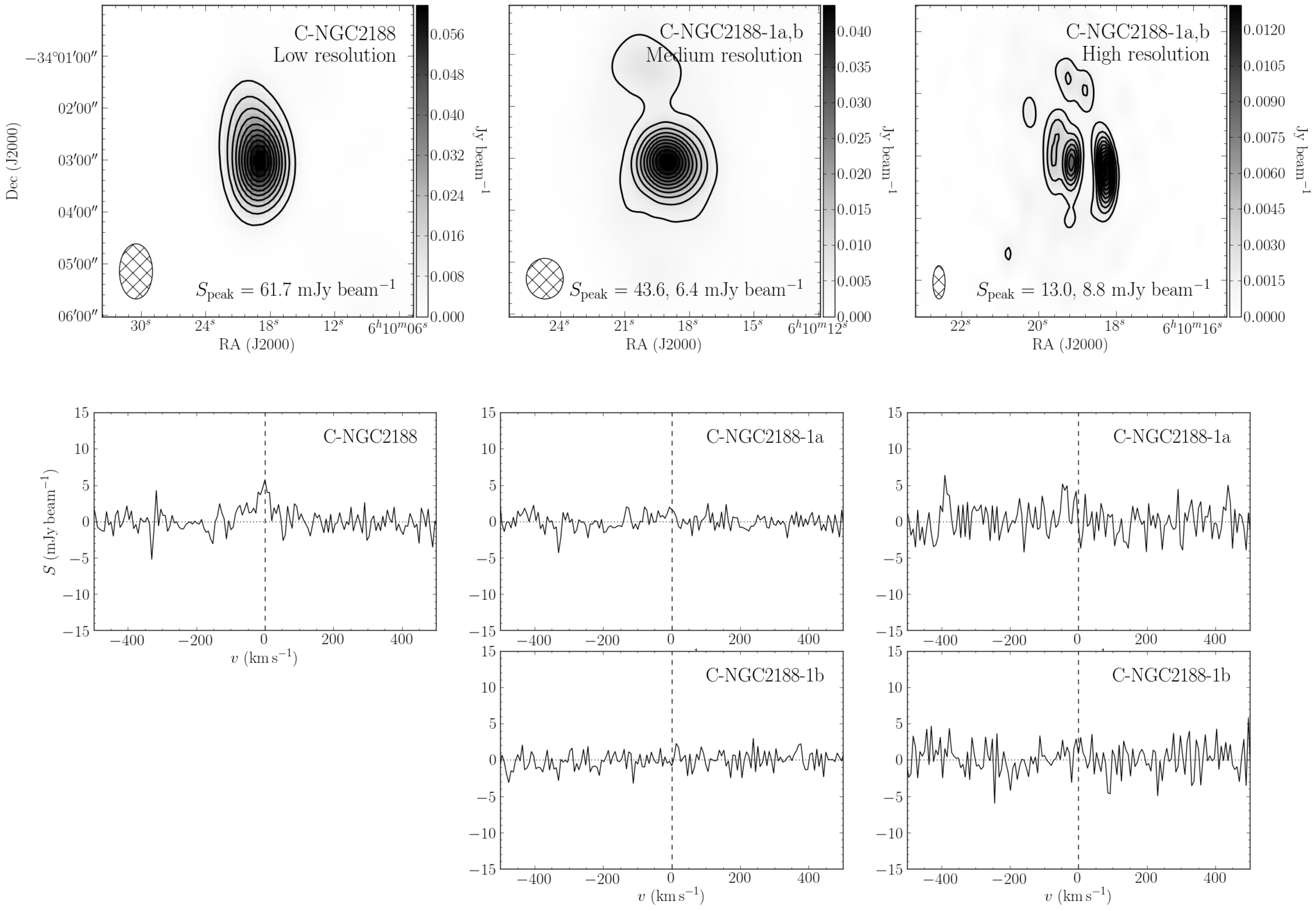}
\includegraphics[width=\linewidth]{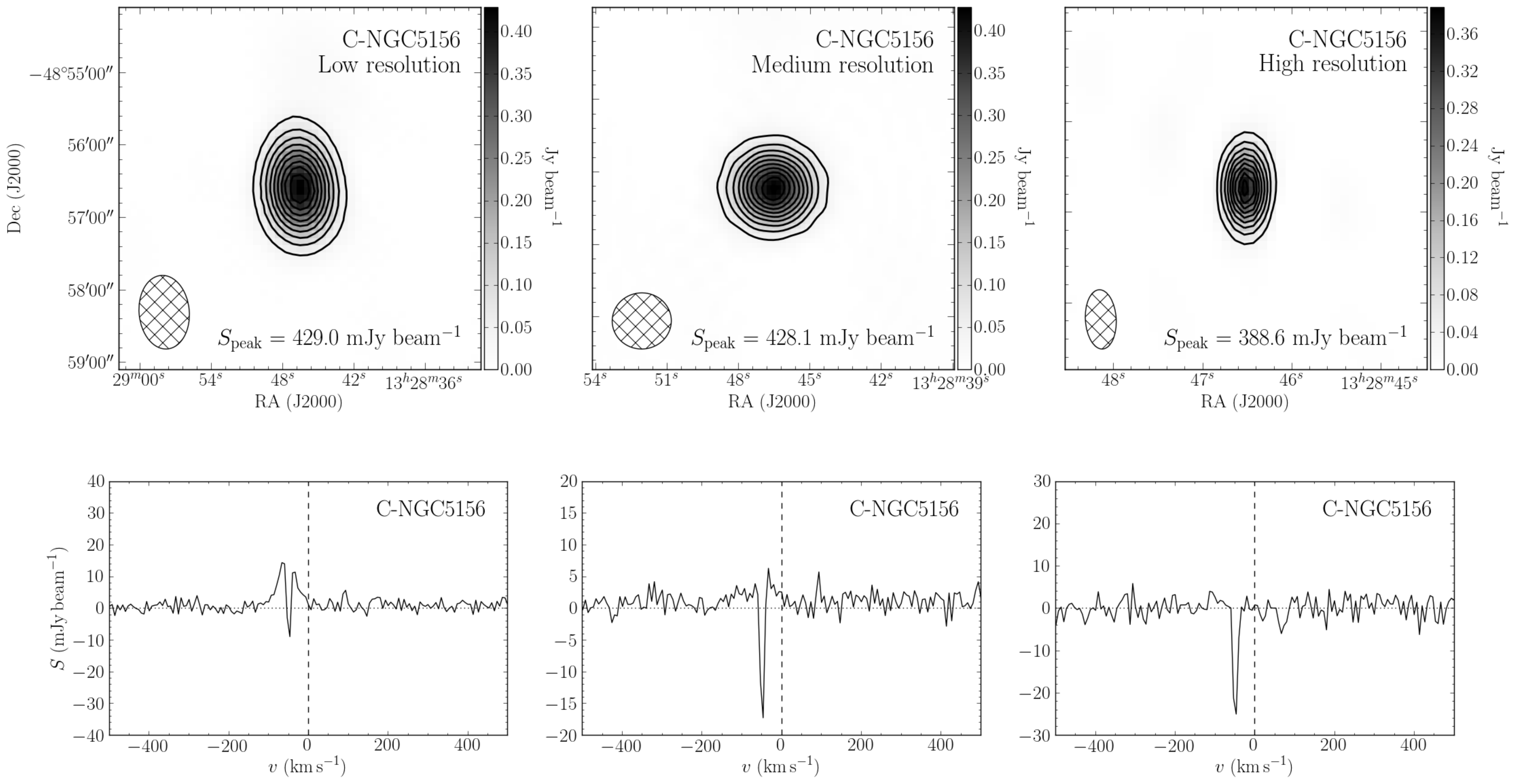}
\contcaption{}
\end{figure*}

\begin{figure*}
\includegraphics[width=\linewidth]{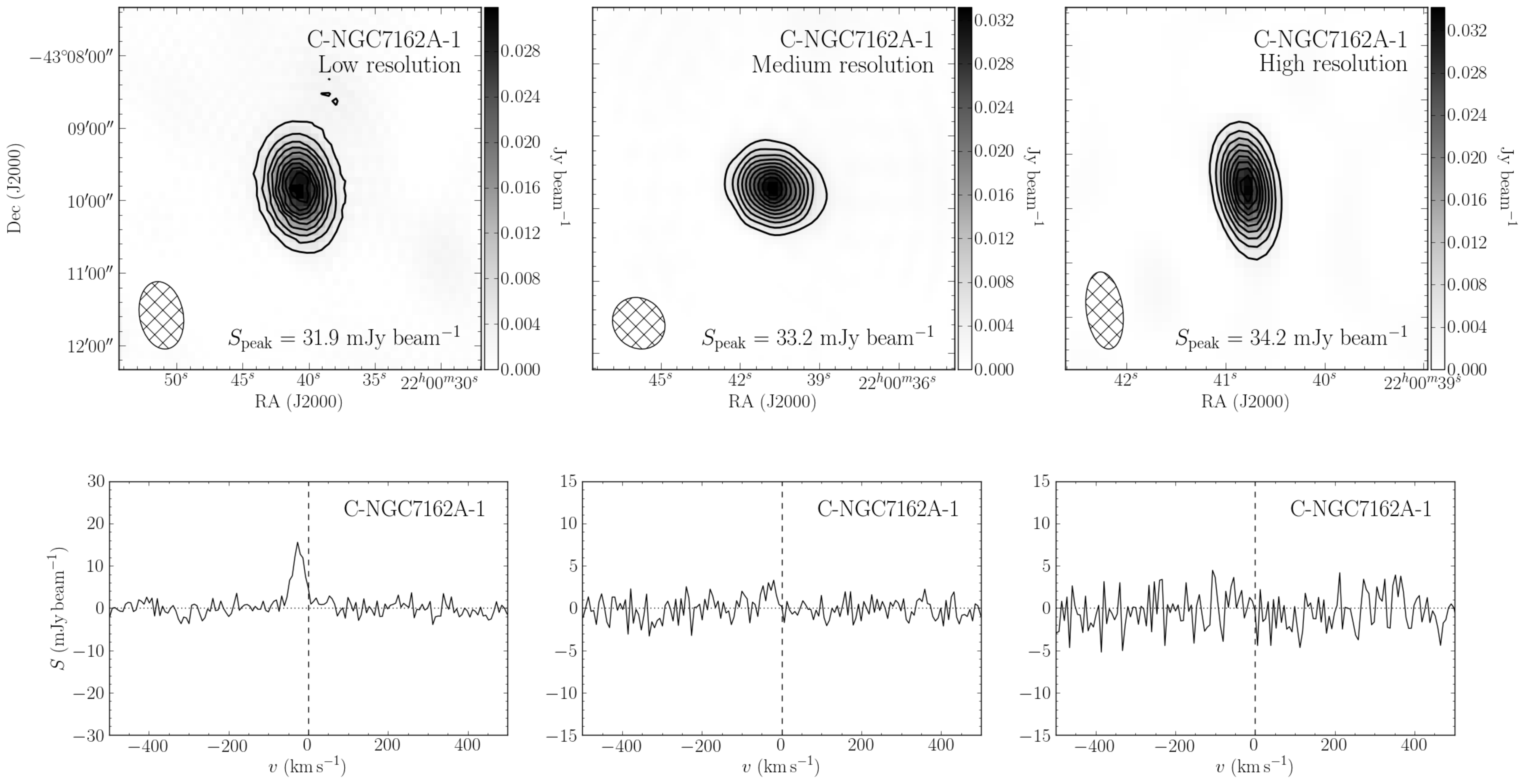}
\includegraphics[width=\linewidth]{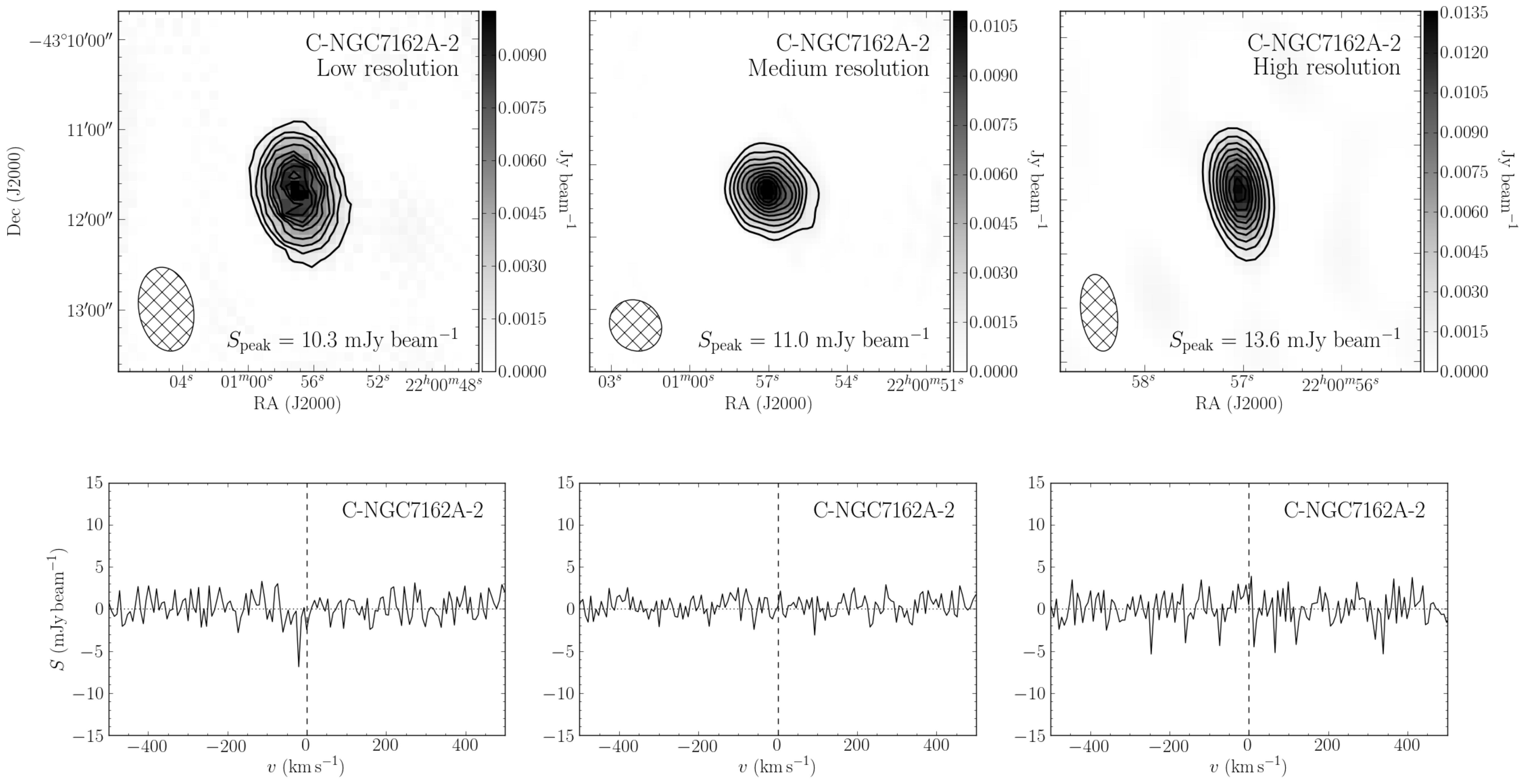}
\contcaption{}
\end{figure*}

\bsp

\label{lastpage}

\end{document}